\documentclass[fleqn,usenatbib]{mnras}

\usepackage{newtxtext,newtxmath}

\usepackage[T1]{fontenc}

\DeclareRobustCommand{\VAN}[3]{#2}
\let\VANthebibliography\thebibliography
\def\thebibliography{\DeclareRobustCommand{\VAN}[3]{##3}\VANthebibliography}

\usepackage{graphicx}	
\usepackage{amsmath}	
\usepackage{xspace}

\usepackage{comment}

\usepackage{adjustbox}  
\usepackage{bm}

\makeatletter 
  \patchcmd{\NAT@citex}
    {\@citea\NAT@hyper@{%
      \NAT@nmfmt{\NAT@nm}%
      \hyper@natlinkbreak{\NAT@aysep\NAT@spacechar}{\@citeb\@extra@b@citeb}%
      \NAT@date}}
    {\@citea\NAT@nmfmt{\NAT@nm}%
    \NAT@aysep\NAT@spacechar\NAT@hyper@{\NAT@date}}{}{}

  \patchcmd{\NAT@citex}
    {\@citea\NAT@hyper@{%
      \NAT@nmfmt{\NAT@nm}%
      \hyper@natlinkbreak{\NAT@spacechar\NAT@@open\if*#1*\else#1\NAT@spacechar\fi}%
        {\@citeb\@extra@b@citeb}%
      \NAT@date}}
    {\@citea\NAT@nmfmt{\NAT@nm}%
    \NAT@spacechar\NAT@@open\if*#1*\else#1\NAT@spacechar\fi\NAT@hyper@{\NAT@date}}
    {}{}
\makeatother

\newcommand{\msun}{{\,\rm M_\odot}}
\newcommand{\Lsun}{{\,\rm L_\odot}}
\newcommand{\kms}{\,{\rm km}\,{\rm s}^{-1}}
\newcommand{\cm}{\,{\rm cm}}
\newcommand{\erg}{\,{\rm erg}}

\newcommand{\Myr}{\,{\rm Myr}}

\newcommand{\pkpc}{\,{\rm pkpc}}
\newcommand{\Mpc}{\,{\rm Mpc}}

\newcommand{\mmag}{\,{\rm mag}}
\newcommand{\kev}{\,{\rm keV}}
\newcommand{\keV}{\,{\rm keV}}

\newcommand{\Hz}{\,{\rm Hz}}

\newcommand{\luminasim}{\textsc{Lumina}\xspace}

\newcommand{\arepo}{\textsc{Arepo}\xspace}
\newcommand{\areport}{\textsc{Arepo-rt}\xspace}
\newcommand{\illustris}{Illustris\xspace}
\newcommand{\illustrisTNG}{IllustrisTNG\xspace}
\newcommand{\tng}{\textsc{TNG100}\xspace}
\newcommand{\mtng}{\textsc{Mtng}\xspace}

\newcommand{\HI}{\ion{H}{I}\xspace}

\newcommand{\HeI}{\ion{He}{I}\xspace}
\newcommand{\HeII}{\ion{He}{II}\xspace}
\newcommand{\HeIII}{\ion{He}{III}\xspace}

\newcommand\orcid[1]{\href{https://orcid.org/#1}{\adjustbox{trim={-.15\width} 0 {-.15\width} 0,clip}{\includegraphics[height=9pt]{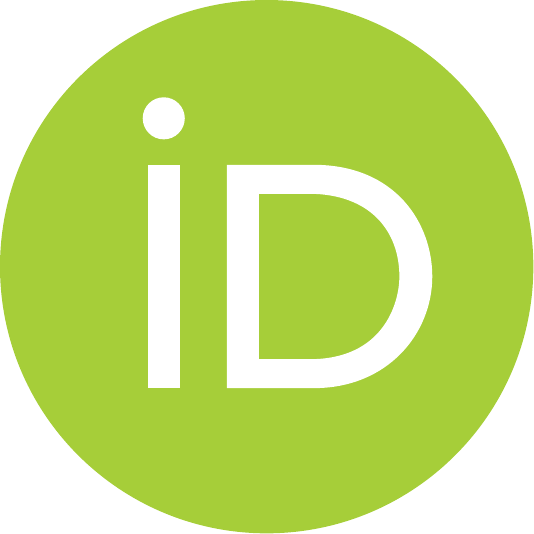}}}}

\def\aap{A\&A}

\def\apjl{ApJ}

\def\apjs{ApJS}

\title[AGN luminosity functions in \luminasim]{The \luminasim Project: The Demographics of Active Galactic Nuclei from Quasars to Little Red Dots at $z\geq 3$}

\author[Shen et al.]{\parbox{17.5cm}{
Xuejian Shen\orcid{0000-0002-6196-823X},$^{1}$\thanks{E-mail: \href{mailto:xuejian@mit.edu}{xuejian@mit.edu}}
Oliver Zier\orcid{0000-0003-1811-8915},$^{2}$
Aaron Smith\orcid{0000-0002-2838-9033},$^{3}$
Rongrong Liu\orcid{0000-0003-0685-3525},$^{2}$
Rahul Kannan\orcid{0000-0001-6092-2187},$^{4}$
Teodora-Elena Bulichi\orcid{0000-0001-8174-6389},$^{1}$
Sonja M. Koehler\orcid{0009-0008-0814-3328},$^{2}$ 
Volker Springel\orcid{0000-0001-5976-4599},$^{5}$ 
Mark Vogelsberger\orcid{0000-0001-8593-7692},$^{1,6}$
Lars Hernquist\orcid{0000-0001-6950-1629},$^{2}$
Rohan P. Naidu\orcid{0000-0003-3997-5705},$^{1}$\thanks{NASA Hubble Fellow}
Anna de Graaff\orcid{0000-0002-2380-9801},$^{2,7}$\thanks{Clay Fellow}
Elia Pizzati\orcid{0000-0002-9712-0038},$^{2}$\thanks{NASA Einstein Fellow}
David M. Alexander\orcid{0000-0002-5896-6313},$^{8}$
Luis C. Ho\orcid{0000-0001-6947-5846},$^{9,10}$
Vasily Kokorev\orcid{0000-0002-5588-9156},$^{11,12}$
Gene Leung\orcid{0000-0002-9393-6507},$^{1}$
Anna-Christina Eilers\orcid{0000-0003-2895-6218},$^{1}$
and 
Ryan C. Hickox\orcid{0000-0003-1468-9526}$^{13}$
}\vspace*{0.2cm}\\%
$^{1}$ Department of Physics, Kavli Institute for Astrophysics and Space Research, Massachusetts Institute of Technology, Cambridge, MA 02139, USA\\%
$^{2}$ Center for Astrophysics $|$ Harvard $\&$ Smithsonian, 60 Garden Street, Cambridge, MA 02138, USA\\%
$^{3}$ Department of Physics, The University of Texas at Dallas, Richardson, TX 75080, USA \\%
$^{4}$ Department of Physics and Astronomy, York University, 4700 Keele Street, Toronto, ON M3J 1P3, Canada\\%
$^{5}$ Max Planck Institute for Astrophysics, Karl-Schwarzschild-Str. 1, D-85741 Garching, Germany \\
$^{6}$ Fachbereich Physik, Philipps Universit\"at Marburg, D-35032 Marburg, Germany \\
$^{7}$ Max Planck Institute for Astronomy, Königstuhl 17, D-69117 Heidelberg, Germany \\
$^{8}$ Centre for Extragalactic Astronomy, Department of Physics, Durham University, Durham DH1 3LE, UK\\
$^{9}$ Kavli Institute for Astronomy and Astrophysics, Peking University, Beijing 100871, China \\
$^{10}$ Department of Astronomy, School of Physics, Peking University, Beijing 100871, China \\
$^{11}$ Department of Astronomy, The University of Texas at Austin, Austin, TX 78712, USA \\
$^{12}$ Cosmic Frontier Center, The University of Texas at Austin, Austin, TX 78712, USA \\
$^{13}$ Department of Physics and Astronomy, Dartmouth College, 6127 Wilder Laboratory, Hanover, NH 03755, USA
}

\date{Accepted XXX. Received YYY; in original form ZZZ}

\pubyear{\the\year{}}

\begin{document}
\label{firstpage}
\pagerange{\pageref{firstpage}--\pageref{lastpage}}
\maketitle

\begin{abstract}
High-redshift active galactic nuclei (AGN) are powerful probes of early black-hole growth, galaxy formation, and the evolving intergalactic medium (IGM). We use \luminasim, a cosmological radiation-hydrodynamic simulation spanning the epochs of hydrogen and helium reionization, which combines a large $(500\,{\rm cMpc})^3$ volume with $2\times 6000^3$ resolution elements, to explore high-redshift AGN. The simulation self-consistently follows hundreds of millions of galaxies and supermassive black holes (SMBHs), together with their impact on the IGM. We exploit this uniquely large dynamic range to predict multi-band AGN luminosity functions (LFs) at $z \geq 3$, from hard X-rays to the mid-infrared. These predictions encompass both moderately luminous quasars and the faint ``Little Red Dots'' (LRDs) uncovered by \textit{JWST}. We develop an empirical model that maps simulated SMBHs onto observed AGN using bolometric and extinction/absorption corrections for canonical AGN and LRDs, and in which SMBHs with $M_{\rm BH}\leq 10\,M_{\rm seed} \sim 10^{7}\msun$ stay in the LRD phase with a duty cycle of $30\%$. This simple framework reproduces the observed LFs and clustering of LRDs. Meanwhile, the pre-\textit{JWST} quasar LF constraints are recovered, although we find that a $\sim 0.3$ dex log-normal scatter in bolometric luminosity is required to reproduce the bright end. We quantify the redshift evolution of AGN and LRD number densities, and their contributions to the integrated BH mass densities. The same AGN population is the dominant driver for the \HeII reionization modelled self-consistently in \luminasim. This paves the way for general population-synthesis models of high-redshift AGN, including LRDs, in a unified cosmological framework.
\end{abstract}

\begin{keywords}
quasars: supermassive black holes -- galaxies: high-redshift -- galaxies: active -- methods: numerical
\end{keywords}


\section{Introduction}

Active galactic nuclei (AGN), and luminous quasars in particular, are the observable manifestation of gas accretion onto supermassive black holes (SMBHs) at the centres of most galaxies. The accreted gas forms a hot, dissipative disc \citep[e.g.,][]{Shakura1973,Rees1984}, whose high radiative efficiency makes such systems among the most luminous persistent sources in the Universe and allows them to be detected out to $z\gtrsim7$ \citep[e.g.,][]{Mortlock2011,Venemans2015,Banados2018} and now to $z\gtrsim 10$ with \textit{JWST} \citep{Bogdan2024,Napolitano2025,Maiolino2025}. The cosmic demography of AGN therefore directly traces the assembly of the SMBH population, as most of the local SMBH mass density was built up during the radiatively efficient accretion phases that we observe as AGN \citep{Soltan1982}. AGN are also among the most important sources of high-energy radiation in the Universe, contributing significantly to the cosmic infrared (IR) and X-ray backgrounds. Their hard extreme ultraviolet (EUV) emission is believed to drive the reionization of singly ionized helium \citep[\HeII; e.g.,][]{Miralda1998,Madau1999,FG2008,McQuinn2009,Haardt2012,Worseck2014} and may also contribute to the reionization of hydrogen \citep[e.g.,][]{MadauHaardt2015,Madau2024,Asthana2025}, although star-forming galaxies are believed to dominate the hydrogen-ionizing photon budget in most current models \citep[e.g.,][]{FG2008Letter,Kuhlen2012,Robertson2015,Haardt2015,Onoue2017,Shen2020}. 

Beyond their role as energy sources, AGN are tightly coupled to the evolution of their host galaxies. The masses of SMBHs are correlated with the masses, luminosities, and velocity dispersions of their host bulges \citep[e.g.,][]{Magorrian1998,Ferrarese2000,Gebhardt2000,Gultekin2009}. AGN feedback is widely invoked to regulate or quench star-formation in massive galaxies in theoretical models \citep[e.g.,][]{Springel2005,DiMatteo2005,Croton2006,Hopkins2006}, with evidence from observed AGN-driven outflows \citep{Feruglio2010,Cicone2014}, and to resolve the classical ``cooling-flow'' problem \citep[e.g.,][]{Cowie1977,Fabian1977,Fabian1994}. Recent reviews summarize the breadth of these connections \citep[][and references therein]{Alexander2012,Fabian2012,Kormendy2013,Heckman2014,Alexander2025}. 

Studying the cosmic evolution of the AGN population is therefore central to both cosmology and galaxy formation. The measurement of AGN abundance, i.e. luminosity functions (LFs), has a decades-long history in the rest-frame optical/UV \citep[e.g.,][]{Schmidt1968,Boyle1988,Warren1994,Fan2001,Richards2006b,McGreer2013}, soft X-ray \citep[e.g.,][]{Maccacaro1991,Boyle1993,Page1997,Miyaji2000,Hasinger2005}, hard X-ray \citep[e.g.,][]{Ueda2003,LaFranca2005,Barger2005,Ueda2014,Aird2015a}, and IR \citep[e.g.,][]{Brown2006,Matute2006,Assef2011,Lacy2015} bands. These measurements have revealed strong redshift evolution of AGN LFs \citep[both in normalization and in shape; e.g.,][]{Cowie1996,Barger2005,Hasinger2005,Aird2015a,Kulkarni2018}, typical AGN spectral shape \citep[e.g.,][]{Richards2006a,Steffen2006,Lusso2010,Lusso2016}, and the obscuring column-density distribution \citep[e.g.,][]{Steffen2003,Ueda2003,Sazonov2004,Barger2005,Ueda2014,Ananna2019}. Syntheses of these multi-band measurements provided a comprehensive pre-\textit{JWST} view of the AGN population from $z\simeq 0$ to $z\simeq 7$ \citep{Hopkins2007,Shen2020}.

Despite this rich, established picture, the rapid decline of the AGN number density beyond $z\simeq 6$ made it particularly difficult to constrain AGN LFs at the highest redshifts, which correspond to early SMBH growth phases. Pre-\textit{JWST} wide-field optical and near-IR surveys identified $\gtrsim 500$ luminous quasars at $z\gtrsim 5.3$ \citep[e.g.,][see \citealt{Fan2023} for a review]{Willott2010,Venemans2013,Jiang2016,Matsuoka2018} and pushed the redshift frontier to $z\gtrsim 7$ \citep[e.g.,][]{Mortlock2011,Banados2018,Yang2020,Wang2021,Fujimoto2022}. These observations provided key probes of early SMBH growth and of the hydrogen reionization history through their absorption spectra \citep[e.g.,][]{Miralda1998,Madau2000,Fan2002,Fan2006}. However, these samples were dominated by the rare, hyperluminous tip of the AGN population, and the demographics of moderate- and low-luminosity AGN at high redshift, which likely dominate the integrated SMBH mass-growth budget, remained essentially unconstrained.

\textit{JWST} has dramatically transformed the landscape of AGN studies at high redshifts. Deep \textit{JWST} NIRSpec and NIRCam Grism surveys have already uncovered hundreds of broad-line AGN (BLAGN) at $4\lesssim z\lesssim 11$ \citep[e.g.,][]{Harikane2023-agn,Maiolino2024,Greene2024,Matthee2024,Kokorev2024}, with spectroscopically identified candidates extending to $z>10$ \citep{Bogdan2024,Napolitano2025,Maiolino2025}. Meanwhile, deep \textit{JWST} MIRI and X-ray surveys are beginning to probe the obscured high-redshift AGN population \citep[e.g.,][]{Bulichi2026,Hviding2026,Ling2026}. Faint AGN appear to be far more abundant in the early Universe than predicted by extrapolations of pre-\textit{JWST} quasar LFs, with potentially important implications for the high-redshift ionizing photon budget and build-up of the cosmic SMBH mass density \citep{Madau2024,Asthana2025,Inayoshi2025-review}.

Among the most striking \textit{JWST} discoveries is the population of ``Little Red Dots'' (LRDs), compact sources with characteristic ``V-shaped'' spectral energy distributions (SEDs) that feature a blue UV continuum, a steep red optical continuum, and a turnover near the Balmer limit \citep[e.g.,][]{Matthee2024,Kokorev2024,Akins2025,Kocevski2025,Setton2025b}. Spectroscopic follow-up has revealed broad Balmer lines in the majority of the observed LRDs \citep[e.g.,][]{Greene2024,Matthee2024,Maiolino2024,Hviding2025}, supporting an AGN origin for at least the dominant red continuum component. Yet LRDs differ in essentially every other respect from canonical Type-I/Type-II AGN: they are systematically X-ray weak or undetected \citep[e.g.,][]{Yue2024,Ananna2024,Maiolino2025}, faint or absent in the mid- and far-IR \citep[e.g.,][]{Setton2025a,Casey2025}, and exhibit little
or no continuum variability \citep[at least in the majority; e.g.,][but see also \citealt{Zhang2025}]{Hayes2024,Kokubo2025,Tee2025,Zhang2025a,Liu2026}. Many show prominent Balmer breaks and Balmer absorption features \citep[e.g.,][]{Naidu2025,DeGraaff2025b,Setton2025b,Ji2025} that are difficult to explain by simple dust-reddened Type-I AGN. The same sources are often inferred to host BHs that are over-massive relative to their host stellar masses \citep[e.g.,][]{Pacucci2023,Maiolino2024,Inayoshi2025-review,Durodola2025}, although the interpretation depends sensitively on the assumed line-broadening mechanism \citep[e.g.,][]{Naidu2025,Brazzini2025,Chang2026,Ji2026,Rusakov2026} and selection effects \citep[e.g.,][]{Li2025}. Whether LRDs represent a transient, early-growth phase of more typical SMBHs, a fundamentally distinct population, or a heterogeneous mixture of both remains an open question.

Interpreting these new observational findings, however, is complicated by the fact that observations in any single band are subject to selection effects, host-galaxy contamination, and obscuration in a wavelength-dependent manner. Although AGN are intrinsically very luminous in the optical/UV, dust extinction along certain viewing angles \citep[e.g.,][]{Antonucci1993,Urry1995} can render them difficult to detect, and the optical/UV emission of heavily obscured AGN is easily contaminated by stellar UV light from their host galaxies \citep[e.g., see review of][]{Hickox2018}. In the X-ray, which is much less affected by dust, the search for Compton-thick (CTK) AGN, which are heavily attenuated and constitute $20\%-50\%$ of the AGN population \citep[e.g.,][]{Burlon2011,Ricci2015,Carroll2021} and remain highly incomplete in current surveys. Mid-IR observations from the ground are limited by atmospheric absorption and contamination by warm dust in host galaxies. In the far-IR to millimetre regime ($30\micron-10\,{\rm mm}$), emission from cold dust in host galaxies can dominate over the AGN, blunting AGN identification. LRDs clearly highlight this problem: if we had relied solely on X-ray or IR data, nearly all of these BHs would have gone undetected. Furthermore, LF measurements based on a single survey are limited in their luminosity coverage and volume probed and are subject to non-negligible biases and uncertainties in completeness corrections. It is therefore essential to develop a framework that links observations across multiple bands, providing a more comprehensive picture of AGN demographics over cosmic time.

Specifically for LRDs, the physical interpretation is evolving rapidly. One emerging class of models invokes accreting BHs embedded in extremely dense, optically thick gas envelopes that reprocess the emergent emission from the central engine into a ``star-like'' (but not identical to stars) optical spectrum \citep[e.g.,][]{Inayoshi2025,Naidu2025,DeGraaff2025b,Kido2025,Rusakov2026}, possibly forming a quasi-star configuration \citep[e.g.,][]{Begelman2006,Begelman2008,Volonteri2010,Begelman2026}. A complementary scenario invokes super-Eddington accretion onto modest-mass seed BHs as the explanation for X-ray weakness, and the lack of variability in LRDs \citep[e.g.,][]{Lambrides2024,PacucciNarayan2024,Madau2024b,Lupi2024,Inayoshi2025-superedd}. Whether LRDs trace heavy seeds (e.g., direct-collapse BHs), light seeds undergoing efficient early growth, or some combination of both remains a major theoretical question that connects directly to early SMBH-seeding scenarios, whether through heavy direct-collapse seeds \citep[e.g.,][]{Bromm2003,Begelman2006,Lodato2006}, runaway collisions in dense star clusters \citep[e.g.,][]{Devecchi2009}, or dark matter-related channels \citep[e.g.,][]{Xiao2021,ShenT2025,Jiang2026}. A unified cosmological framework that can simultaneously address the abundance and large-scale environments of \textit{JWST}-selected faint AGN together with luminous quasars is therefore needed.

Theoretical efforts to interpret the high-redshift AGN population have progressed in parallel with \textit{JWST} observations. Large-volume cosmological hydrodynamic simulations, such as \illustris \citep{Vogelsberger2013,IllustrisIntro,IllustrisNature}, EAGLE \citep{Schaye2015,Crain2015}, \illustrisTNG \citep{Pillepich2018first,Pillepich2018Model,Springel2018}, SIMBA \citep{Dave2019}, Horizon-AGN \citep{Dubois2016,Volonteri2016}, BlueTides \citep{Feng2016}, ASTRID \citep{Bird2022,Ni2022}, MillenniumTNG \citep{Pakmor2023}, and FLAMINGO \citep{Schaye2023flamingo}, have established a coherent framework for modelling the growth and feedback of SMBHs alongside their co-evolution with host galaxies in a cosmological context. Cross-comparison studies \citep[e.g.,][]{Habouzit2021,Habouzit2022} highlight that the predicted high-redshift AGN LFs and BH-host scaling relations are sensitive to the choice of seeding prescription, sub-resolution recipes for BH accretion and AGN feedback, motivating more theoretical experiments. Smaller-volume zoom and more dedicated suites, such as BRAHMA \citep{Bhowmick2024,Bhowmick2024-overmassive}, NewHorizon \citep{Dubois2021}, and OBELISK \citep{Trebitsch2021}, systematically explored seed channels, accretion regimes, and feedback prescriptions, while semi-analytic and empirical frameworks, such as CAT \citep{Trinca2022,Trinca2024}, DELPHI \citep{Dayal2014,Dayal2019}, TRINITY \citep{Zhang2023}, L-Galaxies \citep{Bonoli2025,HC2026-Lgal}, and the Santa Cruz SAM \citep{Yung2021}, efficiently explored the high-redshift SMBH model parameter space. 

\begin{figure*}
    \centering
    \includegraphics[width=1.04\linewidth]{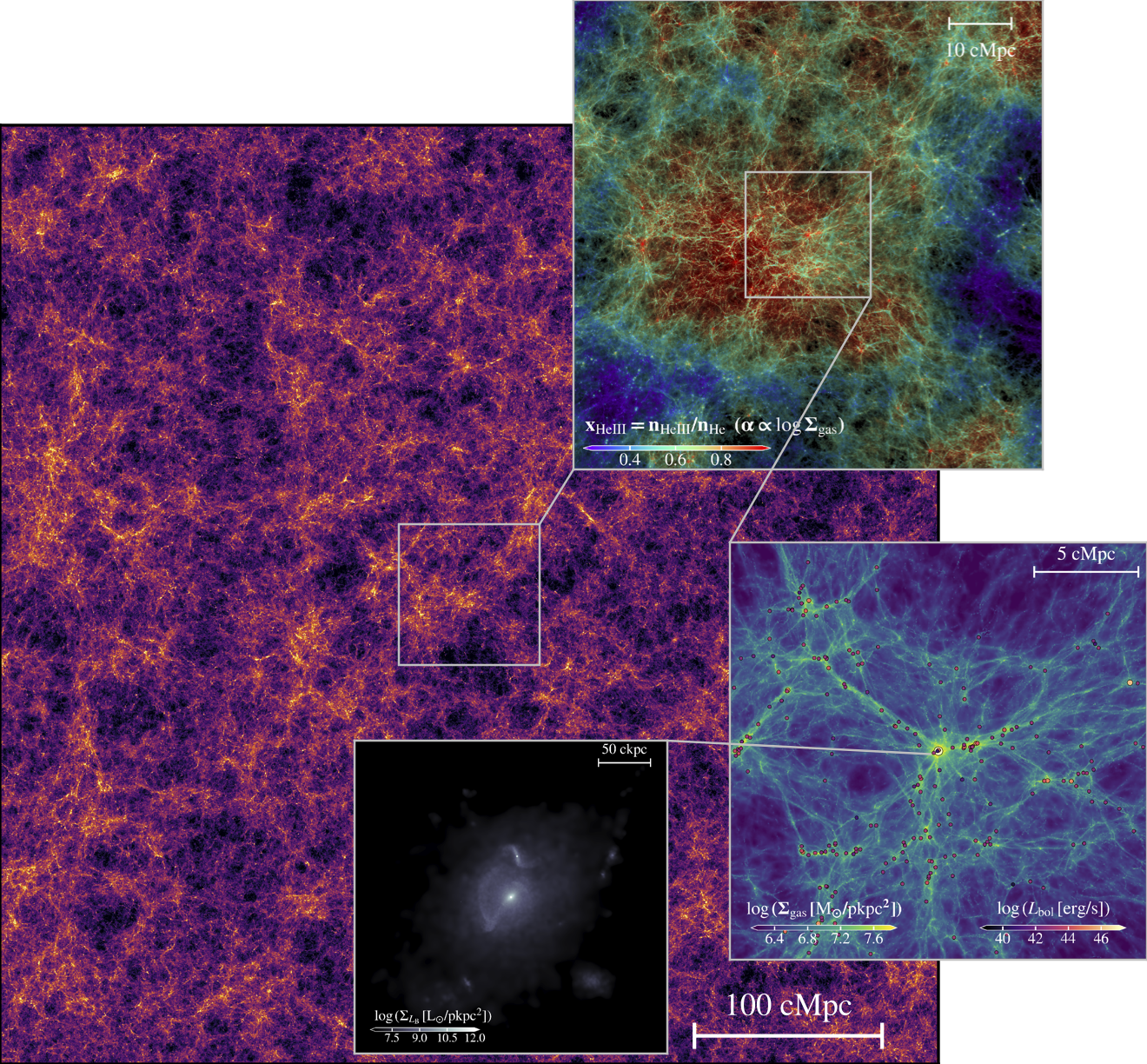}
    \caption{DM surface density field of a slice of the simulation volume centred on the most luminous AGN at $z\simeq 4$. The thickness of the slice is $150\,{\rm cMpc}$. We first zoom into a $75\times 75\,{\rm cMpc}^{2}$ subfield and present the blending of the mass-weighted \HeIII fraction and the gas surface density field. This illustrates the \HeIII bubble driven by the hard ionizing radiation from the central AGN. We further zoom in on the $25\times 25\,{\rm cMpc}^{2}$ neighbourhood around the central AGN and show the gas surface density map with individual circles showing locations of nearby AGN, colour-coded by their bolometric luminosities. The brightest central AGN has $L_{\rm bol}\sim 10^{47}\erg\,{\rm s}^{-1}$. Finally, we zoom into the host galaxy of this AGN and show a surface brightness map of the galaxy and AGN (as saturated point sources) in the B band. One can spot the two AGN in the centres of a merging pair of galaxies. This overview highlights the exceptionally broad dynamical range and wealth of physical information captured by \luminasim.}
    \label{fig:image1}
\end{figure*}

\begin{figure*}
    \centering
    \includegraphics[width=\linewidth]{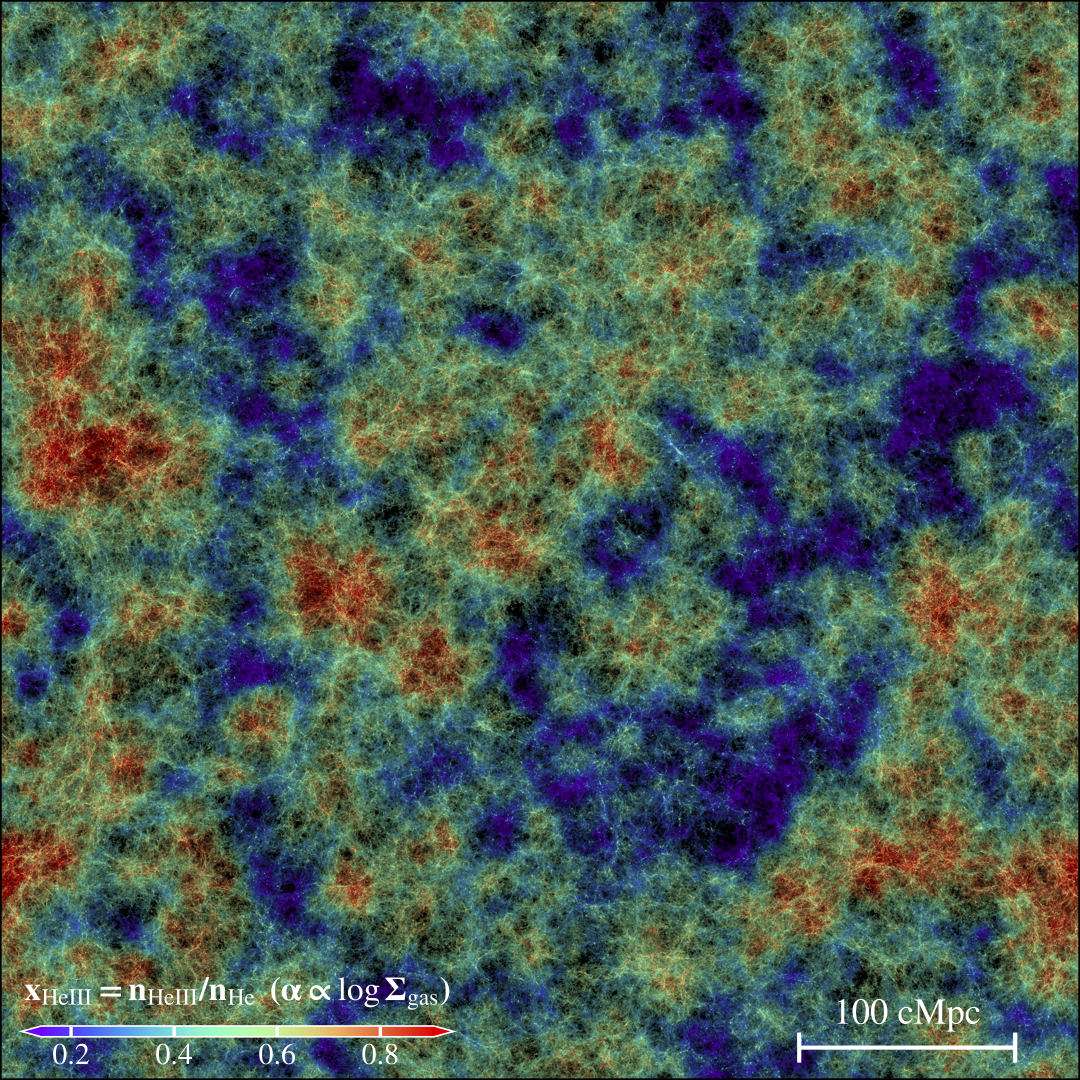}
    \caption{A blending of the gas surface density map and the mass-weighted fraction of \HeIII for a slice of the simulation volume centred on the most luminous AGN at $z\simeq 4$. The \HeIII fraction is colour-mapped while the gas surface density dictates the transparency. The thickness of the slice is $150\,{\rm cMpc}$. This image illustrates the large-scale topology of \HeII reionization driven mostly by AGN in \luminasim.}
    \label{fig:image2}
\end{figure*}

In this paper, we use \luminasim\ --- a large-volume $(500\,{\rm cMpc})^3$ cosmological radiation-hydrodynamic simulation with an extreme dynamical range of $2\times 6000^3$ adaptive resolution elements that follows galaxies, AGN, and the intergalactic medium (IGM) through hydrogen and helium reionization down to $z=3$ --- to make predictions for multi-band  AGN LFs at $z\geq 3$ across the electromagnetic spectrum, from hard X-rays to the mid-IR. \luminasim is designed to play a complementary role to the theoretical efforts described above. The large volume samples bright quasars together with a statistically representative, well-resolved population of galaxies and faint AGN, enabling joint predictions for AGN, host galaxies, and their distributions on large scales coherently. In addition, \luminasim is the first simulation in this regime to include on-the-fly multi-frequency radiative transfer through both hydrogen and \HeII reionization, allowing self-consistent modelling of the ionization and thermal state of the IGM impacted by galaxies and AGN. This combination enables the unified, multi-band view of high-redshift AGN in the cosmological context that we develop in this paper. In Figure~\ref{fig:image1}, we show a multi-scale view of \luminasim at $z\simeq 4$ from the large-scale dark matter (DM) distribution to a \HeIII bubble driven by a bright AGN to the smaller-scale gas distribution and BHs, eventually to the host galaxy of the AGN. In Figure~\ref{fig:image2}, we give a more complete overview of the topology of \HeII reionization at $z\simeq 4$, featuring large $\sim 10-100\,{\rm cMpc}$ size \HeIII bubbles created by AGN along with islands of \HeII in voids. 

Based on \luminasim, we develop an empirical model to bridge BH particles in the simulation with observed AGN, including the puzzling LRD population, and then use it to test whether the simulated SMBH population can simultaneously reproduce several important observables of high-redshift AGN, including LFs, large-scale clustering measurements, and integrated cosmological constraints. The paper is organized as follows. Section~\ref{sec:sim} summarizes the \luminasim simulation, including its galaxy formation and SMBH model, the on-the-fly radiative transfer, and the structure-finding pipeline. Section~\ref{sec:smbh-agn} describes the empirical model that maps simulated SMBHs to observed AGN. Section~\ref{sec:results} presents the predicted multi-band AGN LFs at $z\simeq 3-6$, the large-scale clustering of LRDs, and the role of AGN in the cosmological context. Section~\ref{sec:discussion} discusses the underlying assumptions and limitations of our framework, and compares it to other theoretical models. Conclusions are presented in Section~\ref{sec:conclusions}.

\begin{figure*}
    \centering
    \includegraphics[width=1.0\linewidth]{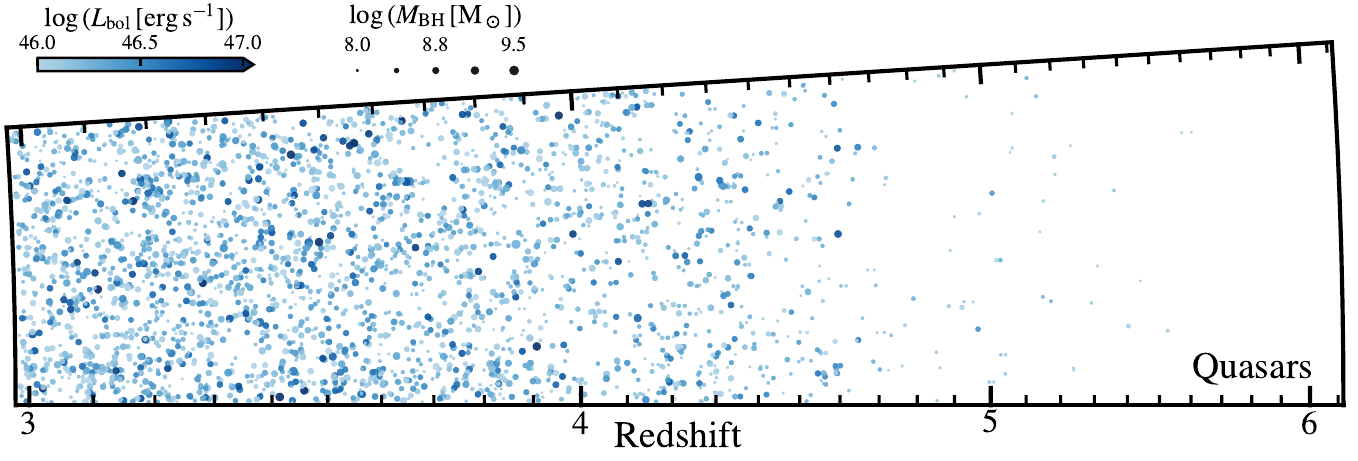}
    \includegraphics[width=1.0\linewidth]{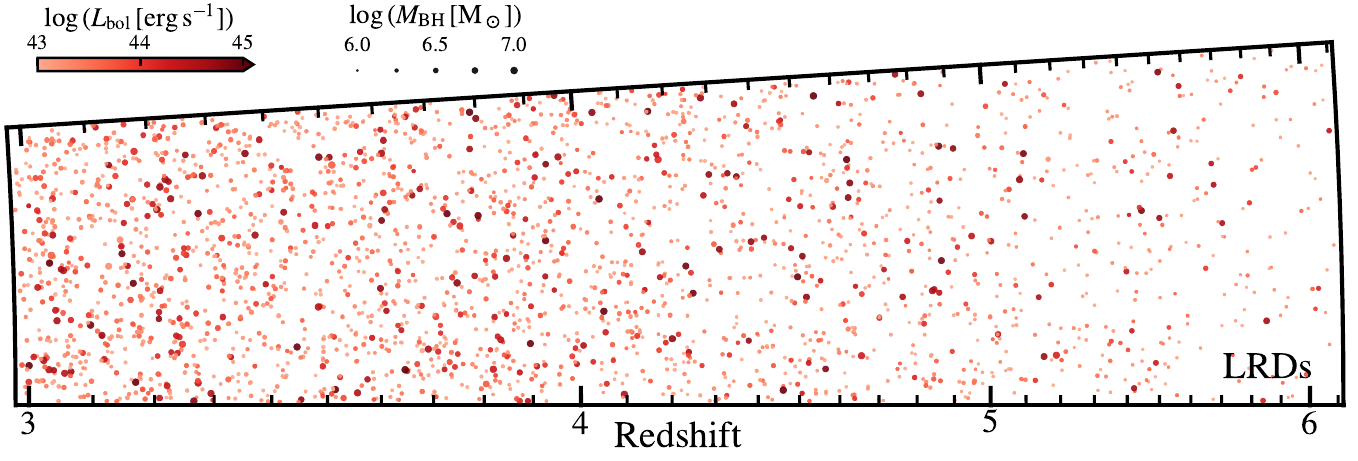}
    \caption{BH lightcone wedges from \luminasim. The wedge shows a $3.6^\circ \times 3.6^\circ$ pencil-beam (full opening angle $0.063$ rad) cut from the lightcone (corrected for redshift space distortions) from $z\simeq3$ to $z \simeq 6$. Each filled circle marks a BH with sightline comoving distance plotted along the long axis and the transverse position along the short axis. Marker colour encodes the bolometric luminosity, while marker size encodes the BH mass. In the top panel, we show the distribution of massive bright AGN in \luminasim (considered as moderately luminous quasars) with $L_{\rm bol}\geq 10^{46}\erg\,{\rm s}^{-1}$ and $M_{\rm BH}\geq 10^{8}\msun$. In the bottom panel, we show the LRDs selected according to our empirical model at $M_{\rm BH}\lesssim 10^{7}\msun$ and with $L_{\rm bol}\geq 10^{43}\erg\,{\rm s}^{-1}$ (downsampled by a factor of $100$ to avoid overcrowding the figure). It demonstrates the ability of \luminasim in connecting SMBHs with their evolving large-scale environments. The quasars are notably rarer at higher redshifts, where LRDs start to dominate.}
    \label{fig:lightcone}
\end{figure*}

\begin{table}
\centering
\addtolength{\tabcolsep}{15.0pt}
\begin{tabular}{ll}
\hline
Parameter & Value \\
\hline
\hline
$\Omega_{\rm m}$   & 0.3096 \\ 
$100\times\Omega_{\rm b}$  & 4.897 \\ 
$\Omega_\Lambda$   & 0.6903 \\
$H_0\,[\kms \Mpc^{-1}]$ & $67.66 $ \\
$\sigma_8$         & $0.8102$  \\
$n_{\rm s}$        & $0.9665$\\
$z_{\rm init}$     & 49 \\
$T_{\rm init}\,[\mathrm{K}]$ & $54.17$ \\
$T_{\rm CMB,0}\,[\mathrm{K}]$    & $2.7255$ \\
$T_{\nu,0}\,[\mathrm{K}]$        & $1.9524$ \\
$N_{\rm neutrino}$ & 3 \\
\hline
\end{tabular}
\caption{Key cosmological parameters from the TT,TE,EE+lowE + lensing + BAO data set of \protect\cite{Planck2020} used in \luminasim. We list the matter density $\Omega_{\rm m}$, baryon density $\Omega_{\rm b}$, dark-energy density $\Omega_\Lambda$, the Hubble constant $H_0$, the amplitude of fluctuation $\sigma_8$ and spectral index $n_s$ of the primordial power spectrum, the starting redshift of our simulations $z_{\rm init}$, the initial baryon temperature $T_{\rm init}$ \citep{Fixsen2009}, the present-day CMB temperature $T_{\rm CMB,0}$, the present-day temperature of the cosmic neutrino background $T_{\nu,0}$, and the number of massless neutrino species used $N_{\rm neutrino}$. We absorb the slightly larger effective number of relativistic species, $N_{\rm eff}=3.046$, into the effective neutrino background temperature.}
\label{tab:cosmo}
\end{table}

\section{Simulation}
\label{sec:sim}

\luminasim is a cosmological radiation-hydrodynamic simulation that evolves $2\times 6000^{3}$ resolution elements in an $L_{\mathrm{box}}=500\,\mathrm{cMpc}$ box from $z_{\rm init}=49$ down to $z=3$, spanning the epochs of hydrogen and \HeII reionization. The DM mass resolution is $m_{\rm DM}=1.9\times 10^7\msun$, and the baryonic mass resolution is $m_{\rm b}=3.6\times 10^6\msun$. DM particles have a constant Plummer-equivalent gravitational softening of $1.2\,h^{-1}\,{\rm ckpc}$ and the gas cells are adaptively softened with a minimum of $0.3\,h^{-1}\,{\rm ckpc}$. This combination of volume and resolution provides the dynamic range required to follow simultaneously the galaxy population that drives hydrogen reionization and the rare, bright quasars that drive \HeII reionization. The simulation is run with the moving-mesh code \arepo \citep{Springel2010, Pakmor2016, weinberger2020}, that solves the cosmological Euler equations on an unstructured moving Voronoi mesh using a finite volume Godunov method \citep{Godunov1959}. Gravity is solved with an adaptive Tree-PM scheme \citep{Bagla2002,Bagla2003}. It is coupled to the GPU-accelerated radiative-transfer module \areport \citep{Kannan2019, Zier2024}. Similar to \mtng \citep{Pakmor2023}, magnetic fields are not evolved. We adopt a flat $\Lambda$CDM cosmology consistent with \textit{Planck} 2018 \citep{Planck2020}, which we summarize in Table~\ref{tab:cosmo}. Initial conditions are generated from separate baryon and CDM transfer functions computed with \textsc{camb} \citep{camb2000,camb2012}, which captures the delayed growth of baryonic perturbations relative to CDM at high redshift as well as the associated large-scale streaming velocity \citep{Eisenstein1998,Tseliakhovich2010,Fialkov2012}. Below we summarize the physics ingredients most relevant to the AGN modelling presented in this paper, and refer the reader to \citet{Zier2026} for a complete description of the simulation campaign.

\subsection{Galaxy Formation Model}

\luminasim adopts a slightly modified version of the well-tested \illustrisTNG galaxy formation model \citep{Weinberger2017,Pillepich2018Model}, which is itself an updated version of the original \illustris framework \citep{Vogelsberger2013,IllustrisIntro,IllustrisNature}. All galaxy-formation and BH model parameters remain unchanged from the original \illustrisTNG model (except for the cooling routines of low-density gas as described in (ii) below and the fact that magnetic fields are not evolved). The model was calibrated at $(m_{\rm dm}\simeq 1.2\times10^7\msun,\,m_{\rm b}\simeq 2.3\times 10^6\msun)$ resolution against a set of local galaxy observables and the cosmic star-formation rate density \citep{Pillepich2018Model}. No re-calibration was performed for \luminasim\footnote{We note that the mass resolution of \luminasim is only a factor of $1.6$ coarser than the calibration run in \citet{Pillepich2018Model}, and \citet{Zier2026} shows that the resulting properties of the galaxy population are consistent with TNG100-1 (with $1.6$ times better mass resolution than the calibration run) across redshifts.}, except for the stellar escape fraction required for radiative transfer (see Section~\ref{subsec:rt} below). We have shown in \citet{Zier2026} that \luminasim produces consistent results for galaxy stellar mass functions, star-forming main sequence, the stellar mass--halo mass relations, and the mass metallicity relations compared to \illustrisTNG.

In the following, we briefly summarize the relevant components of this model and highlight the modifications adopted for \luminasim. The \illustrisTNG model includes:
\begin{enumerate}
  \item A two-phase effective equation-of-state (EOS) model for the interstellar medium (ISM), incorporating sub-resolution pressurization from stellar feedback \citep{Springel2003}. The EOS acts as a temperature floor for gas above the density threshold $n_{\rm H} \ge 0.106\,\mathrm{cm}^{-3} \equiv n_{\rm H, SF}$, while gas cells heated above the EOS are modelled as a single hot phase.

  \item Primordial and metal-line cooling of gas \citep{Smith2008,Wiersma2009} below $n_{\rm H, SF}$ in the presence of a redshift-dependent, spatially uniform UV radiation background \citep{FG2009} with corrections for self-shielding \citep{Katz1992,Katz1996,Rahmati2013}. In \luminasim, these equilibrium cooling routines for low-density gas are replaced by a non-equilibrium chemical network coupled to the radiative-transfer solver (see Section~\ref{subsec:rt} below).

  \item Star-formation in dense ($n_{\rm H} \geq n_{\rm H, SF}$) gas, following the empirical volumetric Kennicutt--Schmidt relation, with stellar particles spawned stochastically as single stellar populations following the \citet{Chabrier2003} initial mass function (IMF). Metal enrichment from asymptotic-giant-branch (AGB) stars and supernovae (SNe) follows \citet{Pillepich2018Model}, except that we track only the total gas metallicity rather than individual elemental species.

  \item Galactic winds driven by SNe, implemented with a non-local thermal-plus-kinetic scheme. Randomly selected star-forming gas cells are converted into wind particles that decouple from the hydrodynamics, leave the dense ISM, and re-couple to the Voronoi mesh at low density. The mass and energy loading factors, described in detail by \citet{Pillepich2018Model}, depend on the SNe rate, the local DM velocity dispersion, redshift, and gas metallicity.
\end{enumerate}

\subsection{Seeding, accretion, and feedback of SMBHs}
\label{subsec:smbh}

The BH model largely follows \citet{Weinberger2017}. BHs are seeded in haloes identified by a friends-of-friends (FoF) algorithm once the halo mass first exceeds $5\times10^{10}\,h^{-1}\msun$ and no BH is yet present. The gas cell with the lowest gravitational potential is converted into a BH particle with an initial mass of $M_{\rm seed}=8\times10^{5}\,h^{-1}\msun$.

BHs grow by accreting from nearby gas cells at the Bondi--Hoyle rate \citep{Bondi1944,Bondi1952}
\begin{equation}
\dot{M}_{\rm Bondi} = \dfrac{4\,\pi\,G^2\,M_{\rm BH}^2\,\rho_{\rm gas}}{c_{\rm s}^3} \, ,
\label{eq:bondiRate}
\end{equation}
where $G$ is the gravitational constant, $M_{\rm BH}$ is the BH mass, $\rho_{\rm gas}$ is the ambient gas density, and $c_{\rm s}$ is the local sound speed. Ambient quantities are estimated with a smoothed-particle-hydrodynamics(SPH)-like kernel over the $128$ nearest gas cells. A rough estimate of this kernel size can be obtained from the effective gas cell size corresponding to the local gas density,
\begin{equation}
h_{\rm acc} \simeq 128^{1/3} \left(\frac{3\,m_{\rm b}}{4\,\pi\,\rho}\right)^{1/3}
\simeq 1.5\left(\frac{n_{\rm H}}{\mathrm{cm^{-3}}}\right)^{-1/3}\mathrm{pkpc}.
\label{eq:spatialResolution}
\end{equation}
It varies from $\sim 0.7\pkpc$ in dense nuclear gas ($n_{\rm H}\simeq 10\,\mathrm{cm^{-3}}$) to roughly $\sim 3\pkpc$ at the star-formation threshold. For example, at $z\simeq 3$, the typical BH-accretion kernel size is $1.20^{+0.58}_{-0.44}\,\pkpc$ in \luminasim. This can become comparable to the characteristic sizes of low-mass galaxies at these redshifts. Although the kernel weighting favours the central gas, the resulting accretion rates still represent galaxy-scale estimates of BH fuelling rather than resolved nuclear accretion flows. To prevent unphysically high Bondi rates in low-density environments, which can arise from excessively large smoothing lengths, we suppress the accretion rate in the high-accretion (quasar) mode discussed below by a factor $\left(P_{\rm ext}/P_{\rm ref} \right)^2$ whenever the kernel-weighted external pressure $P_{\rm ext}$ falls below the reference pressure $P_{\rm ref}$ defined in \citet[][Equation~23]{Vogelsberger2013}. Because \luminasim does not evolve magnetic fields, $P_{\rm ext}$ here does not include the magnetic-pressure contribution adopted in the original \illustrisTNG simulations, which can otherwise enhance the accretion rate in low-density, magnetically dominated regions.

BH accretion rate is capped at the Eddington limit
\begin{equation}
\dot{M}_{\rm Edd} \equiv \dfrac{4\,\pi\, G\, M_{\rm BH}\, m_{\rm p}}{\epsilon_{\rm rad}\,\sigma_{\rm T}\,c} \, ,
\label{eq:eddingtonRate}
\end{equation}
where $m_{\rm p}$ is the proton mass, $c$ is the speed of light, $\sigma_{\rm T}$ is the Thomson scattering cross-section, and $\epsilon_{\rm rad}=0.2$ is the radiative efficiency, so that the effective accretion rate is $\dot{M}_{\rm BH}=\min(\dot{M}_{\rm Bondi},\,\dot{M}_{\rm Edd})$. BHs further grow through mergers with neighbouring BHs as soon as one enters the accretion kernel of the other \citep{Sijacki2015} and, to mimic the effect of unresolved dynamical friction, are repositioned at each globally synchronized time step to the local minimum of the gravitational potential. This frequent repositioning is essential to ensure that BHs remain near the centres of their host galaxies, where they can efficiently accrete gas and avoid premature growth suppression.

AGN feedback is implemented in a dual-mode prescription. At high accretion rates (the ``quasar'' mode), the feedback is a pure thermal energy injection
\begin{equation}
\dot{E}_{\rm therm} = \epsilon_{{\rm f, high}}\,\epsilon_{\rm rad}\,\dot{M}_{\rm BH}\,c^2 \, ,
\label{eq:etherm}
\end{equation}
with efficiency \(\epsilon_{{\rm f, high}}=0.1\). The energy is injected in a kernel-weighted manner into the gas cells of the same neighbourhood used for the accretion estimate. At low accretion rates (the ``radio'' mode), the feedback is injected in a kinetic form,
\begin{equation}
\dot{E}_{\rm kin} = \epsilon_{{\rm f, kin}}\;\dot{M}_{\rm BH}\,c^2 \, ,
\end{equation}
with the efficiency
\begin{equation}
\epsilon_{{\rm f, kin}} = \min\left(\dfrac{n_{\rm H}}{0.05\,n_{\rm H, SF}},\,0.2\right) \, .
\end{equation}
This is implemented as a pure isotropic kernel-weighted momentum injection into the gas cells of the same neighbourhood used for the accretion estimate in the rest frame of the BH particle. Different from the continuous injection in the thermal feedback, the kinetic mode feedback is pulsed: the feedback energy is accumulated until it exceeds a minimum injection energy $E_{\rm inj,min} = (f_{\rm re} / 2)\,\sigma_{\rm DM}^2\, m_{\rm enc}$, with the pulsing parameter $f_{\rm re} = 20$, $\sigma_{\rm DM}$ as the local one-dimensional DM velocity dispersion, and $m_{\rm enc}$ as the gas mass in the kernel.

The transition between the ``quasar'' and ``radio'' modes occurs at the mass-dependent critical Eddington ratio
\begin{equation}
\lambda_{\rm Edd} = \frac{\dot{M}_{\rm BH}}{\dot{M}_{\rm Edd}} = \min\!\left[\,0.002\left(\frac{M_{\rm BH}}{10^8\,M_\odot}\right)^{\!2},\,0.1\,\right] \, ,
\end{equation}
so that more massive BHs can more easily enter the kinetic mode at fixed Eddington ratios. Throughout this paper, the bolometric luminosity of an accreting BH is taken to be
\begin{equation}
    L_{\rm bol} = \epsilon_\mathrm{rad}\,(1-\epsilon_{f,\mathrm{high}})\,\dot{M}_{\rm BH}\,c^2 \, ,
    \label{eq:lbol-mdot}
\end{equation}
i.e.\ the radiated power after the energy allocated to AGN feedback in the quasar mode has been subtracted, consistent with the AGN source treatment in the radiative transfer described in the following section. Following the \illustris/\illustrisTNG prescription, we suppress radiation from BHs accreting at Eddington ratios $\lambda_{\rm Edd} < 2\times 10^{-3}$, where we assume an advection-dominated accretion flow \citep[ADAF; e.g.,][]{NarayanYi1994}. We note that the transition to the ``radio'' mode feedback is distinct from the transition to radiatively inefficient (ADAF) accretion. In particular, the former adopts a phenomenological BH-mass-dependent threshold, which introduces an additional dependence of the characteristic galaxy-quenching mass scale on the subgrid feedback model.

All the free parameters related to the BH physics in this Section were inherited from the \illustrisTNG model, which has been calibrated against a range of local Universe observables of galaxies and SMBHs \citep{Weinberger2017}. No recalibration is involved in this work. These parameters include the choice of $\epsilon_{\rm rad}=0.2$, which is larger than the value $0.1$ typically assumed in other works. Nevertheless, it is within the $0.057-0.32$ range in the classical \citet{Novikov1973} model of thin-disc accretion for BH spins between $0$ and $0.998$. It was inherited from the \illustris simulation \citep{Sijacki2015}, motivated by the findings of \citet{Yu2002}.

\subsection{Radiative transfer}
\label{subsec:rt}

We solve the radiation-transport equations on the moving Voronoi mesh using a moment-based finite-volume method with the M1 closure \citep{Levermore1984,Dubroca1999,Kannan2019,Zier2024}. For $z>4.75$, the radiation field is discretized into six frequency bins: three photoionizing bands associated with \HI, \HeI, and \HeII, plus three higher-energy X-ray bins. At $z<4.75$, after hydrogen reionization is complete, the radiation field below $54.42\,{\rm eV}$ is replaced with the spatially uniform UV background of \citet{FG2009}\footnote{This transition to uniform background is justified by the disappearance of the last neutral HI island in the volume \citep{Zier2026}. It could suppress the small fluctuations in the radiation field and temperature in the IGM. For the observables studied in this paper, the impact is negligible.}, and only a single high-energy bin relevant to \HeII reionization is evolved on the fly. As shown in \citet{Zier2026}, AGN dominate the photon production in this remaining bin, so radiation from stellar particles and from other X-ray sources described below is deactivated at $z<4.75$. Absorption is coupled to a non-equilibrium thermo-chemical network for primordial gas \citep{Kannan2019}, while metal-line cooling follows the tables of \citet{Vogelsberger2013} and is not coupled to the local radiation field. We employ the reduced-speed-of-light approximation \citep{Gnedin2001} with $\tilde{c}=0.2\,c$, which has been shown to yield a converged reionization history for our setup \citep{Thesan1,ThesanEnrico,ThesanAaron}. Radiation transport and chemistry are subcycled relative to the hydrodynamics by a factor of $64$ during hydrogen reionization and by a factor of $256$ for $z<4.75$. During these RT subcycles, the Voronoi mesh is held fixed to enable efficient GPU acceleration. We include three classes of radiation sources.
\begin{enumerate}
    \item \textit{Stellar populations.} The ionizing output of stellar particles is computed with the \textsc{bpass} v2.2.1 binary population-synthesis models \citep{BPASS2017,Stanway2018}, assuming a Chabrier IMF with an upper-mass cutoff of $100\,\msun$. Unresolved ISM absorption is captured by a constant stellar escape fraction, $f_{\mathrm{esc},\star}=0.18$, calibrated against the observed timing of hydrogen reionization (see \citealt{Zier2026}).

    \item \textit{Accreting SMBHs.} We adopt the canonical Type-I AGN spectrum of \citet{Shen2020}, which combines the optical/UV/IR template of \citet{Krawczyk2013}, the extreme-UV power law of \citet{Lusso2015}, and an X-ray component normalized through X-ray-optical luminosity scaling relations \citep{Steffen2006,Lusso2010}. A more detailed description is given in Section~\ref{subsec:sed} below. The hard EUV tail of this template provides the photons that drive the \HeII reionization signal analysed in Section~\ref{subsec:cosmo}. A luminosity-dependent effective escape fraction following \citet{Vogelsberger2013} is applied at the source
    \begin{equation}
        f_{\rm esc} \equiv \frac{\dot{N}^{\rm esc}_{\gamma}}{\dot{N}_{\gamma}}
        = 0.3\left(\frac{L_{\rm bol}}{10^{46}\erg\,{\rm s}^{-1}}\right)^{0.07},
    \end{equation}
    which encapsulates phenomenologically the absorption/extinction of unresolved circumnuclear gas and dust\footnote{In reality, the photoionization opacity is wavelength-dependent \citep{Verner1996}. However, for the large column densities characteristic of AGN obscuration, individual clouds/torus are expected to remain optically thick throughout much of the H- and He-ionizing EUV bands. In this regime, the effective escape fraction is controlled primarily by the geometry and covering factor of the unresolved obscurer. We therefore neglect an explicit wavelength dependence of $f_{\rm esc}$ within the EUV wavelengths. This approximation becomes less appropriate at X-ray energies, which motivates the different treatment in Section~\ref{subsec:extinction}.} along with the uncertainties in the intrinsic EUV spectra of AGN.

    \item \textit{X-ray emission from star-forming processes.} We include X-ray emission from shock-heated ISM gas and high-mass X-ray binaries (HMXBs). For the ISM component, we adopt a thermal bremsstrahlung SED with $k_{\rm B}T_{\rm ISM}=240\,{\rm eV}$ \citep{2012bMineo-HotISM,Pacucci2014,Eide2018}. For HMXBs, we use the SEDs of \citet{fragos2013energy,fragos2016erratum}. The normalizations scale with the local star-formation rate, and, for HMXBs, additionally with metallicity, following \citet{2012bMineo-HotISM} and \citet{Madau2017}.
\end{enumerate}

To couple the radiation field cleanly to the TNG galaxy-formation model, we use the transparent equation of state introduced by \citet{bulichi2025high}. Gas denser than the star-formation threshold, $n_{\rm H,SF}$, remains optically transparent in the RT calculation: radiation is transported through the cell, but its absorption and heating are not coupled to the gas thermochemistry, which instead continues to follow the equilibrium cooling model. This gives a clean numerical interpretation of $f_{\mathrm{esc},\star}$ as the escape fraction from the unresolved ISM. Further details are provided in \citet{Zier2026}.

\subsection{Structure finding}
\label{subsec:structure}

DM haloes are identified using the FoF algorithm \citep{Davis1985} with a linking length of $0.2$ times the mean inter-particle separation. Gas, stellar, and BH particles are associated with the FoF group of their nearest DM particle if they lie within one linking length. Each FoF group is then processed with the \textsc{subfind-hbt} algorithm \citep{springel2001populating,Gadget4}, which iteratively identifies gravitationally bound substructures (subhaloes) and, unlike the original \textsc{subfind}, tracks subhalo membership across snapshots, improving the identification of infalling systems through mergers \citep{hbt2012}. Following common practice, we identify subhaloes as galaxies, designating the most massive subhalo of each FoF group as the central and the remainder as satellites. Because FoF masses can be inflated by percolation via ``particle bridges'', we report spherical-overdensity quantities of DM haloes: $R_{200\mathrm{c}}$, defined as the radius enclosing an average density of $200$ times the critical density $\rho_{\mathrm{c}}(z)$, and the corresponding enclosed mass $M_{200\mathrm{c}}$. All BH particles in the simulation enter the AGN observable predictions. The host galaxy of a BH is identified as the subhalo to which it is gravitationally bound. In practice, multiple BHs can appear in the same subhalo. This is because the search for potential minima for BH repositioning and mergers is done locally using the same kernel used for BH accretion. BHs can therefore take a few global timesteps to merge at the centre of the subhalo.

\luminasim records the full set of properties for every BH particle at high time cadence, enabling the construction of AGN light cones alongside other quantities intended for sightline-resolved surveys \citep{Smith2026}. Figure~\ref{fig:lightcone} shows a representative $3.6^\circ \times 3.6^\circ$ pencil-beam realization from $z\simeq3$ to $z\simeq6$. Each point is placed according to its sightline comoving distance and transverse position, with colour indicating bolometric luminosity and size indicating BH mass. This visualization highlights the wide dynamic range of the simulated AGN population and the ability of \luminasim to connect individual SMBHs to their evolving large-scale environments. Future studies will combine these light-cone catalogues with BH merger trees to study the temporal variability of accretion, feedback, and clustering.

\begin{figure*}
    \centering
    \includegraphics[width=\linewidth]{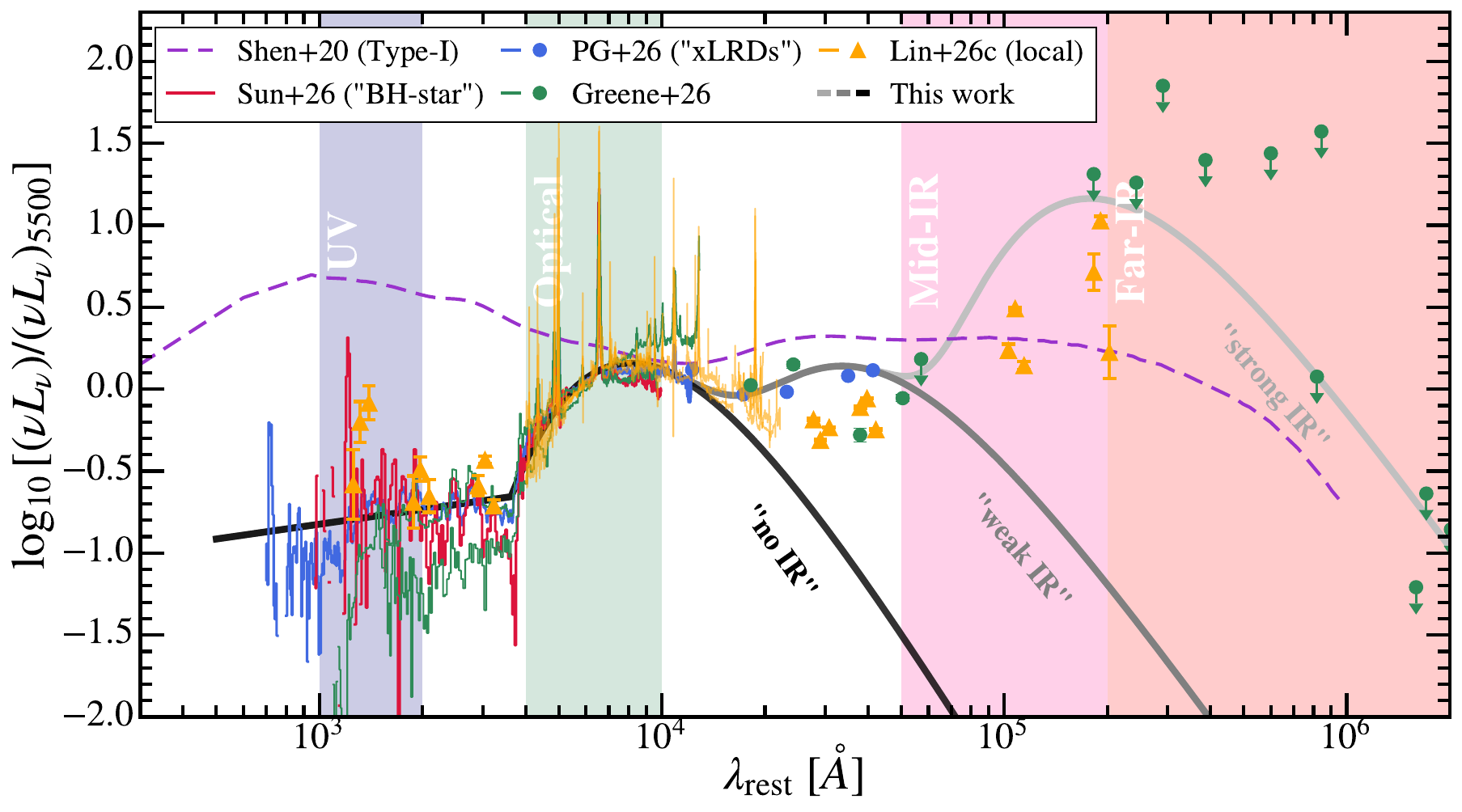}
    \caption{An overview of the AGN SED model adopted. We compile the stacked LRD spectra from \citet{PG2026} (the ``xLRDs'' in their sample) and \citet{Sun2026} (``BH-star'', LRD spectra subtracting host galaxy) as well as individual source spectra from \citet{Greene2026} and \citet{Lin2026-local} (the three local LRD analogues), including their IR constraints. They are normalized at $5500$\AA. A universal SED of LRDs appears in optical-to-near-IR wavelengths, where we find that a modified blackbody with temperature $T\sim 4000\,{\rm K}$ and $\beta \sim 0.5$ gives a reasonable fit. In the UV, we approximate it as a power-law, $\nu\,L_{\nu} \sim \nu^{-0.3}$. In the mid- and far-IR, we attempt adding blackbody components with temperature $T \sim 1000\,{\rm K}$ and $200\,{\rm K}$ (which can be produced by e.g. warm and cold dust) and tune the normalization to the maximum allowed by current observational constraints. In the background, we also show the universal quasar SED assumed in \citet{Shen2020}. The LRDs are distinct from unobscured Type-I AGN with weaker UV and mid-IR emission and more dominant optical-to-near-IR emission. We also assume LRDs are X-ray dark. These together form the basis of the bolometric corrections we derive in Table~\ref{tab:lrd-corr}.}
    \label{fig:sed}
\end{figure*}

\section{An empirical model linking SMBHs to AGN}
\label{sec:smbh-agn}

The BH particles and their mass accretion rates tracked in \luminasim represent the time- and kernel-averaged growth of BHs (plus any unresolved gas or stellar distributions around them), whereas observed AGN are characterized by their multi-band emission along a particular line of sight. Bridging these two descriptions requires a set of choices about how the resolved mean accretion rate is related to the instantaneous bolometric luminosity, how that luminosity is distributed across the electromagnetic spectrum, and how the emerging radiation is post-processed by circumnuclear and host-galaxy material. In this section, we present a compact post-processing framework that makes these choices explicit, with three ingredients: (\textit{i}) a log-normal scatter applied to the instantaneous bolometric luminosity of each BH particle to capture unresolved AGN variability; (\textit{ii}) a pair of empirical SED templates, one for canonical Type-I AGN and one for LRDs, with the mapping between BH particles and SED templates determined by BH mass; and (\textit{iii}) a gas absorption and dust extinction prescription applied to the canonical AGN population. The framework is deliberately constructed to be portable. Although calibrated and applied here to the SMBH population of \luminasim, none of its ingredients depend on the specifics of the simulation, and the same recipe can be applied to other cosmological simulations, semi-analytic models, or empirical halo--BH mapping schemes to generate high-redshift AGN population predictions.

\subsection{Stochasticity of SMBH accretion}
\label{subsec:lbol-scatter}

We first model an intrinsic stochasticity (variability) of SMBH accretion at the sub-resolution level by convolving the raw bolometric LF computed from binned estimations of simulated SMBHs with a Gaussian kernel of dispersion $\sigma_{\rm bol}$. We then reduce the bolometric luminosities systematically by $\Delta \log{L_{\rm bol}} = \ln{10}\,\sigma^2_{\rm bol}/2$ (the mean shift of a log-normal distribution) to ensure conservation of total energy output from AGN. Empirically, we find that a scatter of $\sigma_{\rm bol}=0.3$ dex brings the predicted AGN LFs in \luminasim into good agreement with observational constraints of quasars \citep{Shen2020}. This will be later shown in Section~\ref{subsec:bol}. We keep it as the fiducial choice and explore variations in Section~\ref{subsec:parameters}. We note that the BH accretion rates are capped at the Eddington limit in \luminasim, but the empirical scatter we add here can effectively boost the accretion rates to moderately super-Eddington.

Similar phenomenological treatments have been adopted in other large-volume cosmological hydrodynamic simulations to account for unresolved AGN variability as well. For example, \citet{Rosas-Guevara2016} explored similar log-normal scatters of $0.3-0.5$ dex for the AGN LFs in the EAGLE simulation. \citet{Ding2026} introduced a larger scatter of $0.75$ dex for quasar LFs in the FLAMINGO simulation~\citep{Schaye2023flamingo} to bracket the impact of unresolved small-scale variability. Such a scatter is motivated by the observed variability of AGN across a broad hierarchy of timescales, from hours in the X-ray corona \citep[e.g.,][]{Vaughan2003,GonzalezMartin2012}, through months to years in the optical/UV ``damped-random-walk'' (DRW) regime \citep[e.g.,][]{Kelly2009,Macleod2010}, to the $\sim 0.1-100$ Myr timescales inferred from e.g. broad Eddington-ratio distribution function (ERDF) measured at fixed host-galaxy properties \citep[e.g.,][]{Hickox2014,Aird2018}. Most large-volume cosmological simulations cannot directly resolve the clumpy multiphase gas structures, angular momentum transport, accretion-disc-state transitions, and other relevant small-scale physical processes that underlie these observational phenomena. Recent ultra-high-resolution zoom-in simulations reveal that gas inflow towards SMBHs becomes increasingly intermittent on progressively smaller scales, driven by nuclear gas structure, gravitational torques, stellar/AGN feedback, magnetic fields, and disc-scale instabilities \citep[e.g.,][]{Hopkins2010,Hopkins2011,AA2017a,Hopkins2024}. We defer a more thorough discussion of this prescription to Section~\ref{sec:discussion}.

\subsection{Canonical AGN SED model}
\label{subsec:sed}

For Type-I and Type-II AGN, we adopt a universal model for the intrinsic SED constructed in \citet{Shen2020}. Here we briefly summarize it. For the optical/UV/IR wavelengths, we adopt the SED template in \citet{Krawczyk2013}, which was based on $>100\,000$ luminous broad-line quasars observed at $0.064 < z < 5.46$ that are classified as not strongly affected by dust reddening and obscuration. The dust extinction corrections on the AGN luminosities will be considered separately in the following section. This SED template starts at $\sim 30\micron$ and truncates at $912\text{\AA}$. We extend the SED to the EUV~($\lambda<912\text{\AA}$) using the power-law model $f_{\nu}=\nu^{\alpha_{\nu}}$ with index $\alpha_{\nu}=-1.70$ \citep{Lusso2015}. We truncate this extension at $600\text{\AA}$, where the measurements of \citet{Lusso2015} ended, and directly connect the flux at $600\text{\AA}$ with the X-ray template, which will be discussed below. The EUV SED here is consistent with what we adopted in the radiative transfer calculations in \luminasim. In the IR, we extend the template to $100\micron$ using the \citet{Richards2006a} SED. Dust emission is effectively included in this IR SED template, so we do not explicitly model any dust emission.

The X-ray SED template is modelled as a cut-off power law, $f(E) \propto E^{1-\Gamma}\exp(-E/E_{\rm c})$, with the photon index $\Gamma=1.9$ and the cut-off energy $E_{\rm c}=300\keV$ \citep[e.g.,][]{Dadina2008,Ueda2014,Aird2015a}. An additional reflection component is added using the \textsc{pexrav} model \citep{Magdziarz1995} implemented in \textsc{xspec} \citep{Arnaud1996}, assuming the reflection relative strength $R=1$, the inclination angle $i=60^{\circ}$, and Solar abundances. The normalization of the X-ray SED is determined according to the well-known correlation between X-ray and optical luminosities
\begin{equation}
    \log{L_{\nu}(2\keV)}=\beta \log{L_{\nu}(2500\text{\AA})} + C \, ,
    \label{eq:Lx-L2500}
\end{equation}
where the unit of $L_{\nu}$ is $\erg\,{\rm s}^{-1}\Hz^{-1}$. One can define $\alpha_{\rm ox}$ accordingly as
\begin{equation}
    \alpha_{\rm ox}\equiv \dfrac{\log{L_{\nu}(2\keV)}-\log{L_{\nu}(2500\text{\AA})}}{\log{\nu(2\keV)}-\log{\nu(2500\text{\AA})}}=0.384\log{\left( \dfrac{L_{\nu}(2\keV)}{L_{\nu}(2500\text{\AA})}\right)} \, .
\end{equation}
Then Equation~(\ref{eq:Lx-L2500}) can be rewritten as
\begin{equation}
    \alpha_{\rm ox}= -A\log{\left( \dfrac{L_{\nu}(2500\text{\AA})}{\erg\,{\rm s}^{-1}\Hz^{-1}}\right)} + C^{\prime} \, ,
\end{equation}
where $A=0.384\,(1-\beta)$ and $C^{\prime}=0.384C$. These prefactors have been measured through observations. We choose the parameters measured by \citet{Steffen2006}, $\beta=0.721\pm0.011$ and $C=4.531\pm0.688$, which have been adopted in \citet{Hopkins2007} and \citet{Shen2020}, and are consistent with more recent observations. The X-ray SED is then scaled according to $\alpha_{\rm ox}$ with respect to the optical SED. 

Following \citet{Shen2020}, we define the bolometric luminosity as the integrated luminosity from $30\micron$ to $500\keV$, which represents the total energy budget produced by accretion onto the SMBH. Extending the integration range to $\gtrsim 100\micron$, as done in e.g.\ \citet{Richards2006a}, has a negligible impact on the bolometric luminosity. Some studies instead truncate the integration at $1\micron$ \citep[e.g.,][]{Runnoe2012,Duras2020}. For our SED template, we find that this leads to a factor of $1.3-1.4$ reduction in the bolometric correction, though this remains a subdominant effect for the purposes of this paper. The bolometric correction is defined as the ratio between the bolometric luminosity, $L_{\rm bol}$, and the observed luminosity in a given band, $L_{\rm band}$.

Even Type-I AGN do not share a perfectly universal SED. Real variations in the spectral shape translate into scatter in the bolometric corrections and influence the observed AGN LFs. We follow the same approach as \citet{Shen2020}, generating an ensemble of SEDs by varying the optical/FUV/EUV spectral slopes, the X-ray photon index $\Gamma$, and $\alpha_{\rm ox}$, and evaluating the resulting dispersion in the bolometric corrections. Adopting the fitting functions and best-fit parameters from \citet{Shen2020}, we compute the dispersion of bolometric corrections $\sigma_{\rm corr}(\log{L_{\rm bol}})$, which corresponds to a $\sim 0.1$ dex uncorrelated dispersion in AGN SEDs, consistent with observations. We also adopt the double power-law fit for the dependence of the bolometric corrections on $L_{\rm bol}$ from \citet{Shen2020}, along with the best-fit parameters therein. We note that the optical/UV and X-ray luminosities derived from these bolometric corrections do not yet account for gas absorption and dust extinction, which we discuss in the following sections.

The SED model above is empirical and cannot self-consistently capture systematic variations of AGN spectra with the physical state of the accretion flow. At low Eddington ratios, the thin disc transitions to a radiatively inefficient ADAF \citep[e.g.,][]{NarayanYi1994}, with a weak UV ``big blue bump'', a harder X-ray spectrum, reduced obscuration, and, at the lowest accretion rates, a disappearing broad-line region \citep[e.g.,][]{Ho1999,Ho2008,ElitzurHo2009,She2018}. \citet{Su2026} recently showed that an accretion-disc-based SED model predicts substantially lower hard X-ray luminosities at $L_{\rm bol}\lesssim 10^{38}\,\erg\,{\rm s}^{-1}$ than the \citet{Shen2020} bolometric corrections calibrated on luminous quasars. In our framework, we treat it as electromagnetically silent when $\lambda_{\rm Edd}<2\times 10^{-3}$ (Section~\ref{subsec:smbh}) and it has almost no impact in the luminosity range of high-redshift AGN we study in \luminasim. At the opposite extreme, near or above the Eddington limit, photon advection through a slim disc reduces the radiative efficiency, saturates the UV/optical continuum, and radiation-pressure drives outflows \citep[e.g.,][]{Paczynsky1980,Abramowicz1988,Sadowski2011,KubotaDone2019,Jiang2019}. This regime is plausibly vital for the rapid growth of seed BHs into $z\gtrsim 6$ quasars (Section~\ref{subsec:variability}) and for many observed properties of LRDs \citep[e.g.,][]{PacucciNarayan2024,Inayoshi2025-superedd,Madau2026}, for which we adopt a separate empirical SED in the following section. The $\sigma_{\rm corr}$ absorbs intra-population variations within the canonical regime but not the qualitatively distinct SEDs at either $\lambda_{\rm Edd}$ extreme. A self-consistent $\lambda_{\rm Edd}$-dependent treatment is deferred to future work.

\begin{table*}
    \addtolength{\tabcolsep}{6pt}
    \centering
    \begin{tabular}{lllllll}
    \hline
          & UV & B band & ``Optical'' & H$\alpha$ & Hard X-ray & Mid-IR \\
          & (1450\AA) & ($\sim4400$\AA) & ($\sim5100$\AA) & ($\sim6563$\AA) & ($2$-$10\keV$) & ($\sim15\micron$) \\
    \hline
    \hline
         \textbf{\textit{LRDs}} & \\
         \citet{Greene2026} (no cold dust)  & (26.8)$^{\rm a}$ & (8.5) & 5.4 & 19 & - & - \\
         \citet{Greene2026} (with cold dust)  & (183.6) & (58.6) & 37 & 131 & - & - \\
    \hline
         This work (no IR)  & 12.5 & 4.0 & 2.5 & - & $\infty$ & $\infty$ \\
         \textbf{This work (weak IR)}  & \textbf{23.0} & \textbf{7.3} & \textbf{4.6} & - & $\mathbf{\infty}$ & \textbf{29.1} \\
         \textbf{This work (strong IR)}  & \textbf{140.1} & \textbf{44.3} & \textbf{28.0} & - & $\mathbf{\infty}$ & \textbf{1.7} \\
         \hline
         \hline
         \textbf{\textit{Canonical AGN}} & \\
         \textbf{\citet{Shen2020} (at $\bm{L_{\rm bol}=10^{45}\erg\,{\rm s}^{-1}}$)}  & \textbf{5.4} & \textbf{10.8} & \textbf{11.8} &  & \textbf{34.2} & \textbf{12.6} \\
         \citet{Greene2005}  & & & & 170 &  &  \\
         \citet{Richards2006a} & & & 10.3 &  & \\
         \citet{Runnoe2012,Runnoe2012a}  & 5.7[4.2]$^{\rm b}$ &  & 11.0[8.1] &  &  & 11.6[8.5] \\
         \citet{Duras2020} & & 8.5[5.13] &  &  & 34.7[20.9] &  \\
    \hline
    \end{tabular}
    \caption{Bolometric corrections (defined as $L_{\rm bol}/(\nu\,L_{\nu})$ for single wavelengths and $L_{\rm bol}/L_{\rm band}$ for integrated luminosity in a band) computed for LRDs. We compare the corrections in three scenarios described in Section~\ref{subsec:lrdsed} labelled as: ``no IR'', ``weak IR'', and ``strong IR'', to reflect the uncertainties in the IR. We compare them to the ``average'' bolometric corrections computed in \citet{Greene2026}. The two scenarios highlighted in bold form are the lower and upper limits of bolometric corrections considered in this work. In the bottom, we show the bolometric corrections of canonical AGN at $L_{\rm bol}=10^{45}\erg\,{\rm s}^{-1}$ from literature for comparison.
    \\$^{\rm a}$ The values in parentheses are not directly reported in their original paper. We take their bolometric corrections at $5100$\AA, and convert them to other bands using our LRD SED. $^{\rm b}$ These works assume different integration ranges to compute bolometric luminosities. We use our quasar SED template to correct the values to be consistent with our assumption. Their original values are displayed in the brackets.}
    \label{tab:lrd-corr}
\end{table*}

\subsection{LRD SED model}
\label{subsec:lrdsed}

Compared to the canonical AGN SED model described above, LRDs are notably faint in X-rays given their optical luminosities and often show weak mid-IR and far-IR emission, which is challenging to explain by simple dust-reddening of a standard Type-I AGN \citep[e.g.,][]{Yue2024,Ananna2024,Maiolino2025,Casey2025}. Recent spectroscopic observations have motivated a picture in which many LRDs host accreting SMBHs embedded in extremely dense gas cocoons \citep[e.g.,][]{Naidu2025,DeGraaff2025b,Inayoshi2025,Ji2025,Rusakov2026}, which generate ``star-like'' (but not identical to stars) blackbody SEDs in the optical featuring strong Balmer breaks and Balmer absorption features. Many recent works \citep[e.g.,][]{DeGraaff2025c,Umeda2026,Sun2026} have revealed a surprising quasi-universality of LRD SEDs at least in the optical-to-near-IR. In this section, we will attempt to build an effective universal SED model for LRDs. We note that our approach is deliberately agnostic about the underlying physical configuration of LRDs, whether they are ``star-like'' objects, dust-obscured AGN, or something else entirely. We restrict ourselves to constructing an empirical SED template from current observations and to bracketing its associated uncertainties.

We first take stacked LRD SEDs from \citet{PG2026} and \citet{Sun2026}. \citet{PG2026} classify the LRD population into several spectroscopic flavours based on \textit{JWST}/NIRSpec follow-up. We adopt their stack of extreme LRDs (``xLRDs''), which is the AGN-dominated subset with comparatively minor host-galaxy continuum contamination and interpreted as LRD engines therein. \citet{Sun2026} present a complementary stacked SED after subtracting their best-estimate host-galaxy contribution under the ``BH-star'' picture. In addition, we compile the spectra of individual sources from \citet{Greene2026} (see also \citealt{Setton2025a}) and the three local LRD analogues of \citet{Lin2026-local}, including their mid- and far-IR constraints. These compiled SEDs are shown in Figure~\ref{fig:sed}. A degree of universality emerges across these LRD SEDs at optical-to-near-IR wavelengths, with the remaining diversity concentrated in the mid- and far-IR where current observational constraints are weakest.

Inspired by this universality, we model the optical part of the SED as a modified blackbody
\begin{equation}
    L_\nu \propto \nu^{3+\beta}\,\dfrac{1}{e^{h\nu/k_{\rm B}T}-1} \, ,
\end{equation}
with temperature $T=4000$ K and $\beta=0.5$, broadly consistent with the best-fits of \citet{DeGraaff2025c} and with the $\sim 3000-6000$ K colour temperatures inferred in other works \citep[e.g.,][]{Naidu2025,Setton2025a,Akins2025,Umeda2026}. In the UV, we approximate it as a power law, $\nu\,L_{\nu} \sim \nu^{-0.3}$. In the mid- and far-IR, we attempt to add blackbody ($\beta=0$) components with $T \simeq 1000$ K and $T \simeq 200$ K, with normalisations tuned to the maximum allowed by current observational constraints. The IR components could be produced by warm and cold dust, but are not limited to these sources. 

We consider three scenarios, starting from no mid- and far-IR components at all and then progressively adding the two. They are shown in Figure~\ref{fig:sed}, labelled as ``no IR'', ``weak IR'', and ``strong IR'', respectively. We caution that the mid- and far-IR portion of the LRD SED remains genuinely uncertain: stacking of deep MIRI, ALMA, and Spitzer/Herschel imaging of LRDs yields strong constraints on the hot and cold dust emission \citep[e.g.,][]{Casey2025,Setton2025b,Akins2025,Xiao2025}, while evidence of hot dust emission from a rising near-IR continuum in stacked MIRI SEDs has also been reported \citep{Delvecchio2025}. Our three SED models are designed to bracket this observed diversity, so effectively we allow for a potential mixture of scenarios like ``BH-star'', dust-obscured, and others. We then integrate the SED from $500$ \AA\ to $1000\,\micron$ (assuming that LRDs are completely dark in X-rays, but see also \citealt{Hviding2026}) to obtain the bolometric luminosity. We intentionally include the entire far-IR component in integration for LRDs in order to fully bracket the uncertainties, although cold dust in the host galaxies could be the major source of far-IR radiation, like in canonical AGN. In Table~\ref{tab:lrd-corr}, we summarize the bolometric corrections derived for LRDs in three different scenarios, and they are compared to the average values computed in \citet{Greene2026} and bolometric corrections of canonical AGN. In the rest of the paper, we choose the ``weak IR'' and ``strong IR'' models to make predictions for the lower and upper limits of AGN LFs.

The scatters of the bolometric corrections of LRDs are inferred from the dispersions of the scaling relations between optical, UV, and blackbody luminosities \citep{DeGraaff2025c}. We adopt $\sigma_{\rm corr}\simeq 0.34$ dex in the UV and $\sigma_{\rm corr}\simeq 0.23$ dex in the optical, although these are subdominant to the systematic IR uncertainty discussed above. We do not assign an additional dispersion to the IR component, as that uncertainty is already captured by the IR SED variants.

\begin{figure*}
    \includegraphics[width=0.49\linewidth]{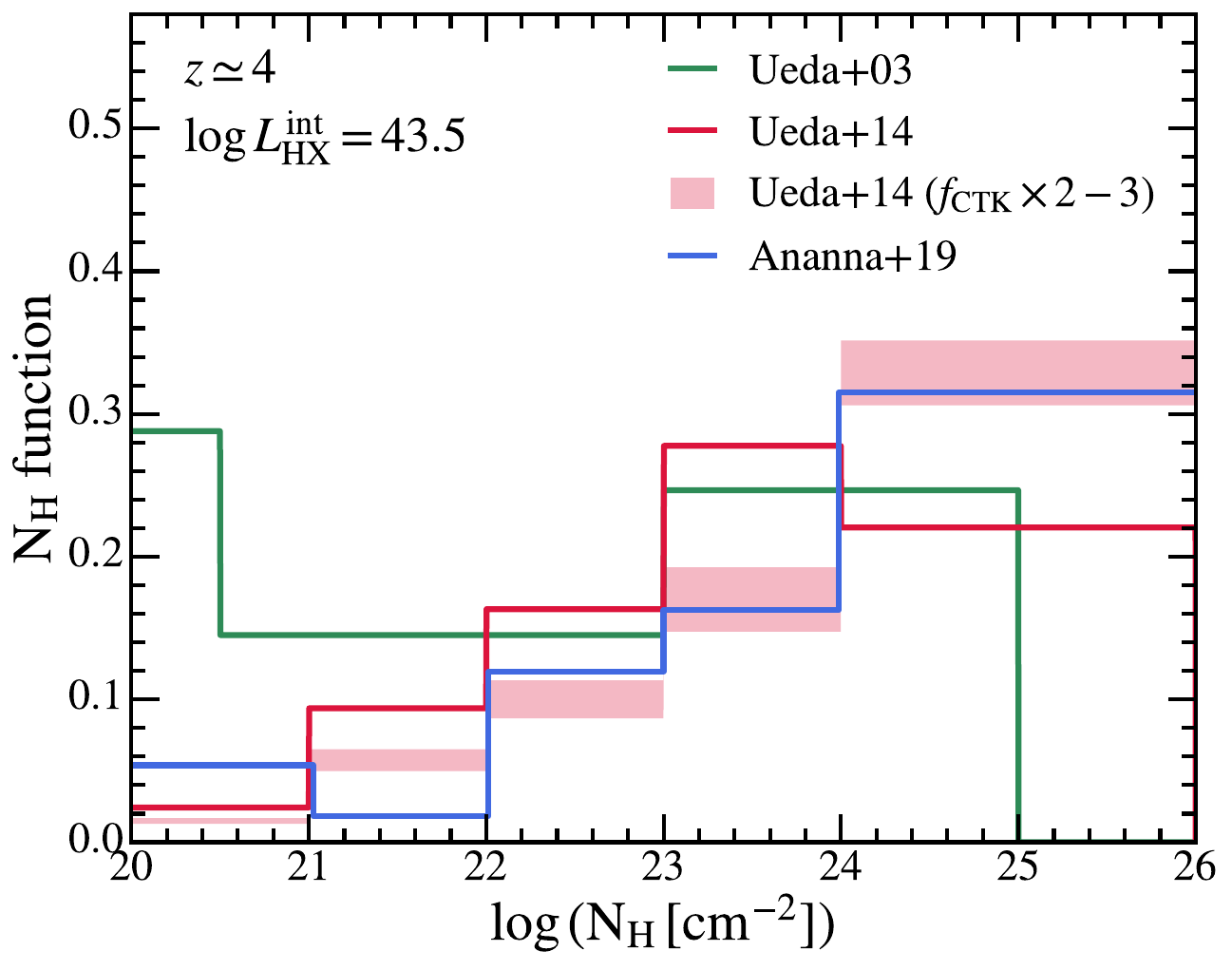}
    \includegraphics[width=0.49\linewidth]{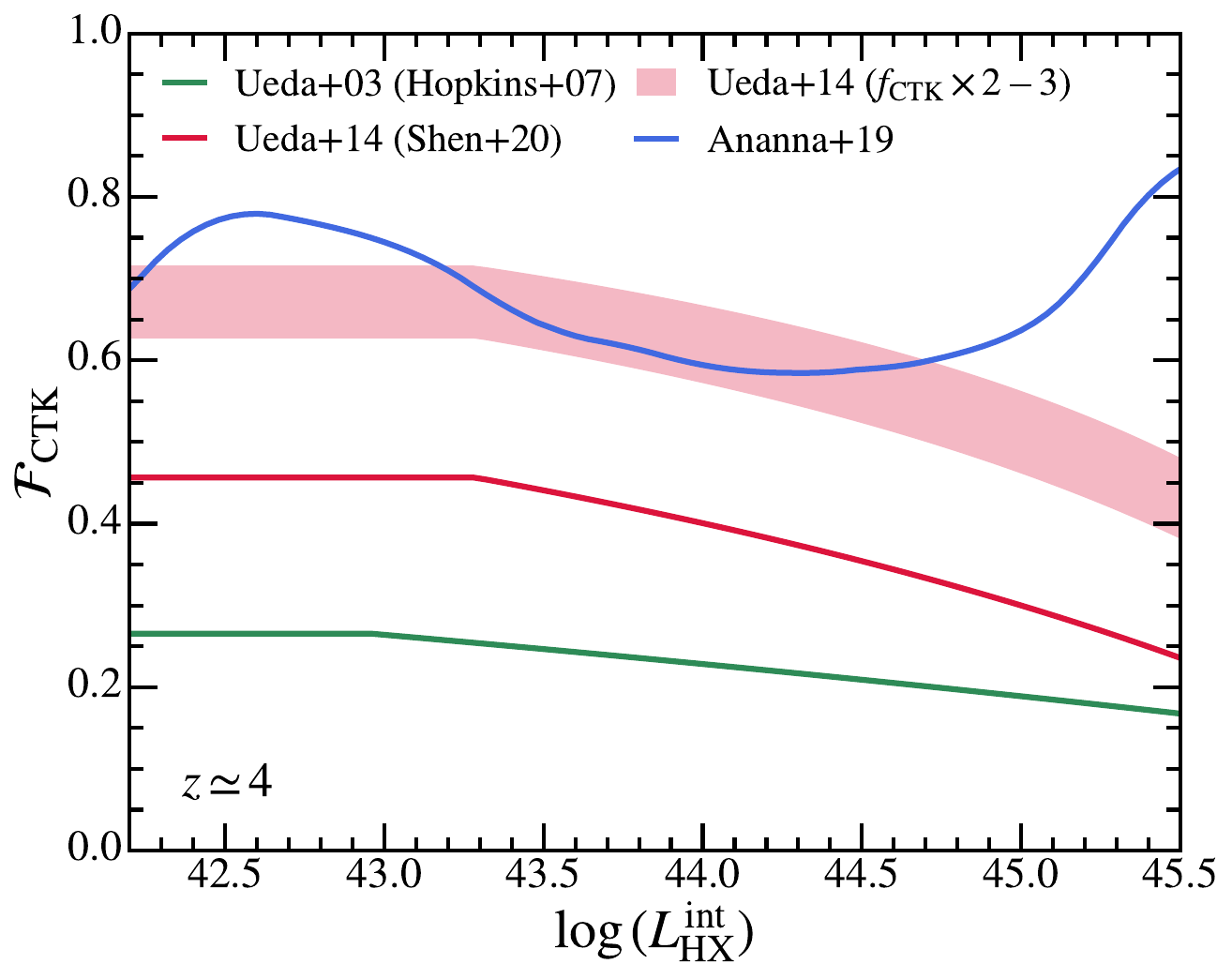}
    \caption{{\it Left:} Hydrogen column density distribution function at $z\simeq 4$ at intrinsic hard X-ray luminosity $L^{\rm int}_{\rm HX} = 10^{43.5}\erg\,{\rm s}^{-1}$. The distribution is normalised over $\log{(N_{\rm H}\,[{\rm cm}^{-2}])= [20, 26]}$. We show the $N_{\rm H}$ distribution from \citet{Ueda2003} (used in \citealt{Hopkins2007}), \citet{Ueda2014} (used in \citealt{Shen2020}), and more recent constraints from \citet{Ananna2019} leveraging the \textit{NuSTAR} data. We show a modified \citet{Ueda2014} distribution, increasing the CTK AGN fraction relative to the obscured CTN population ($f_{\rm CTK}$) by a factor of $2-3$, which roughly agrees with \citet{Ananna2019} at these luminosities at $3 \lesssim z \lesssim 5$. {\it Right:} CTK AGN fraction ($\mathcal{F}_{\rm CTK}$, relative to the whole population, different from $f_{\rm CTK}$) versus $L^{\rm int}_{\rm HX}$ at $z\simeq 4$. We show the inferred fractions from \citet{Ueda2003,Ueda2014} and \citet{Ananna2019} for comparison. The modified \citet{Ueda2014} model by boosting CTK fraction matches roughly the model in \citet{Ananna2019} until reaching $L^{\rm int}_{\rm HX}\sim 10^{45}\erg\,{\rm s}^{-1}$, where the statistics of the observational sample become poor. In this work, the fiducial $f_{\rm CTK}$ boost factor we take is $2.5$.}
    \label{fig:nhdist}
\end{figure*}

\begin{figure*}
    \centering
    \includegraphics[width=1\linewidth, trim={2cm 26cm 4.5cm 0}]{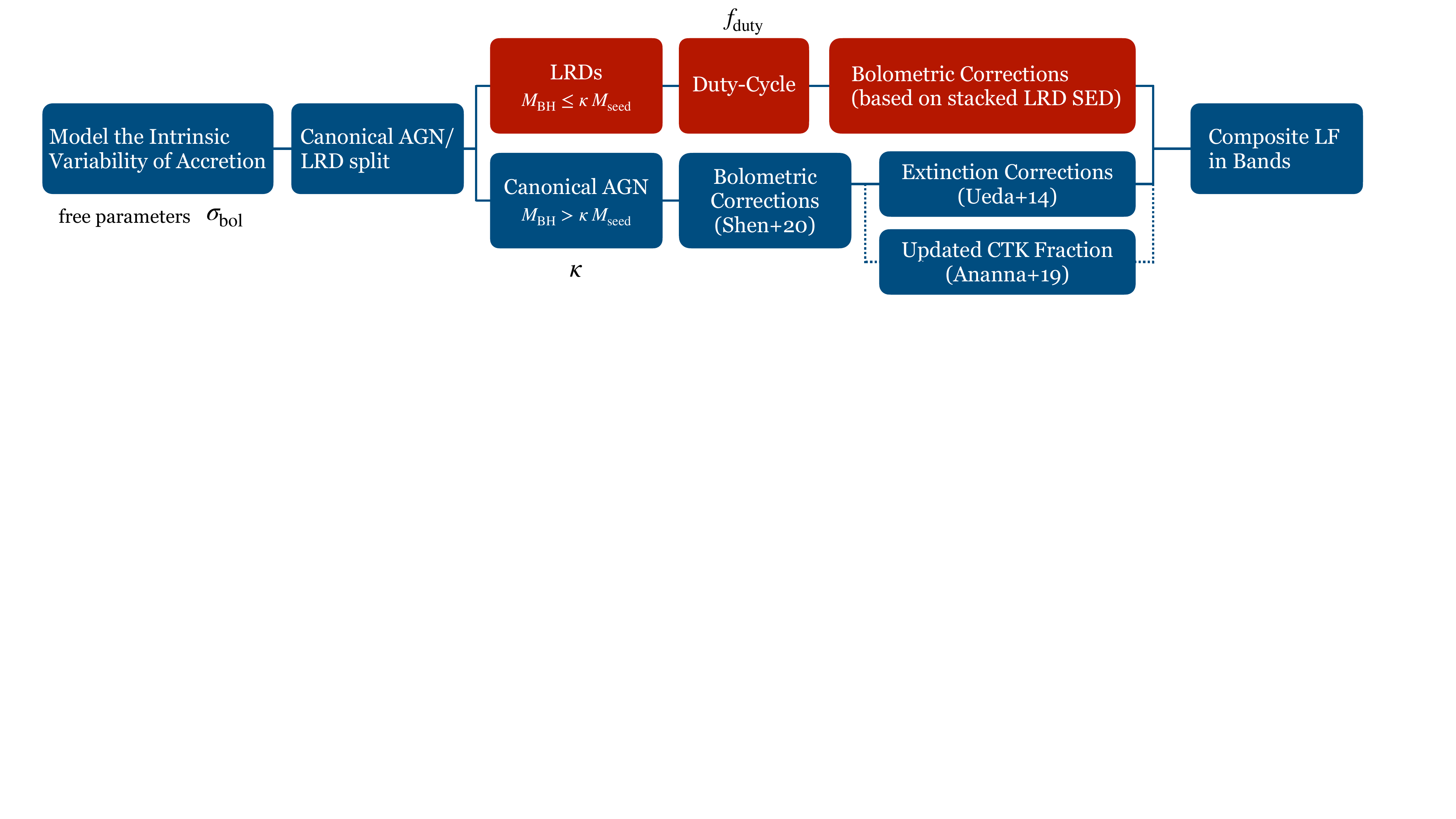}
    \caption{A schematic view of the empirical model for AGN LF. We start from the raw intrinsic bolometric luminosities of SMBHs informed from \luminasim and apply a scatter in bolometric luminosities due to unresolved variability of accretion (see Section~\ref{subsec:lbol-scatter}). Then we perform a split of canonical AGN and the ones in the LRD phase based on the BH particle mass in \luminasim. The former undergoes the standard AGN bolometric and extinction/absorption corrections to obtain the LFs in observed bands, while the latter relies on empirical bolometric corrections based on observed stacked LRD SEDs and has an additional duty-cycle parameter. The three free parameters of this model are highlighted in black.}
    \label{fig:pipeline}
\end{figure*}

\begin{figure}
    \centering
    \includegraphics[width=\linewidth]{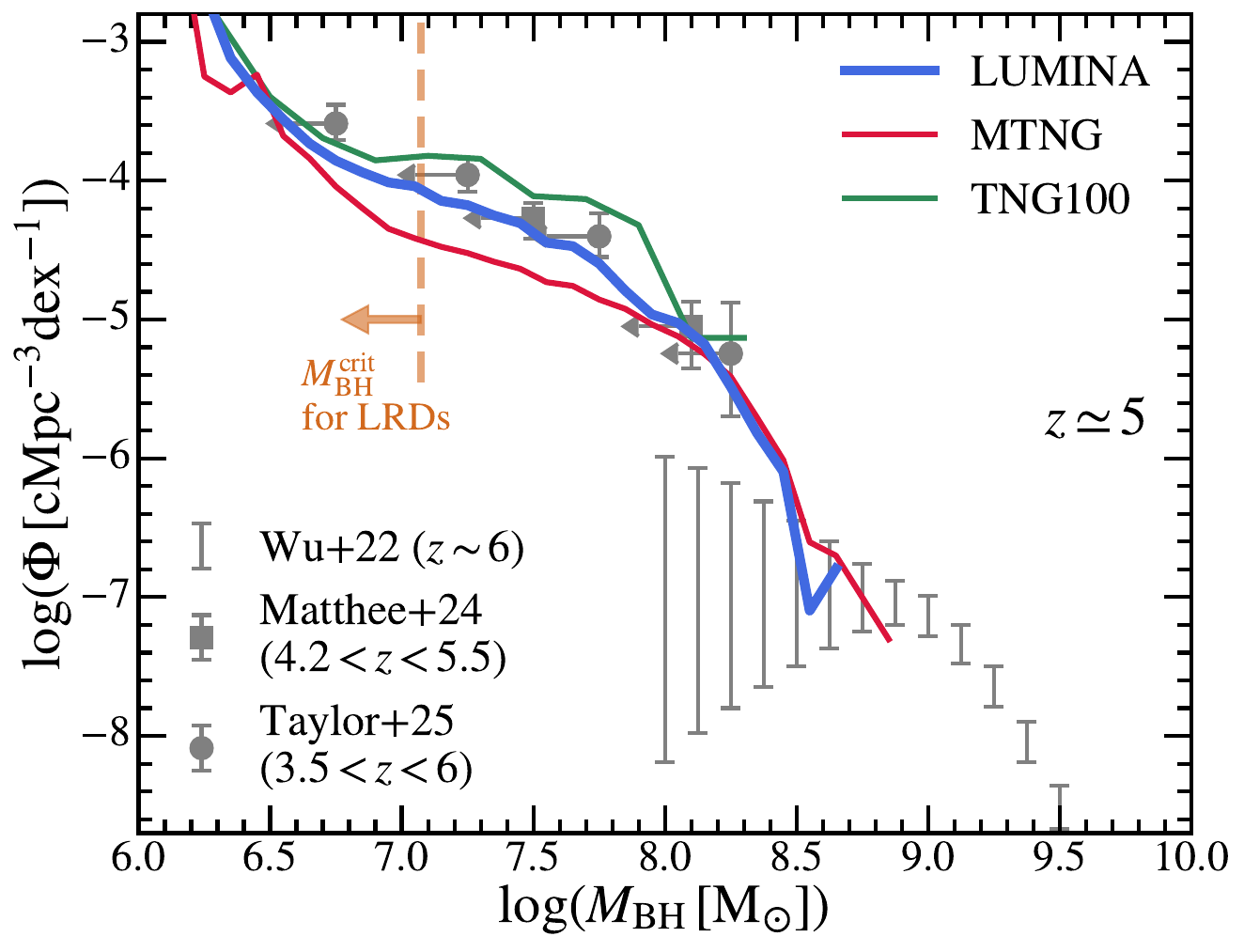}
    \caption{SMBH mass function at $z\simeq 5$ in \luminasim, \tng, and \mtng. We compare the results with observational constraints of BLAGN from \citet{Wu2022} for quasars and \citet{Matthee2024,Taylor2025} for LRDs. The latter should be considered as upper limits due to the large uncertainties in single-epoch BH mass estimates using local scaling relations. Specifically, we highlight the upper mass threshold ($M^{\rm crit}_{\rm BH}=10\,M_{\rm seed}$) we set for the LRD population with the orange dashed line.}
    \label{fig:bhmf}
\end{figure}

\begin{figure*}
    \centering
    \includegraphics[width=0.49\linewidth]{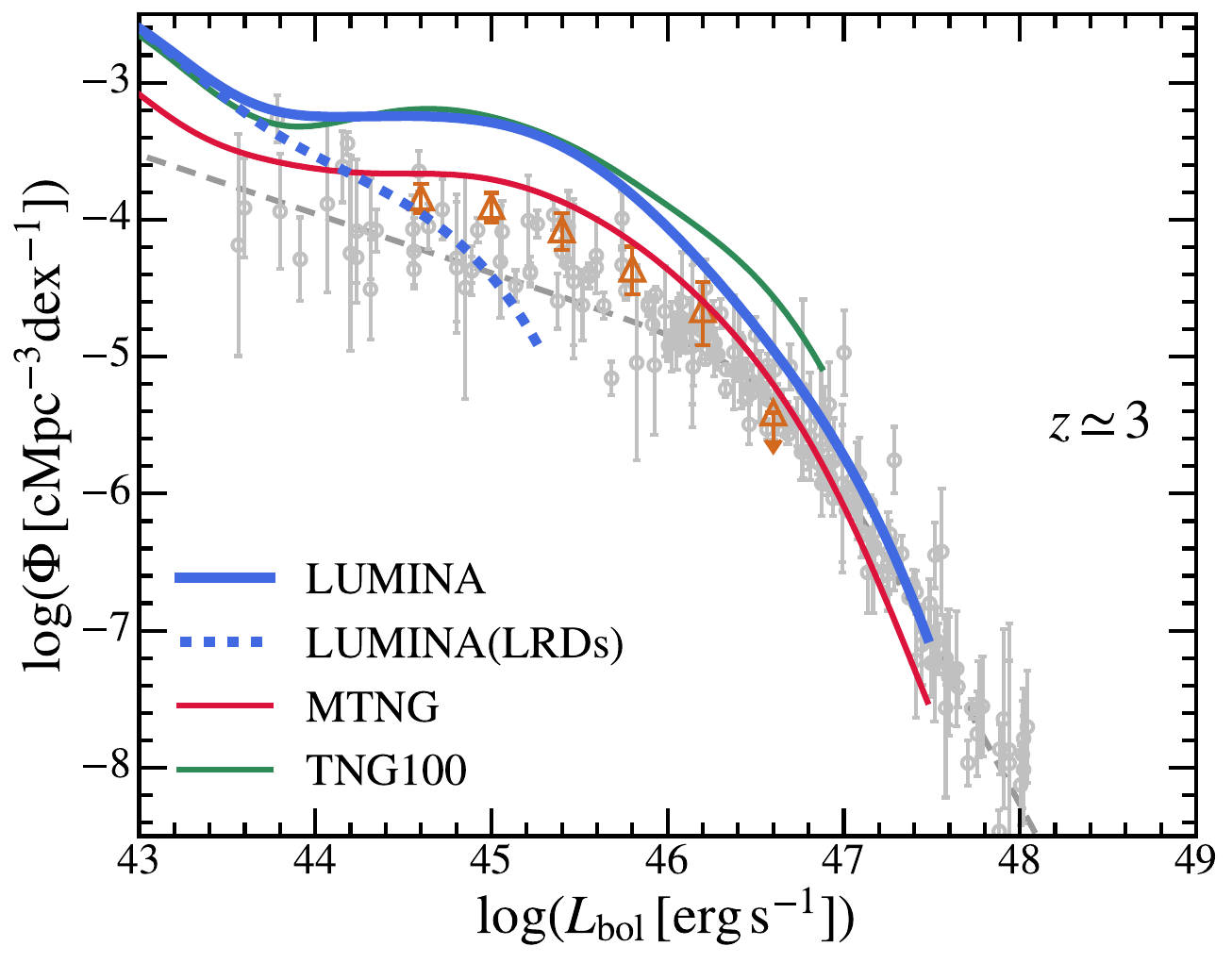}
    \includegraphics[width=0.49\linewidth]{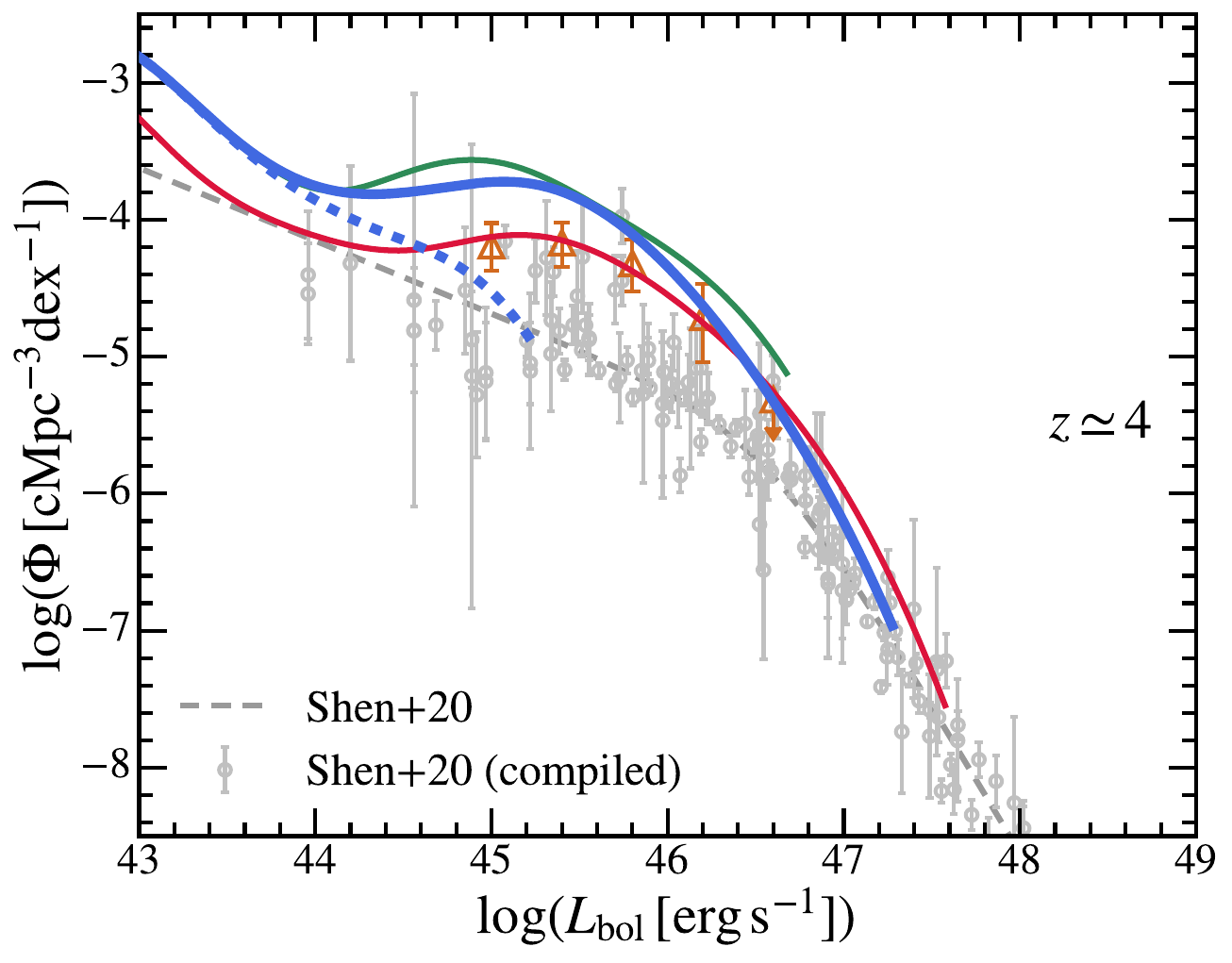}
    \includegraphics[width=0.49\linewidth]{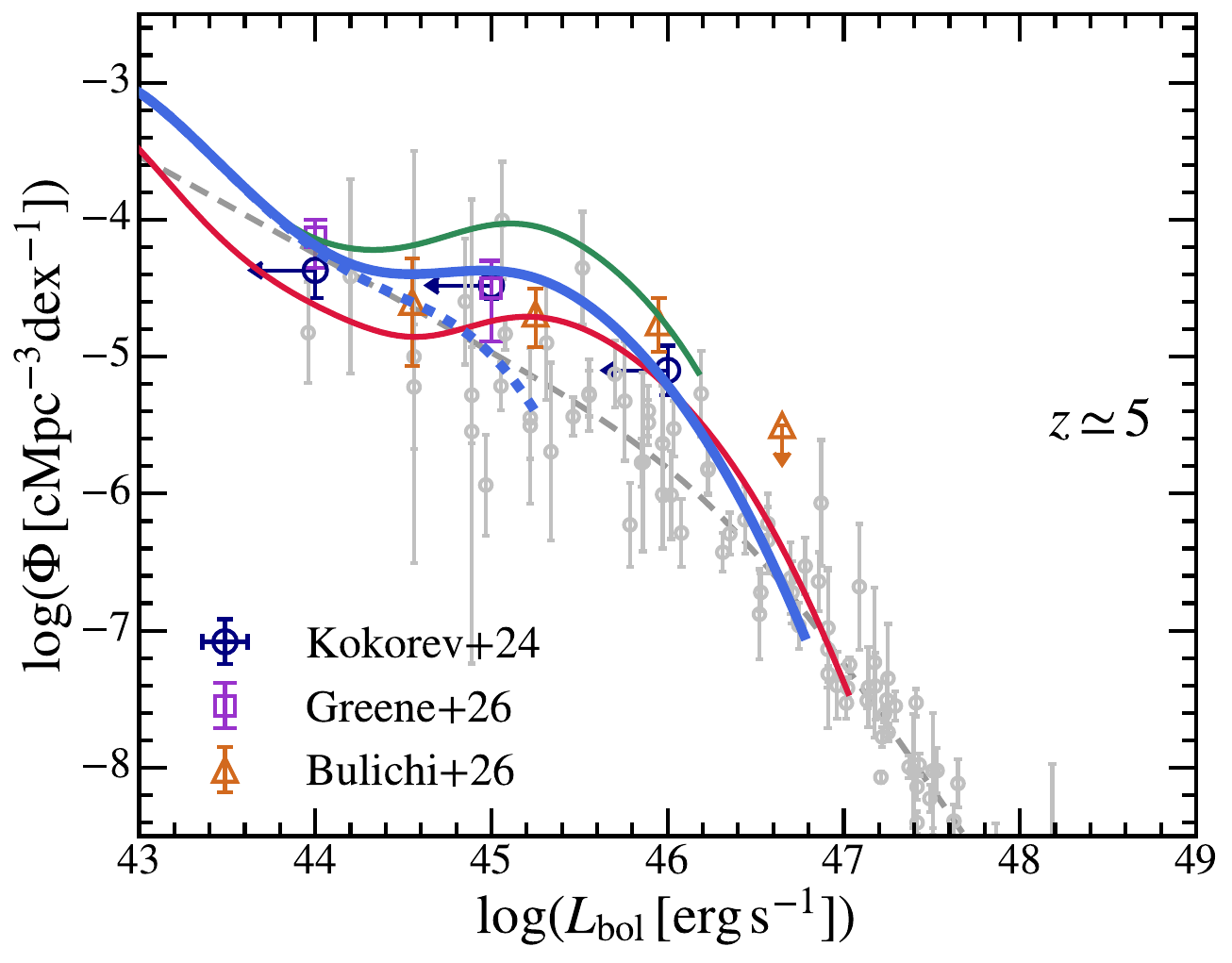}
    \includegraphics[width=0.49\linewidth]{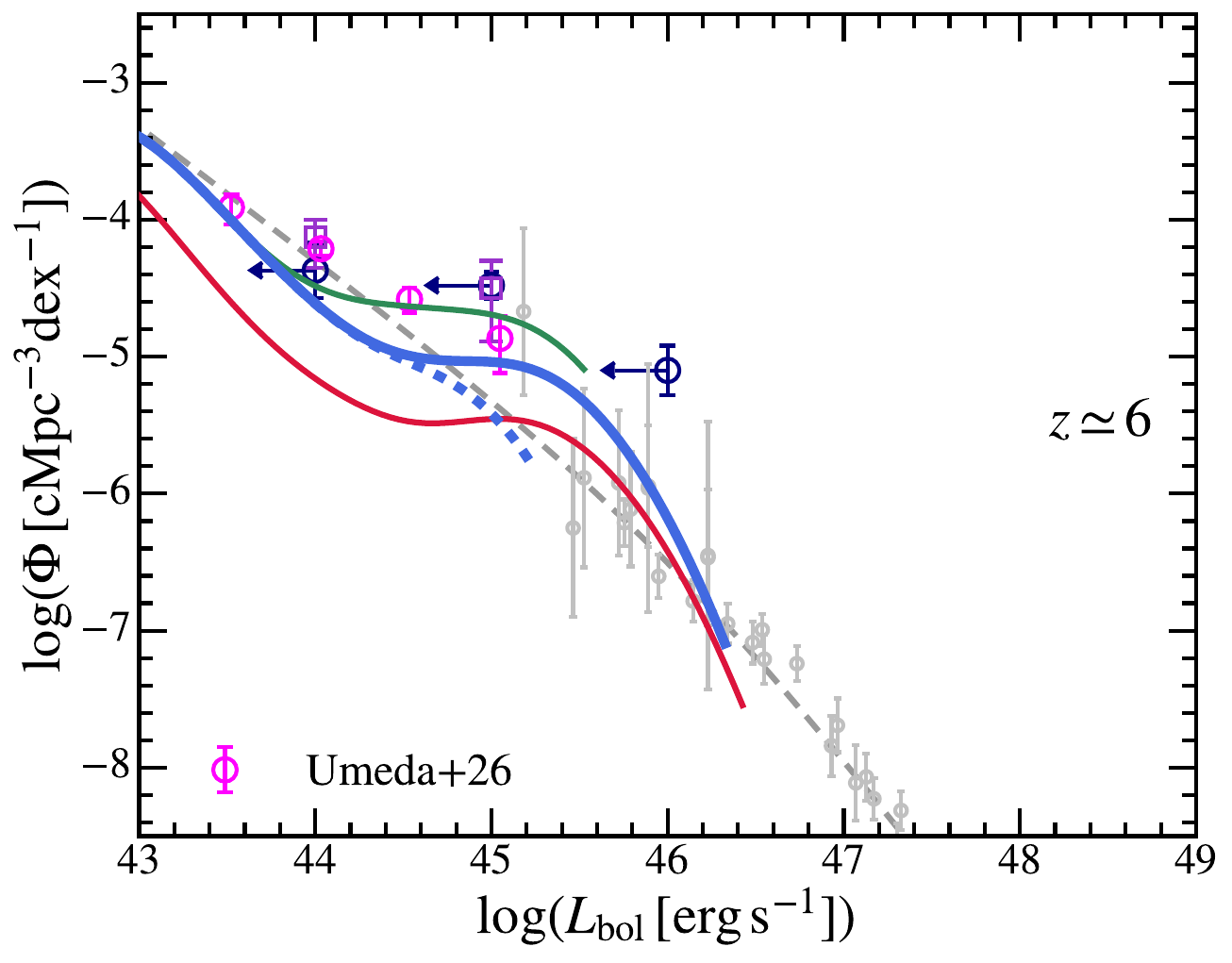}
    \caption{Bolometric LFs of AGN at $z=3-6$. We present the bolometric LFs of SMBHs in \luminasim, \tng, and \mtng. The blue dashed line shows the contribution by LRDs. These simulation predictions are compared to the best-fit quasar LF in \citet{Shen2020} (the ``Global fit A'') and the compiled observational data there (moved to the bolometric plane by preserving the offsets to the best-fit model). We also compare our results to the bolometric LF of LRDs \citep[e.g.,][]{Kokorev2024,Greene2026,Umeda2026} and obscured AGN in the MEOW survey \citep{Bulichi2026}.}
    \label{fig:bolqlf}
\end{figure*}

\subsection{Gas absorption and dust extinction}
\label{subsec:extinction}

The absorption and scattering of surrounding gas and dust further modify the intrinsic emission of AGN. The photoelectric absorption of gas influences X-ray emission, while dust plays a key role in the optical and UV. Here, we first introduce a model for the neutral hydrogen column density ($N_{\rm H}$) distribution, which determines the absorption in the X-ray. $N_{\rm H}$ is then converted to the column density of dust, assuming a dust-to-gas ratio, which determines the dust extinction in the optical and UV. We note that the absorption/extinction model described below is applied to the canonical AGN. For LRDs, we do not apply any additional correction on top of the SED we adopt, since we assume that it has already been reprocessed by the gas (and potentially dust) around the central BH.

Large-volume cosmological simulations cannot resolve the circumnuclear gas distribution and underestimate the gas column density along the line of sight \citep[see the measurement done in][]{Ni2022}. Therefore, an observation-driven empirical model is required. Following \citet{Shen2020}, we adopt the $N_{\rm H}$ distribution from \citet{Ueda2014}, based on measurements of $N_{\rm H}$ and intrinsic hard X-ray luminosity for individual objects in their sample. The model provides the probability distribution of $N_{\rm H}$ as a function of intrinsic hard X-ray luminosity and redshift. We refer readers to \citet{Ueda2014} for the detailed functional form and parameters. In this model, the abundance of CTK $\left(\log{(N_{\rm H}/\cm^{-2})}\geq 24\right)$ AGN is assumed to be the same as absorbed Compton-thin (CTN) $\left(\log{(N_{\rm H}/\cm^{-2})=22-24}\right)$, with both fractions decreasing at higher hard X-ray luminosities and increasing at higher redshift with a plateau at $z\geq 2$. A more recent X-ray AGN synthesis model \citep{Ananna2019} that includes ultra-hard X-ray data from \textit{NuSTAR} suggests an enhanced CTK AGN fraction compared to previous measurements, which is remarkably consistent with \citet{Carnall2023} using an independent method. As shown in Figure~\ref{fig:nhdist}, we find that boosting the CTK AGN fraction relative to the obscured CTN population ($f_{\rm CTK}$) by a factor of $2-3$ in the \citet{Ueda2014} model can mimic these new constraints at $z\simeq 4$. The same agreement is consistently found at $3 \lesssim z \lesssim 5$ where the \citet{Ananna2019} model covers. In this work, the fiducial boost factor we adopt is $2.5$. We need to note that the fraction of CTK AGN (even in the Local Universe) remains observationally uncertain and hotly debated in the literature, with lower values reported in e.g. \citet{Buchner2015,Ricci2015,Boorman2025} that are more consistent with the original \citet{Ueda2014} model. Therefore, the update we apply here should be viewed as a conservative upper bound motivated by new population synthesis models. As will be shown later in Section~\ref{subsec:parameters}, the impact of $f_{\rm CTK}$ on the resulting LFs is in fact quite limited.

Given $N_{\rm H}$, we next apply a multiplicative factor $\exp(-N_{\rm H}\,[\sigma_{\rm pe}(E) + \sigma_{\rm T}])$ to the intrinsic luminosity, where $\sigma_{\rm pe}(E)$ is the photoelectric absorption cross-section from \citet{Morrison1983}, and $\sigma_{\rm T}$ is the Thomson scattering cross-section, the non-relativistic limit of the Compton (Klein-Nishina) cross-section, which is accurate to within a few percent throughout the $2-10\keV$ hard X-ray band of interest. We note that this screen-absorption treatment is a simplification of the full radiative transfer of X-ray photons through circumnuclear material, and can be inaccurate deep in the CTK regime. It does not capture e.g. multi-scattered photons that return to the line of sight in optically thick media, variations of the Compton-reflection continuum produced by scattering off circumnuclear material, and the fluorescent Fe K$\alpha$ line that contributes a small but non-zero fraction of the integrated hard X-ray luminosity. Taken together, these approximations imply that our predicted $2-10\keV$ AGN luminosities are robust for the CTN population but are approximate for the heavily CTK sources. Nevertheless, the impact on the LF is limited since the latter population contributes little to the post-absorption LF. A future treatment based on Monte Carlo radiative-transfer torus models, such as MYTorus \citep{MurphyYaqoob2009}, BNTorus/borus02 \citep{Brightman2011BN,Balokovic2018}, or the more recent clumpy-torus calculations \citep{Tanimoto2019,Buchner2019}, would refine these predictions.

To determine the dust abundance, a dust-to-gas ratio is required along with the extinction curve as observed in the Small Magellanic Cloud \citep[SMC;][]{Pei1992}, motivated by direct fits to individually reddened quasars \citep[e.g.,][but see also \citealt{Gallerani2010}]{Richards2003, Hopkins2004, Krawczyk2015, Zafar2015} over the flat \citet{Gaskell2004} or Milky Way extinction curve. In addition, the SMC-like dust composition is the natural expectation for the metal-poor environments of high-redshift hosts, which is the regime relevant to this work. Note that this is different from the choice in \citet{Shen2020} (where a Milky Way extinction curve is chosen for purely empirical reasons to improve the consistency between X-ray and UV/optical LFs) but consistent with \citet{Hopkins2007}. Observations have revealed that the gas-phase mass-metallicity relation of galaxies overall decreases with redshift \citep[e.g.,][]{Lilly2013,Zahid2013,Sanders2021}. Assuming that the dust-to-metal ratio remains constant, the decrement in the gas-phase metallicity of AGN host galaxies will lead to a decrement in the dust-to-gas ratio at higher redshift. We choose to adopt a redshift-dependent dust-to-gas ratio that scales as the gas-phase metallicity given by the fit in \citet{Ma2016}, which is consistent with recent \textit{JWST} constraints on high-redshift mass-metallicity relations \citep[e.g.,][]{Nakajima2023,Curti2024,Sarkar2025}. The value of the dust-to-gas ratio in the local Universe still follows \citet{Hopkins2007} with $(A_{\rm B}/N_{\rm H})=8.47 \times10^{-22} \cm^{2}$. 

The absorption and extinction model and the bolometric corrections allow us to link the bolometric LF with the observed one in a certain band. Unless otherwise specified, the AGN LFs presented in this paper include both the obscured and unobscured AGN, and the observed AGN LFs account for gas absorption and dust extinction.

\subsection{An empirical split of LRDs and canonical AGN}
\label{subsec:split}

We assume that a subset of LRDs traces the earliest, still-unresolved growth phase of SMBHs from their seeds \citep[e.g.,][]{Inayoshi2025b}. This interpretation is physically motivated by a broad class of massive-seed scenarios, such as quasi-stars as a phase of direct-collapse BHs \citep[e.g.,][]{Begelman2006,Begelman2008,Volonteri2010,Begelman2026}, the collapse of supermassive stars \citep[e.g.,][]{Begelman2010,Hosokawa2013,Woods2017,Pacucci2025}, and DM-seeded BHs \citep[e.g.,][]{Xiao2021,ShenT2025,Jiang2026}. We introduce a simple empirical mass cut: for BHs with $M_{\rm BH} \leq M^{\rm crit}_{\rm BH} \equiv \kappa M_{\rm seed}$, we assume that the object is in an early-growth phase that is not properly resolved in the simulation. In this regime, the BH particle should be interpreted not as an isolated, fully exposed SMBH, but as a sub-grid tracer of an unresolved nuclear system consisting of the seed BH, its dense ambient gas reservoir, and possibly a surrounding nuclear star cluster in a proto-galactic nucleus. We defer a more detailed discussion of this interpretation to Section~\ref{sec:discussion}. We then assume that this unresolved early-growth population appears observationally as LRDs and assign it the representative LRD SED described above in Section~\ref{subsec:lrdsed}. We adopt the LRD bolometric correction factors in Table~\ref{tab:lrd-corr} and do not apply any additional absorption or extinction corrections, since the assumed LRD SED already represents radiation reprocessed by the surrounding material. We also introduce a duty cycle, $f_{\rm duty}$, such that only a fraction of these low-mass SMBHs are active and observable as LRDs at any given time. The remaining $(1-f_{\rm duty})$ population is treated as electromagnetically silent in the bands we study in this work. In Section~\ref{subsec:parameters}, we discuss the impact of this assumption. BHs above the mass threshold are instead treated as canonical AGN and follow the bolometric correction and absorption/extinction correction described in the Sections above. A summary of this pipeline is shown in Figure~\ref{fig:pipeline}. Unless otherwise stated, we adopt $\kappa=10$ (which results in $M^{\rm crit}_{\rm BH}\sim 10^{7}\msun$) and $f_{\rm duty}=0.3$ as our fiducial values, motivated by comparisons to the observed AGN LFs as will be shown in Section~\ref{sec:results}.

In Figure~\ref{fig:bhmf}, we show the SMBH mass function at $z\simeq 5$ in \luminasim, compared to results from \tng \citep{Springel2018} and \mtng \citep{Pakmor2023}, which adopt the same galaxy formation model but differ in both mass resolution and simulation volume. The \tng volume is $\sim 1/92$ that of \luminasim, but it has $\sim 2.6$ times better mass resolution and includes magnetic fields. In contrast, \mtng has a volume $\sim 3.2$ times larger than \luminasim, but with $\sim 8.6$ times poorer mass resolution. We illustrate the fiducial $M^{\rm crit}_{\rm BH}$ below which we model BHs as LRDs. For comparison, we also show the SMBH mass function from observations for luminous quasars \citep{Wu2022} and LRDs \citep{Matthee2024,Taylor2025}. The LRD constraints should be interpreted as upper limits, given the substantial uncertainties in the broadening mechanism of the Balmer lines \citep[e.g.,][but see also \citealt{Brazzini2025,Brazzini2026,Scholtz2026}]{Naidu2025,Rusakov2026} and in single-epoch virial mass estimators calibrated using normal AGN more generally. We note that we do not interpret the simulated BH particle masses as a one-to-one tracer of the physical SMBH mass, so the comparison here is only to give a qualitative sense of the BH particle mass distribution in the simulation. In the remaining sections of the paper, we will focus on the AGN LFs rather than the SMBH mass function as the primary point of comparison with observations.

\section{Results}
\label{sec:results}

\subsection{Bolometric LFs}
\label{subsec:bol}

We begin with the bolometric LF of AGN. This quantity is particularly well suited for comparison with theoretical models because it is directly tied to the BH mass accretion rates and total energy output predicted by simulations (see Equation~\ref{eq:lbol-mdot}). In Figure~\ref{fig:bolqlf}, we show the bolometric LFs of AGN in \luminasim at $z\simeq 3-6$. For comparison, we also show results from \tng and \mtng. At the luminous end, \luminasim and \mtng show good convergence across redshift. At the faint end, \luminasim is converged with \tng at $z\lesssim 4$, whereas a mild resolution dependence persists at higher redshift, with \tng predicting a higher abundance of low-luminosity AGN. Given this residual resolution dependence, we remain conservative and avoid drawing overly precise quantitative conclusions at the faint end at $\lesssim 0.2$ dex level. Another notable feature is the upturn at the very faint end ($L_{\rm bol}\lesssim 10^{44}\erg\,{\rm s}^{-1}$). A similar feature is also visible in the SMBH mass functions shown in Figure~\ref{fig:bhmf}. These features are likely associated with the relatively large seed BH mass adopted in the simulations, since the Eddington luminosity of our seed BHs happens to be $L_{\rm Edd}\sim 10^{44.2}\erg\,{\rm s}^{-1}$. Newly seeded BHs are inserted directly at $M_{\rm seed}$, and the unresolved early growth phase of lower-mass BHs is therefore not explicitly followed. It is one of the key reasons for us to develop the empirical model for low-mass BHs to encapsulate physics in the unresolved phase. The feature can be amplified by the duty-cycle of LRDs in our model, which suppresses the LRD contribution to the bolometric LFs at the faint end. This preferential suppression for LRDs (rather than normal AGN) results in a bump feature in the bolometric LF at $\sim 10^{45}\erg\,{\rm s}^{-1}$, which roughly corresponds to the Eddington luminosity of a BH at $M^{\rm crit}_{\rm BH}$. We acknowledge that a more physically meaningful split of LRDs and normal AGN could avoid these artifacts and result in better agreement with data.

\begin{figure}
    \centering
    \includegraphics[width=\linewidth]{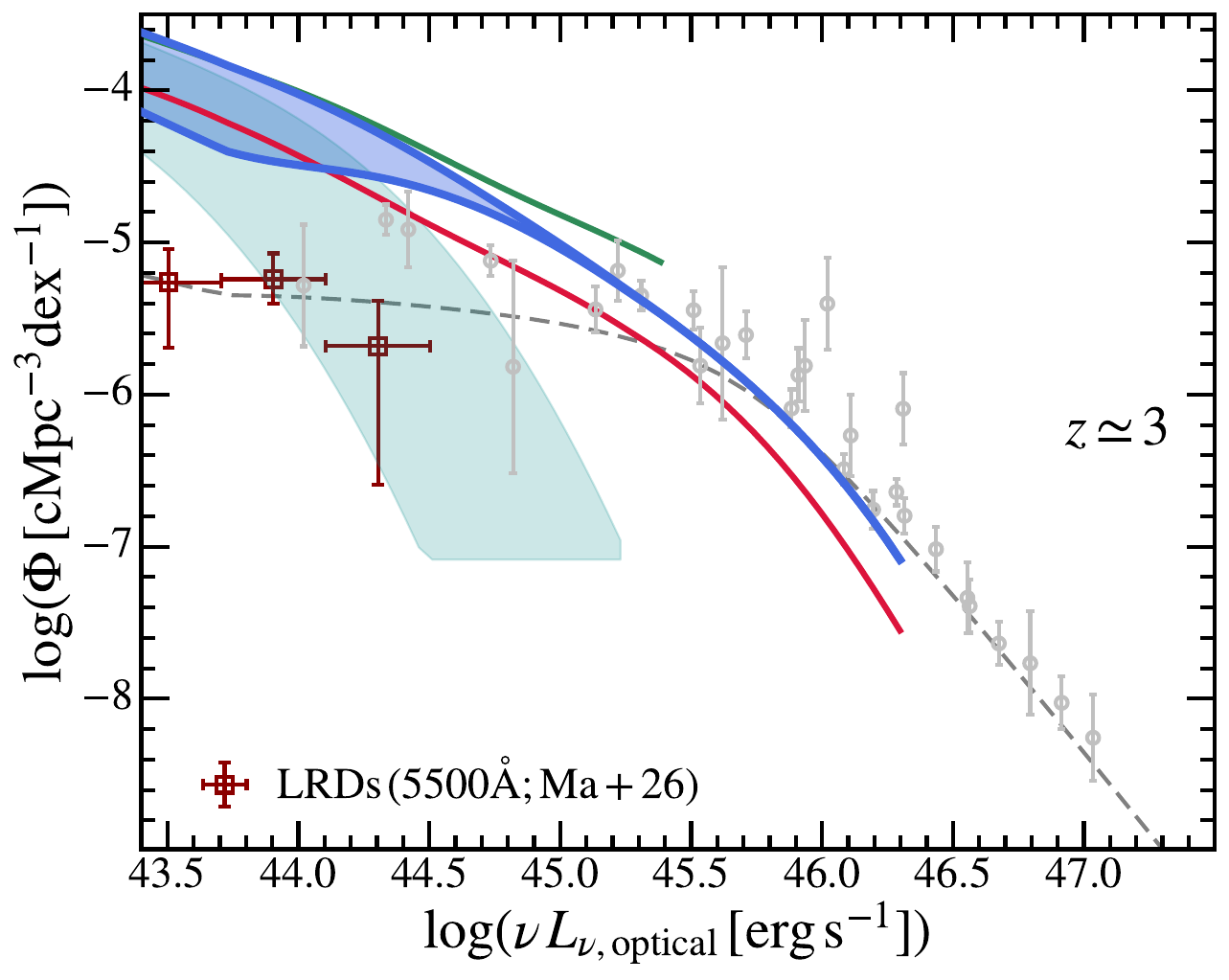}
    \includegraphics[width=\linewidth]{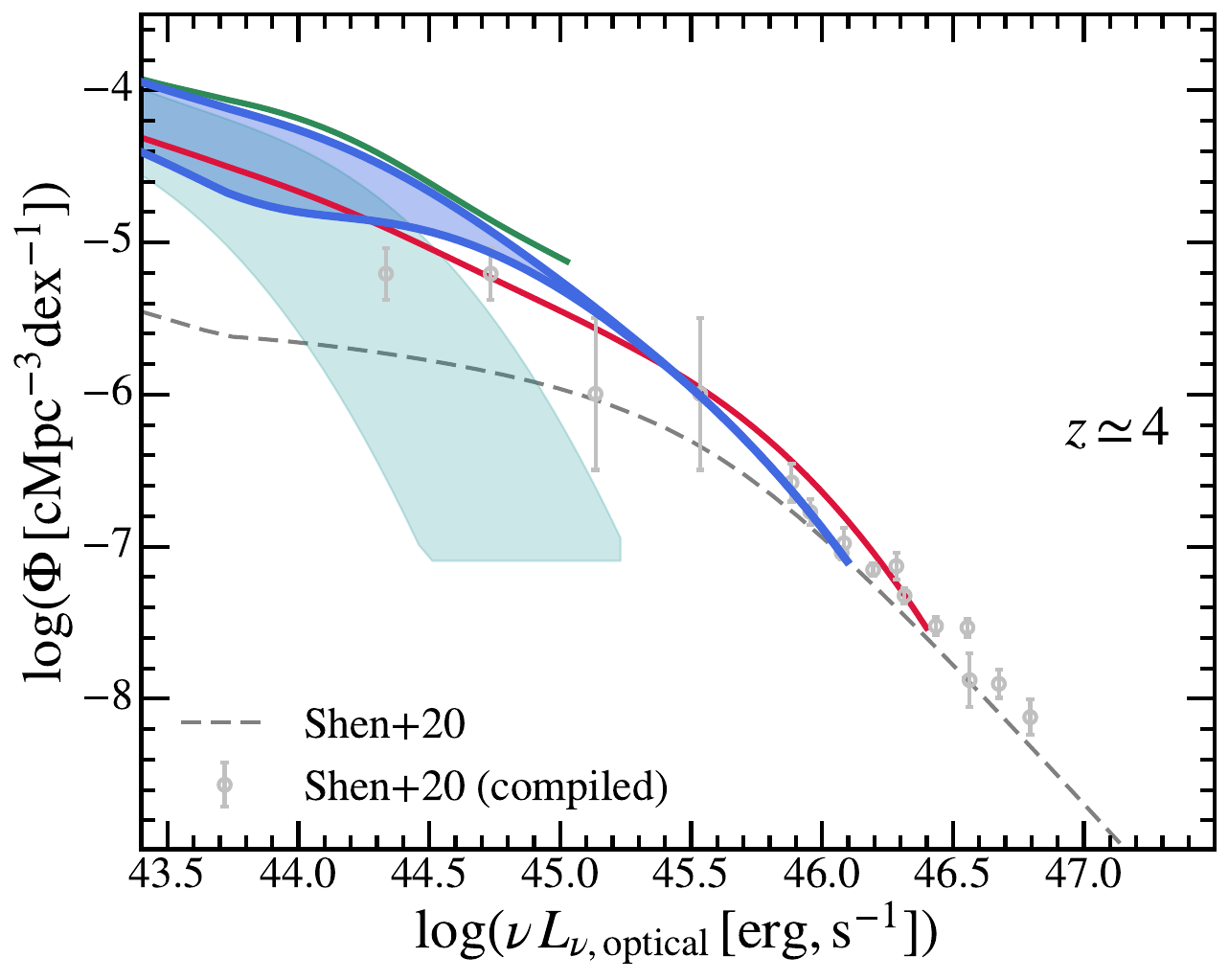}
    \includegraphics[width=\linewidth]{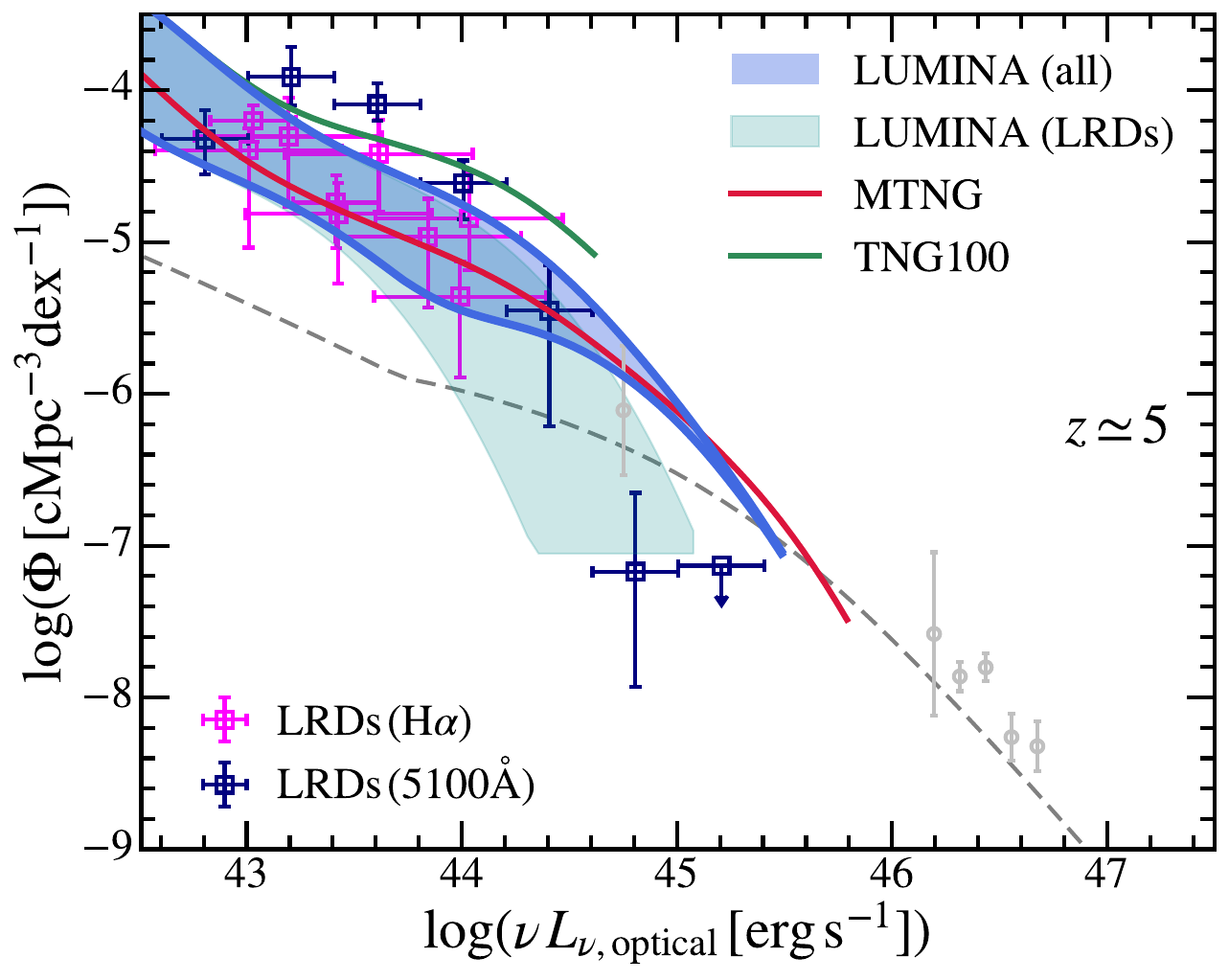}
    \caption{Rest-frame optical (B band, $\sim 4400$\AA) LFs of AGN at $z\simeq 3-5$ from \luminasim, \tng, and \mtng. The shaded region represents the lower and upper limits of the LFs given uncertainties in bolometric corrections of the LRDs. We compare the results to optical observations compiled in \citet{Shen2020} and LRD constraints from \citet{Kokorev2024,Ma2025} (at $5100$\AA) and \citet{Matthee2024,Lin2024,Lin2026} based on H$\alpha$ at $z\sim 5-6$ and the ones from \citet{Ma2026} (at $5500$\AA) at $z\sim 3$. The LRD constraints are all corrected to the B band luminosity using the SED model in Section~\ref{subsec:lrdsed} and the scaling relation in Equation~\ref{eq:5100-halpha}.}
    \label{fig:optlf}
\end{figure}

We compare the simulation predictions to the pre-\textit{JWST} observational constraints compiled by \citet{Shen2020}, which include measurements in the UV, optical, soft and hard X-ray, and mid-IR bands, and the best-fit model therein. These multi-wavelength LFs are mapped onto the bolometric plane using bolometric corrections therein, while preserving the offset in $\log\Phi$ between each observed data point and the best-fit model LF. We choose the ``Global fit A'' from \citet{Shen2020}, which has a relatively steep faint-end slope driven mainly by the UV data in that compilation. We note that, in preparing these observational data, we keep the original \citet{Shen2020} model without any modifications (e.g., the change in $f_{\rm CTK}$) introduced above. With our fiducial choice of $\sigma_{\rm bol}=0.3$ dex, the bolometric LF in \luminasim agrees well with the observational constraints for moderately luminous quasars with $L_{\rm bol}\gtrsim 10^{46}\erg\,{\rm s}^{-1}$. At even higher luminosities, the finite simulation volume becomes insufficient to sample the rarest bright quasars. In Section~\ref{subsec:parameters}, we compare the fiducial LF to versions with different levels of scatter, and also place our results in the context of other simulations in Section~\ref{subsec:model-comparisons}.

At the faint end, we first include recent bolometric LF constraints inferred from observed LRDs. \citet{Kokorev2024} inferred bolometric luminosities of LRDs from their rest-frame optical luminosities assuming standard AGN bolometric corrections and dust corrections \citep[e.g.,][]{Greene2005,Richards2006a}. However, more recent work by \citet{Greene2026} suggests that LRDs have substantially different bolometric corrections from classical unobscured AGN (see our discussion in Section~\ref{subsec:lrdsed} above). We therefore treat the luminosities inferred by \citet{Kokorev2024} as conservative upper limits. Updated constraints on the LRD bolometric LF have been presented by \citet{Greene2026}. Similar conclusions have also been reached by \citet{Umeda2026}, who estimated bolometric luminosities using a BH-envelope model. In addition to LRDs, we further include the mid-IR-based constraints from the \textit{JWST} MIRI Early Obscured AGN Wide Survey \citep[MEOW; Leung in prep.,][]{Bulichi2026}, which provide an independent census of obscured AGN activity at high redshifts.

Overall, recent \textit{JWST} observations probing the faint end of the AGN LFs ($L_{\rm bol}\lesssim 10^{46}\erg\,{\rm s}^{-1}$) suggest an enhanced AGN abundance relative to previous UV and X-ray surveys. These new measurements are broadly consistent with the predictions from \luminasim. Meanwhile, relative to the pre-\textit{JWST} compilation of \citet{Shen2020}, \luminasim predicts a higher abundance of low-luminosity AGN. This discrepancy is expected, since the faint-end LF in \citet{Shen2020} is largely constrained by X-ray observations, which are likely incomplete if a substantial fraction of LRDs and other faint high-redshift AGN are intrinsically X-ray weak or heavily obscured by gas \citep[e.g.,][]{Yue2024, Maiolino2025}. We revisit this issue in Section~\ref{subsec:xray}, where we compare the X-ray LFs directly.

\begin{figure*}
    \centering
    \includegraphics[width=0.49\linewidth]{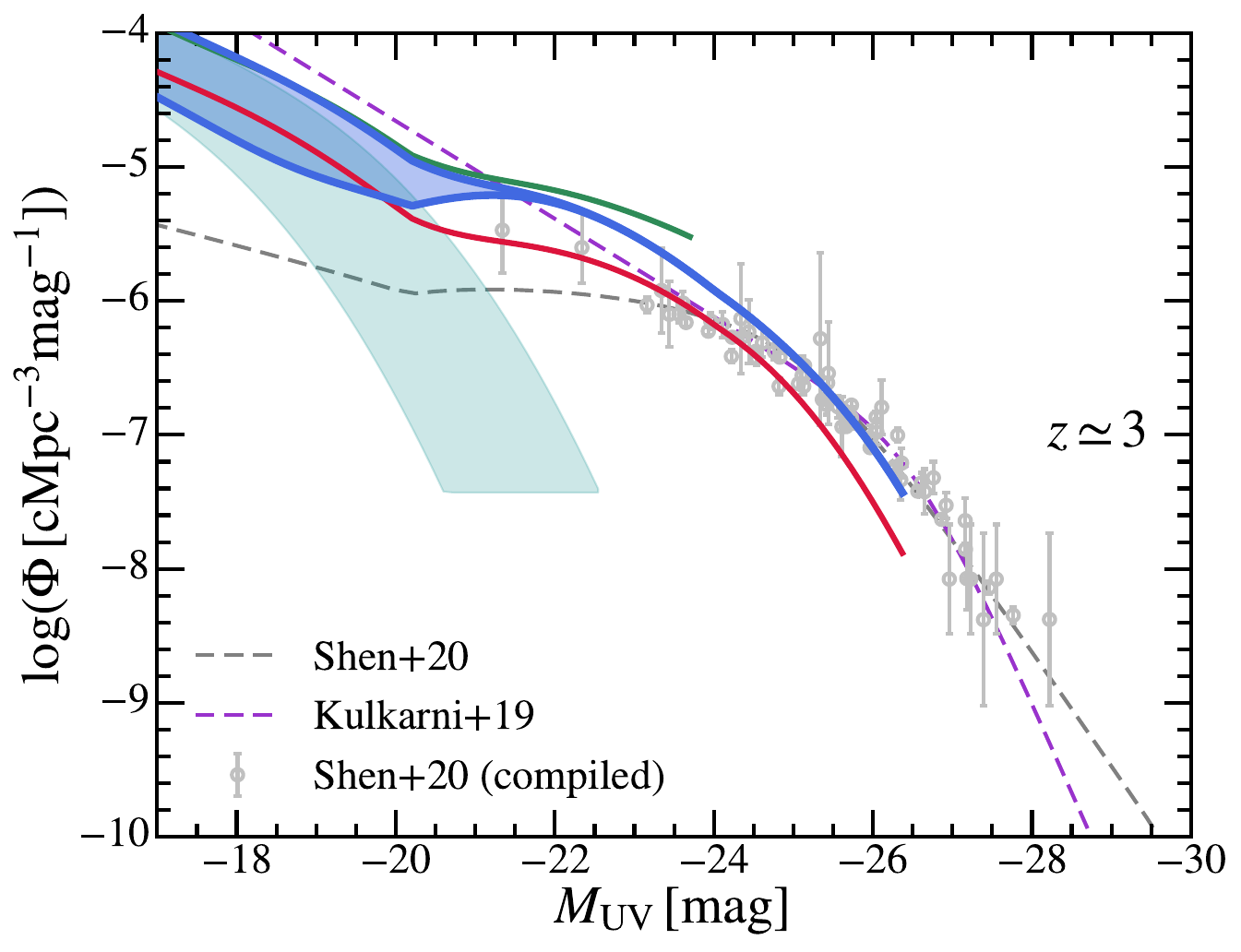}
    \includegraphics[width=0.49\linewidth]{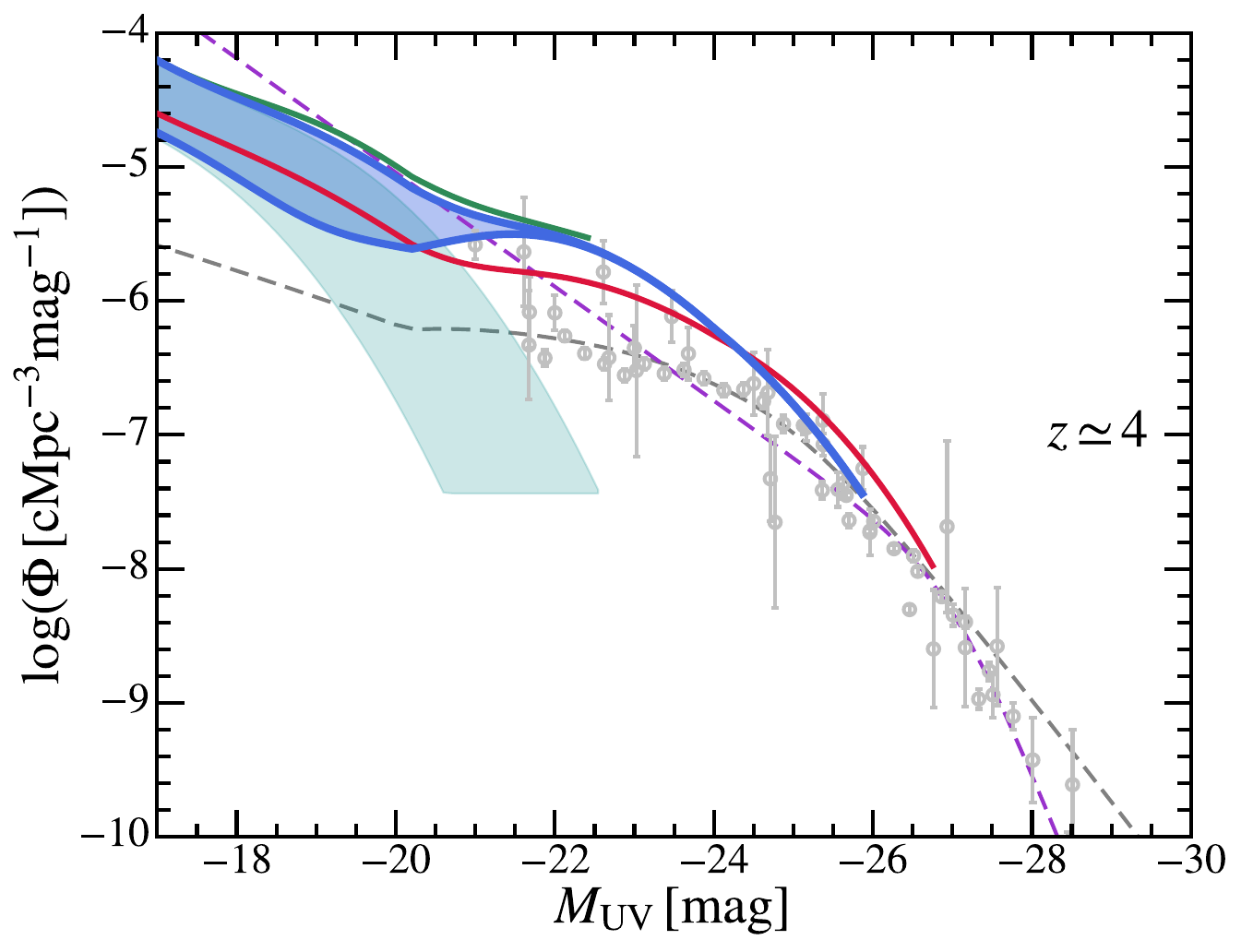}
    \includegraphics[width=0.49\linewidth]{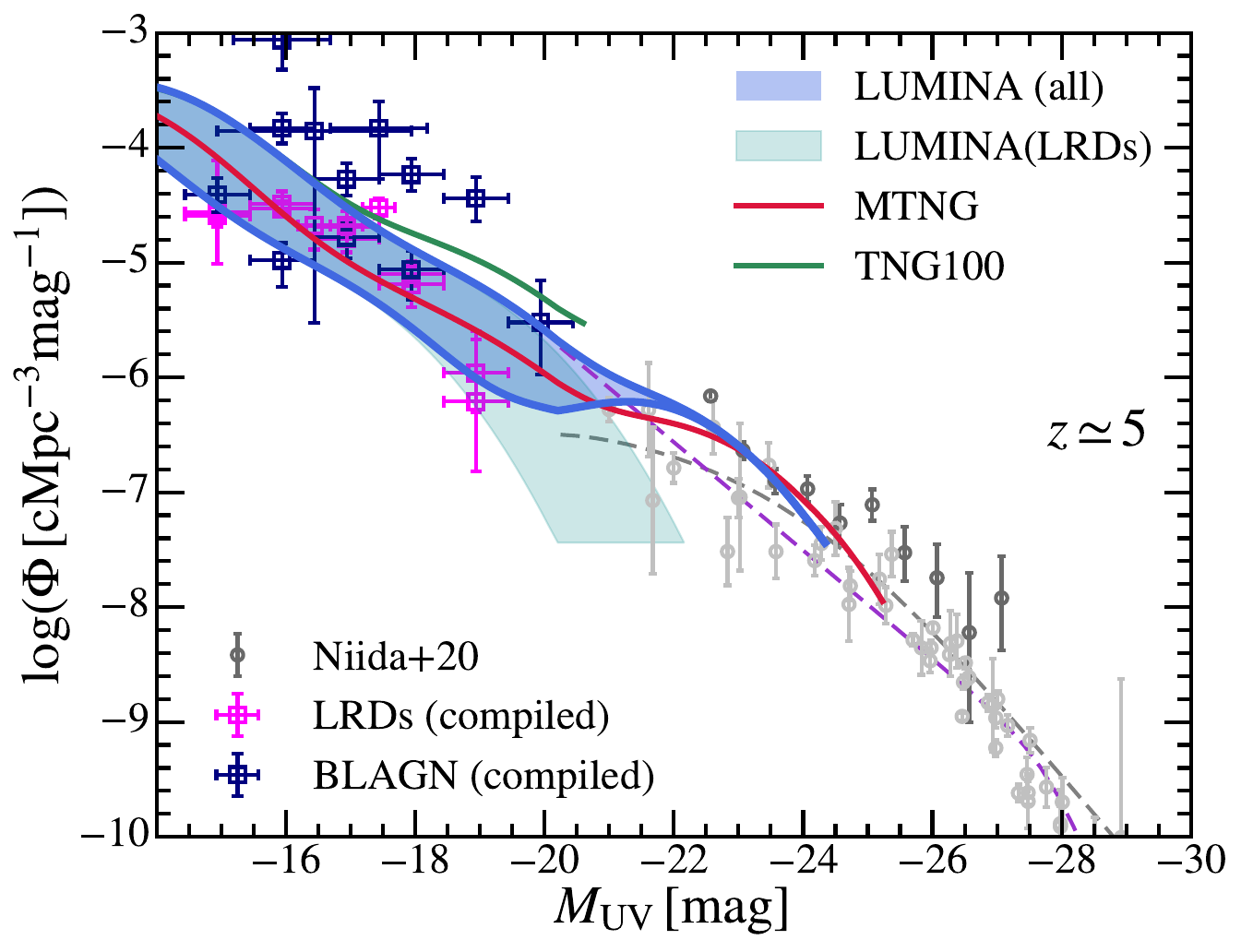}
    \includegraphics[width=0.49\linewidth]{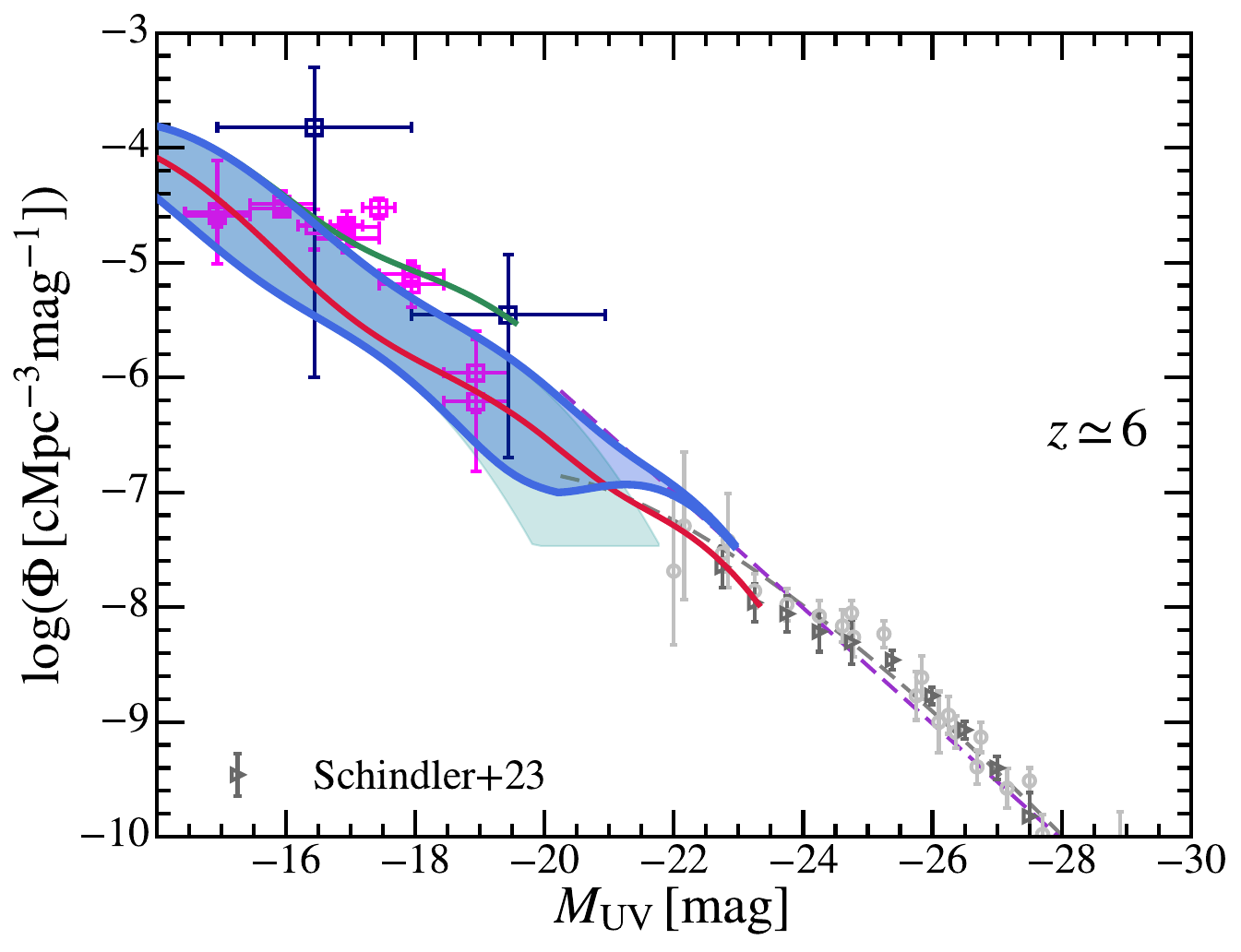}
    \caption{Rest-frame UV LFs of AGN at $z=3-6$ from \luminasim, \tng, and \mtng. We compare them with the UV observations compiled in \citet{Shen2020} and additional constraints from \citet{Niida2020,Schindler2023}. For reference, we also show the best-fit AGN UV LF from \citet{Shen2020} and \citet{Kulkarni2018}. For LRDs, we show two groups of observational constraints: the BLAGN LFs from \citet{Matthee2024,Maiolino2024,Harikane2023-agn,Taylor2025} and the LRD LFs from \citet{Kokorev2024,Greene2024,Kocevski2025}, where the latter have additional explicit/implicit ``V-shape'' SED selection criteria. We assume that AGN account for only $15\%$ of the UV luminosity in LRDs \citep{Sun2026} and shift them to the fainter end accordingly.}
    \label{fig:uvqlf}
\end{figure*}

\begin{figure}
    \centering
    \includegraphics[width=\linewidth]{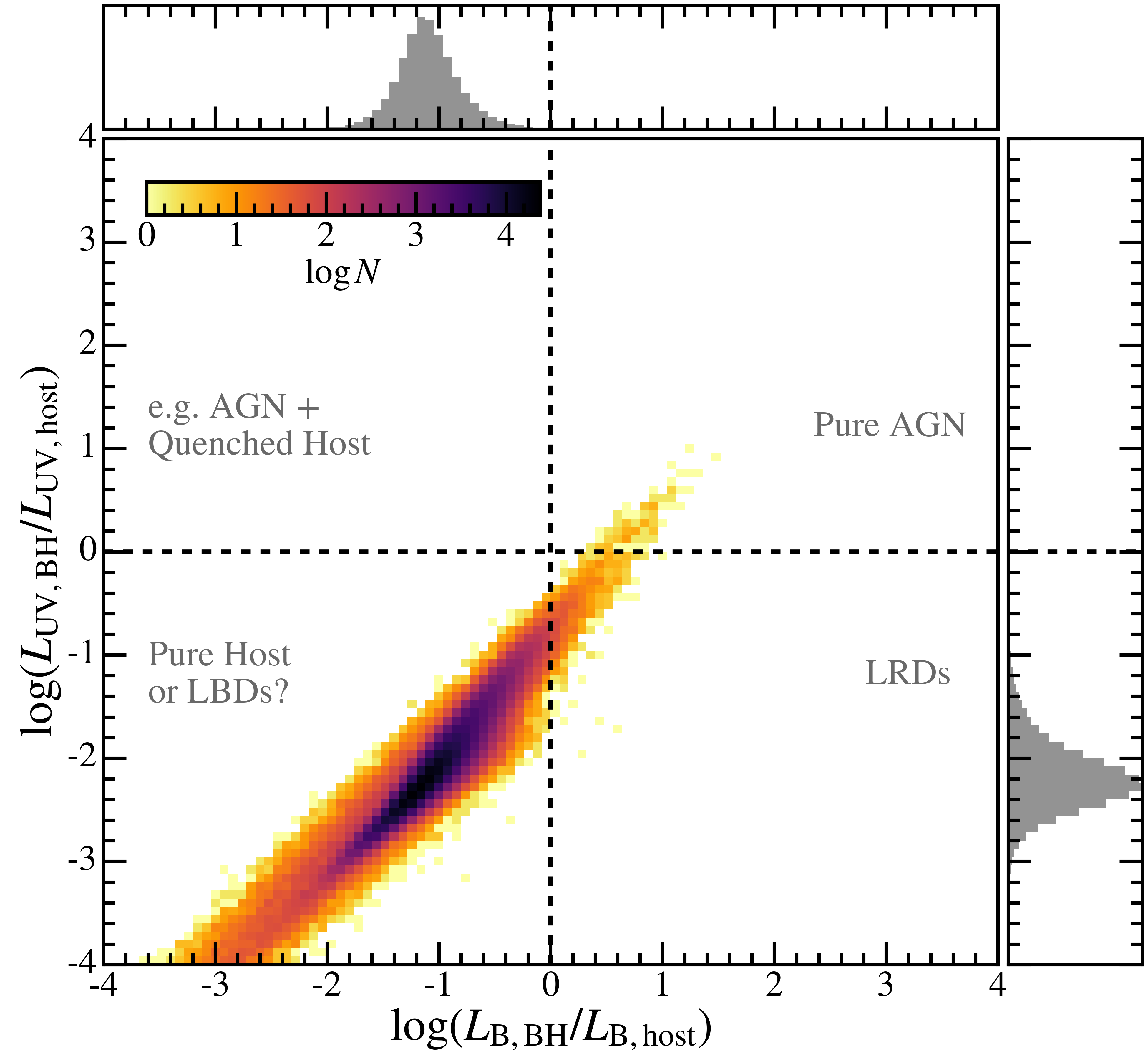}
    \caption{The luminosity ratio between BH and host galaxy in the UV versus in the optical B band at $z\simeq 5$. We show the distribution of BHs below the LRD selection mass threshold $M^{\rm crit}_{\rm BH}$ on this plane. The majority of the BHs are dominated by their host galaxies in the UV and optical, which raises the concern that the fraction of sources that would pass observational selection criteria of LRDs is smaller than the $f_{\rm duty}$ implied. Therefore, a major remaining issue for the simulation is the overbright and overmassive host galaxies of LRDs.}
    \label{fig:bh-host}
\end{figure}

\subsection{Optical and H$\alpha$ LFs}
\label{subsec:opt}

We next consider AGN LFs in specific rest-frame bands, which are less impacted by the varying approaches observational works use to derive bolometric luminosities. We begin with the rest-frame optical, where LRDs are most prominent and where their bolometric corrections are expected to be substantially lower than those of canonical unobscured quasars (see Section~\ref{subsec:lrdsed}). We choose the rest-frame B band ($\lambda\simeq 4400$\,\AA) to do the comparison, mainly to follow the conventions of observations compiled in \citet{Hopkins2007,Shen2020}. This band is also where our bolometric corrections and dispersions for canonical AGN are derived. Alternative optical LFs at $5100$\,\AA\ are shown in Appendix~\ref{appfig:5100lf}. 

In Figure~\ref{fig:optlf}, we show the rest-frame optical LFs of AGN at $z\simeq 3-5$ in \luminasim. Predictions from \tng and \mtng are shown here for reference. For the LRDs in the simulation, we adopt the bolometric correction factors listed in Table~\ref{tab:lrd-corr}. Specifically, we use the values inferred for the ``weak IR'' and ``strong IR'' cases as lower and upper limits, respectively, and present the resulting LFs as a shaded region. To avoid overcrowding the figure, the \tng and \mtng LFs assume the ``weak IR'' model. For comparison, we include the pre-\textit{JWST} observational datasets compiled in \citet{Shen2020}, which all have been converted to the B band. We also include the LRD LF measurements from \citet{Kokorev2024,Ma2025,Ma2026}. These measurements are typically reported at rest-frame $5100$\,\AA\ or $5500$\,\AA, and we convert them to the B band using the LRD SED in Section~\ref{subsec:lrdsed}. The corresponding difference between $\log L_{5100}$ ($\log L_{5500}$) and $\log L_{4400}$ is $0.20$ dex ($0.27$ dex). In addition, we include the broad H$\alpha$ emitter LFs from \citet{Matthee2024,Lin2024,Lin2026}. We convert these measurements to rest-frame optical continuum luminosities using the empirical relation
\begin{equation}
    L_{5100} = 10^{44}\erg\,{\rm s}^{-1}
    \left(\frac{L_{\rm H\alpha}}{10^{43.30}\erg\,{\rm s}^{-1}}\right)^{1/1.04} \, ,
    \label{eq:5100-halpha}
\end{equation}
as found in \citet{DeGraaff2025c} for their full LRD sample. It is a tight correlation with scatter $\sim 0.2$ dex. The resulting H$\alpha$-based constraints are broadly consistent with the continuum luminosity constraints in observations.

The predictions from \luminasim simultaneously reproduce several key features of the observed LRD LFs: the high abundance of sources at $L_{\rm opt}\sim 10^{43}-10^{44} \erg\,{\rm s}^{-1}$, the rapid decline in number density when approaching $L_{\rm opt}\sim 10^{45}\erg\,{\rm s}^{-1}$ at $z\sim 5$ \citep{Ma2025}, and the lower number density of LRDs at $z\sim 3$ \citep{Ma2026}, except for their faintest luminosity bin. At the same time, the total AGN LF connects smoothly to the pre-\textit{JWST} constraints on the canonical AGN population compiled by \citet{Shen2020}. The faint-end optical LF is sensitive to our choices of $\kappa$ and $f_{\rm duty}$. For example, reducing $\kappa$ and $M^{\rm crit}_{\rm BH}$ will reduce the brightest LRDs we predict, and reducing $f_{\rm duty}$ results in a change in the normalization of the LF. Therefore, the agreement with the observed optical LFs should be interpreted partly as a calibration of these effective parameters. In Section~\ref{sec:discussion}, we revisit this point by exploring alternative parameter choices, the associated degeneracies, and their physical implications. In the following sections, we test whether the same empirical model and fiducial parameter choices remain consistent with observations in other wavelength bands.

\begin{figure*}
    \centering
    \includegraphics[width=0.49\linewidth]{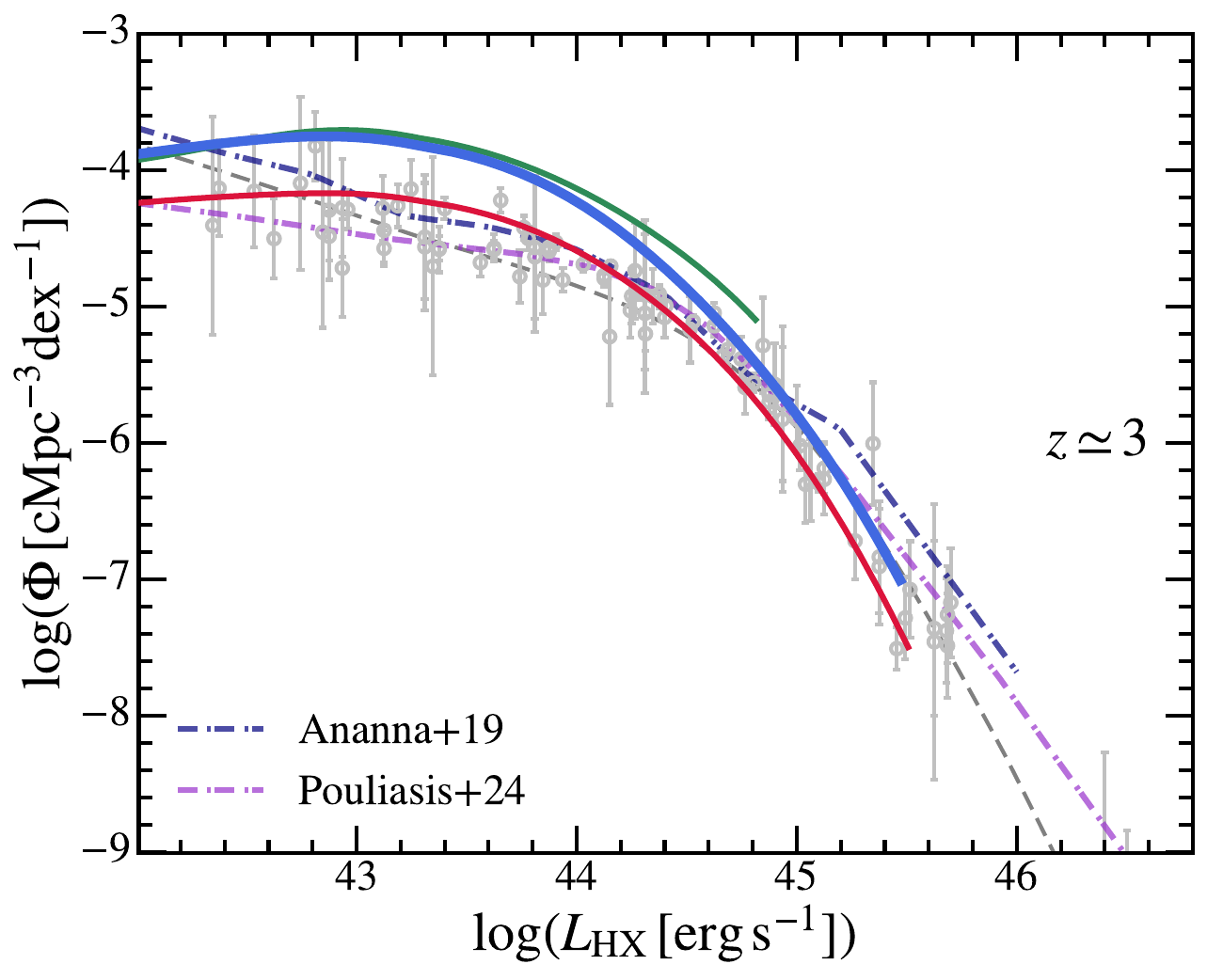}
    \includegraphics[width=0.49\linewidth]{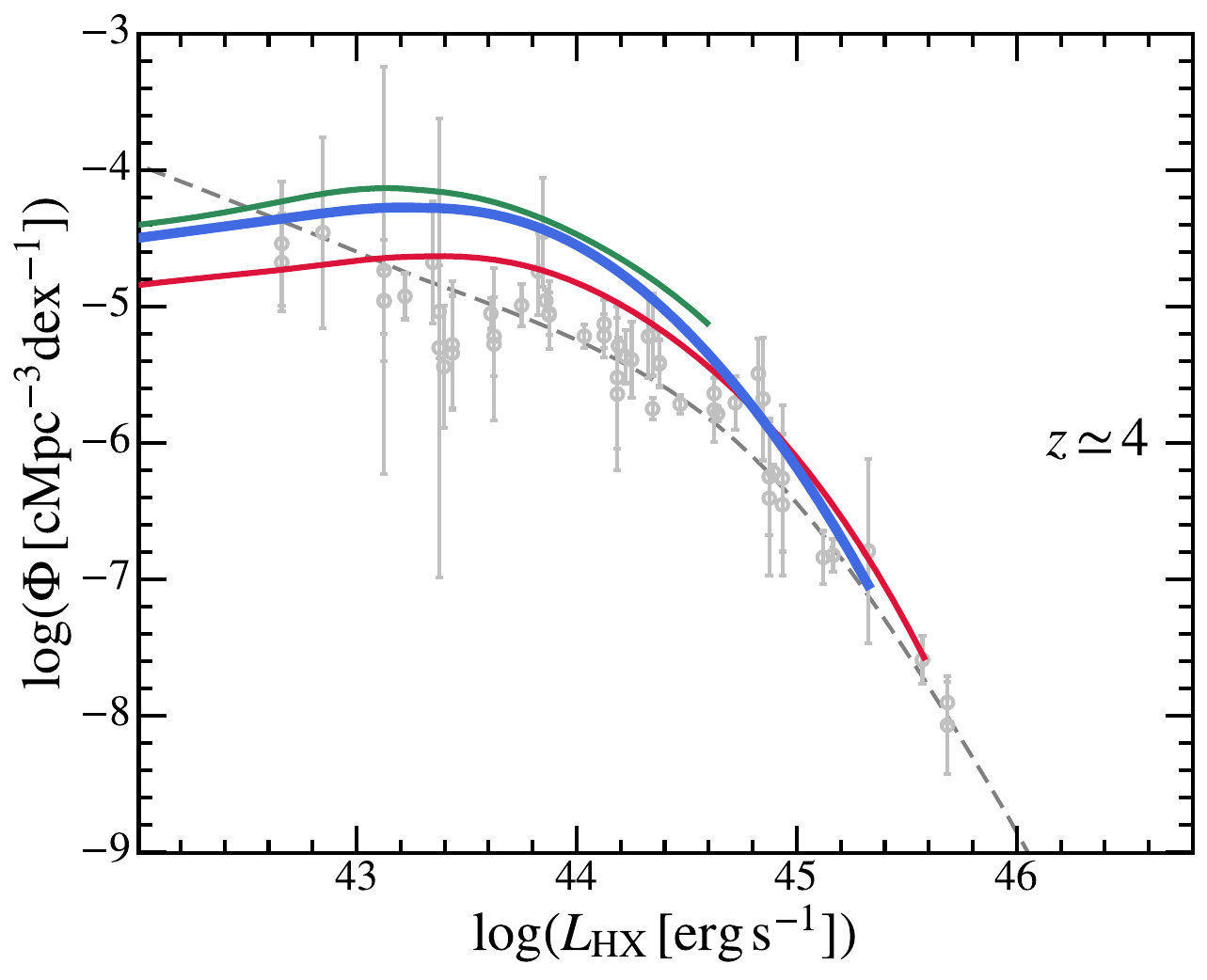}
    \includegraphics[width=0.49\linewidth]{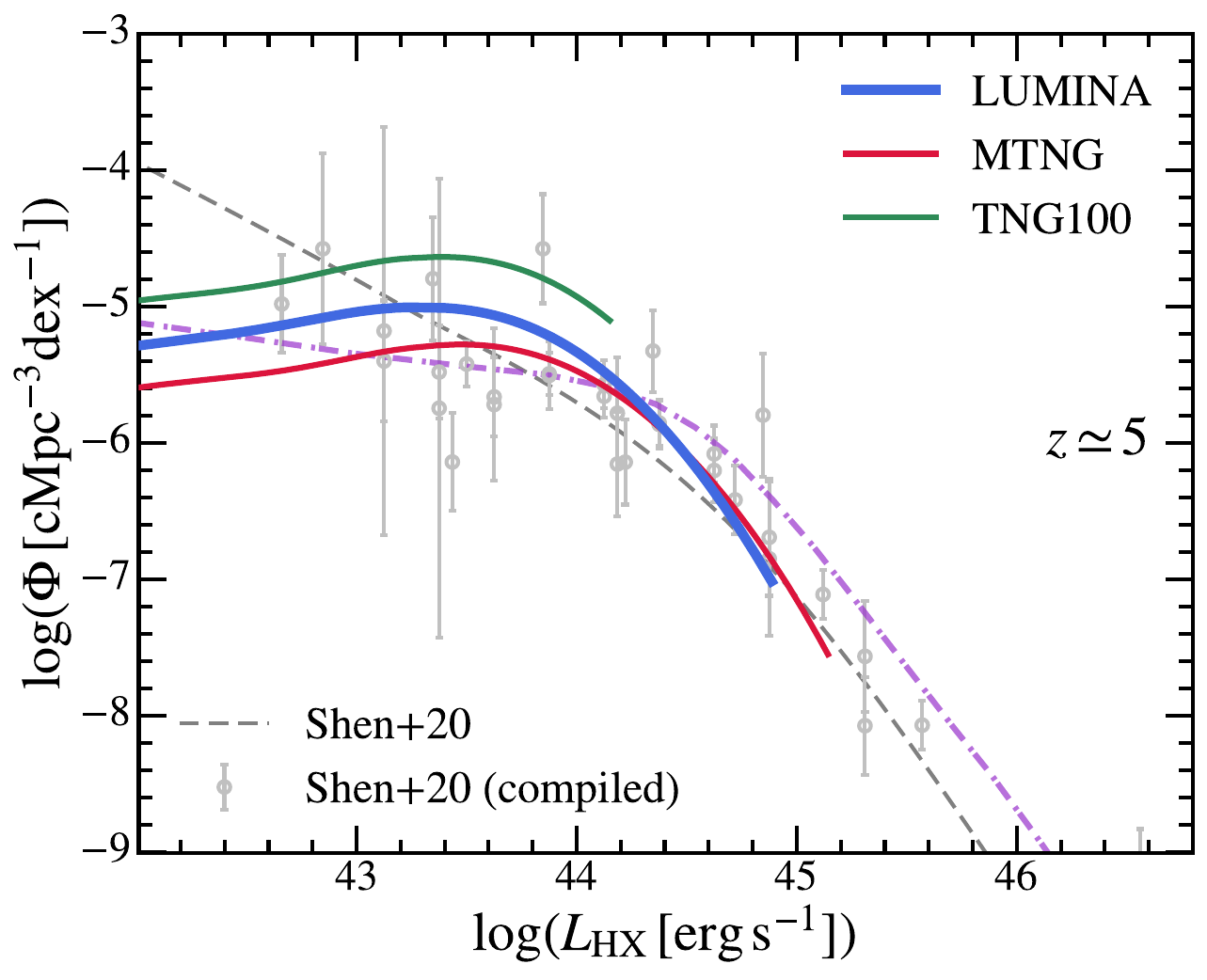}
    \includegraphics[width=0.49\linewidth]{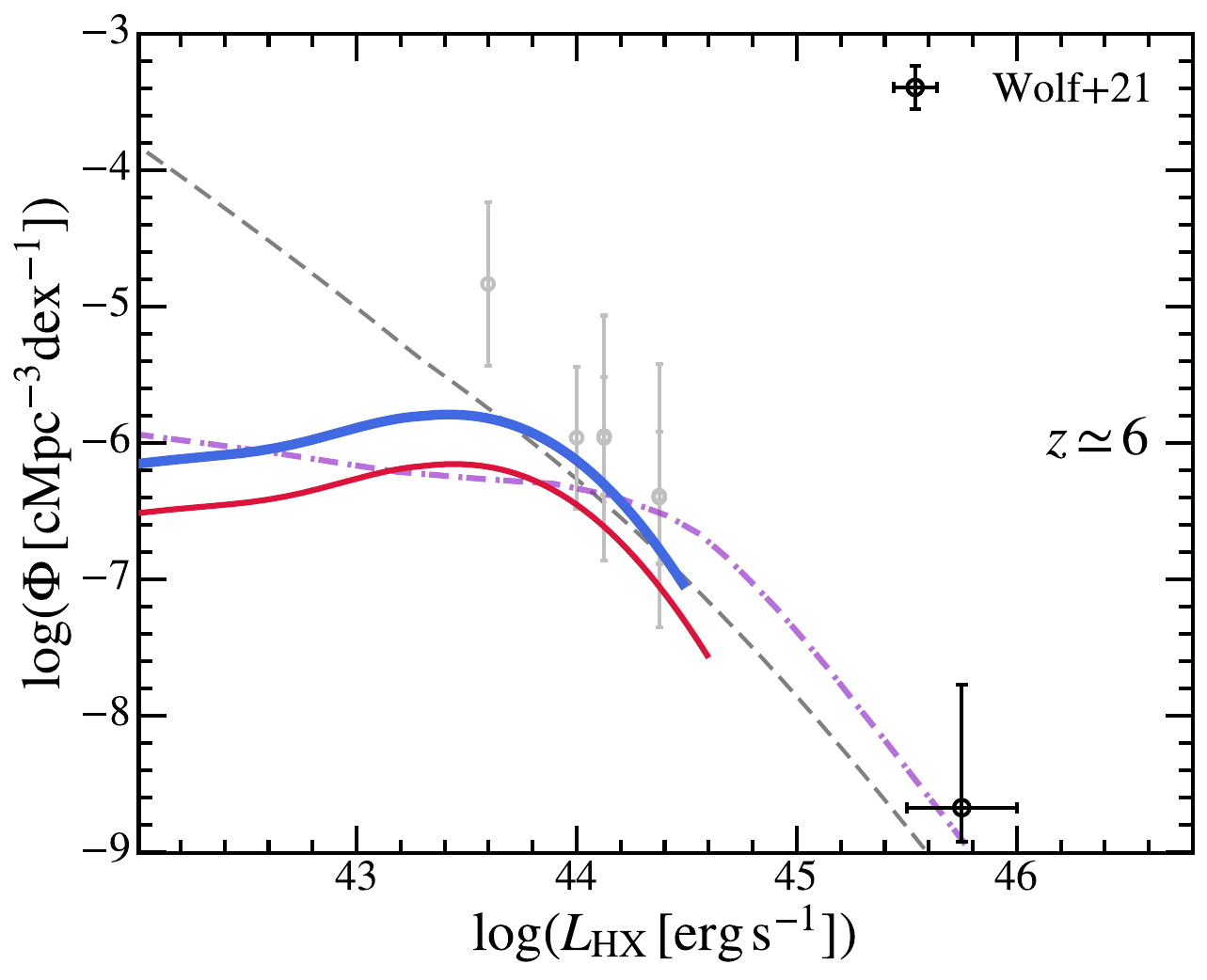}
    \caption{Hard X-ray ($2-10\keV$) LFs of AGN at $z=3-6$ from \luminasim, \tng, and \mtng. We note that the X-ray LFs here are only contributed by the non-LRD population we identify in the simulations, so there is no dependence on the $f_{\rm duty}$ and bolometric corrections of LRDs. We compare the simulation results with X-ray observations compiled in \citet{Shen2020} along with more recent constraints from \citet{Ananna2019,Wolf2021,Pouliasis2024}. At $z\simeq 6$, all BHs in \tng are below $M^{\rm crit}_{\rm BH}$ and therefore no prediction is presented.}
    \label{fig:hxlf}
\end{figure*}

\begin{figure}
    \centering
    \includegraphics[width=\linewidth]{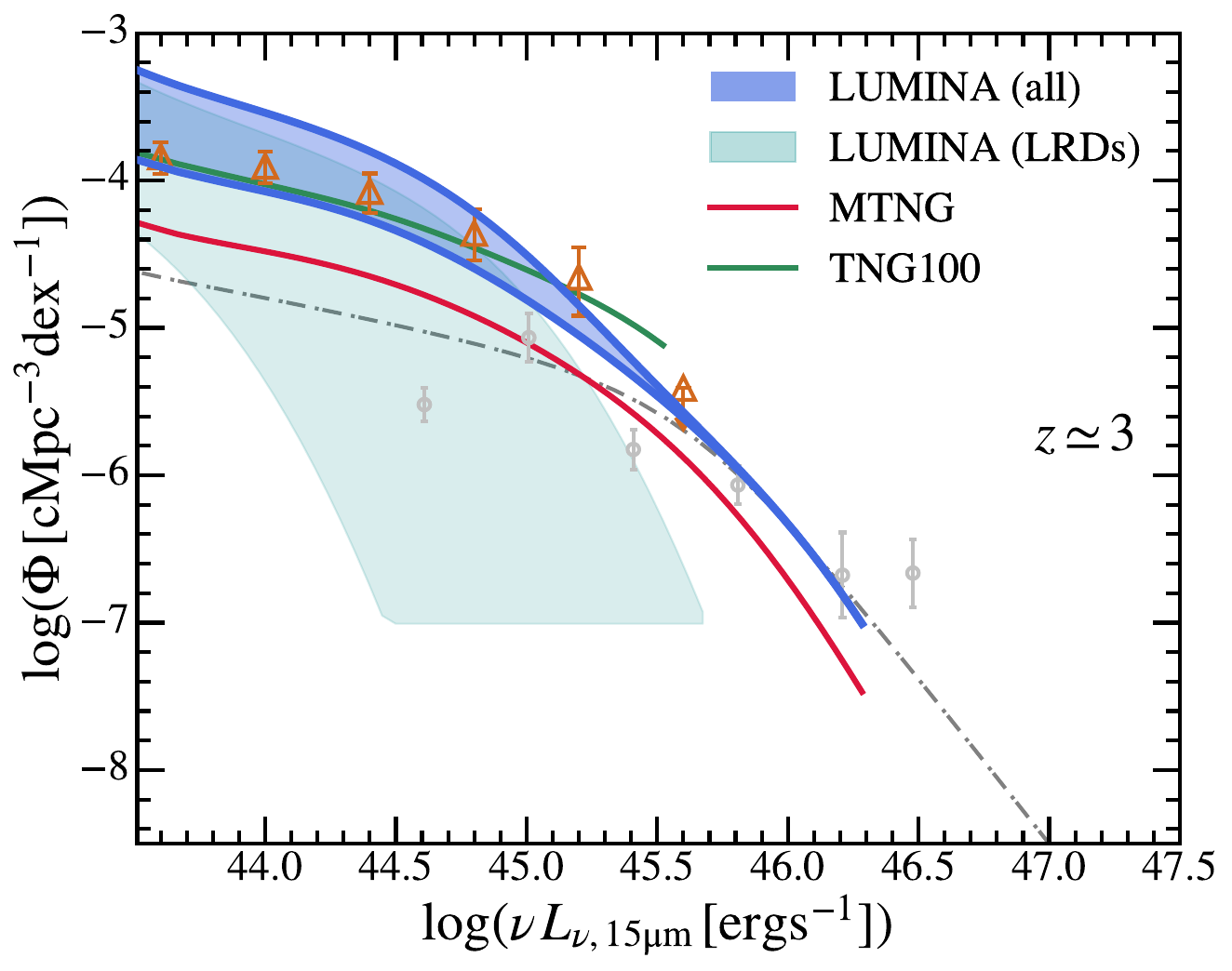}
    \includegraphics[width=\linewidth]{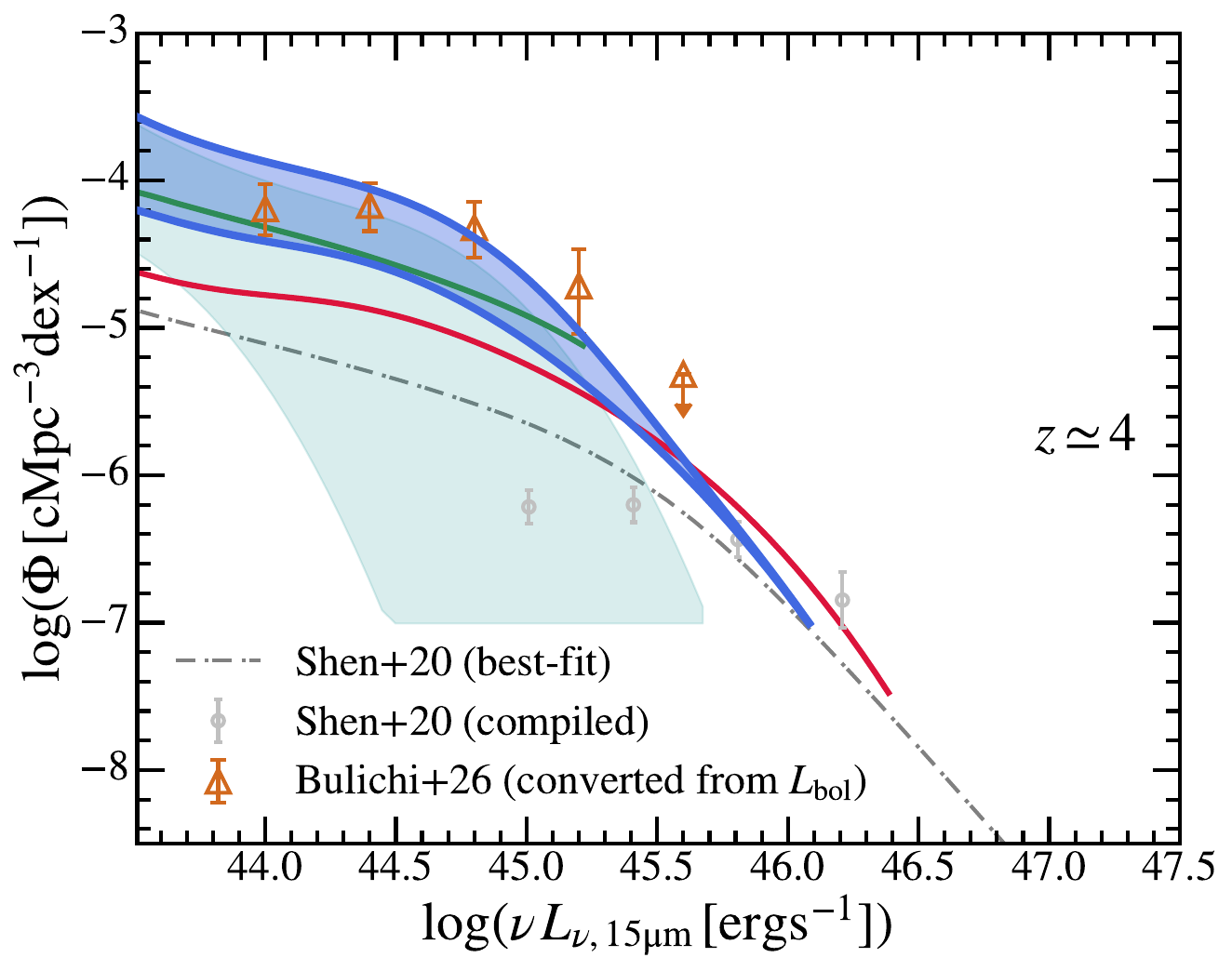}
    \includegraphics[width=\linewidth]{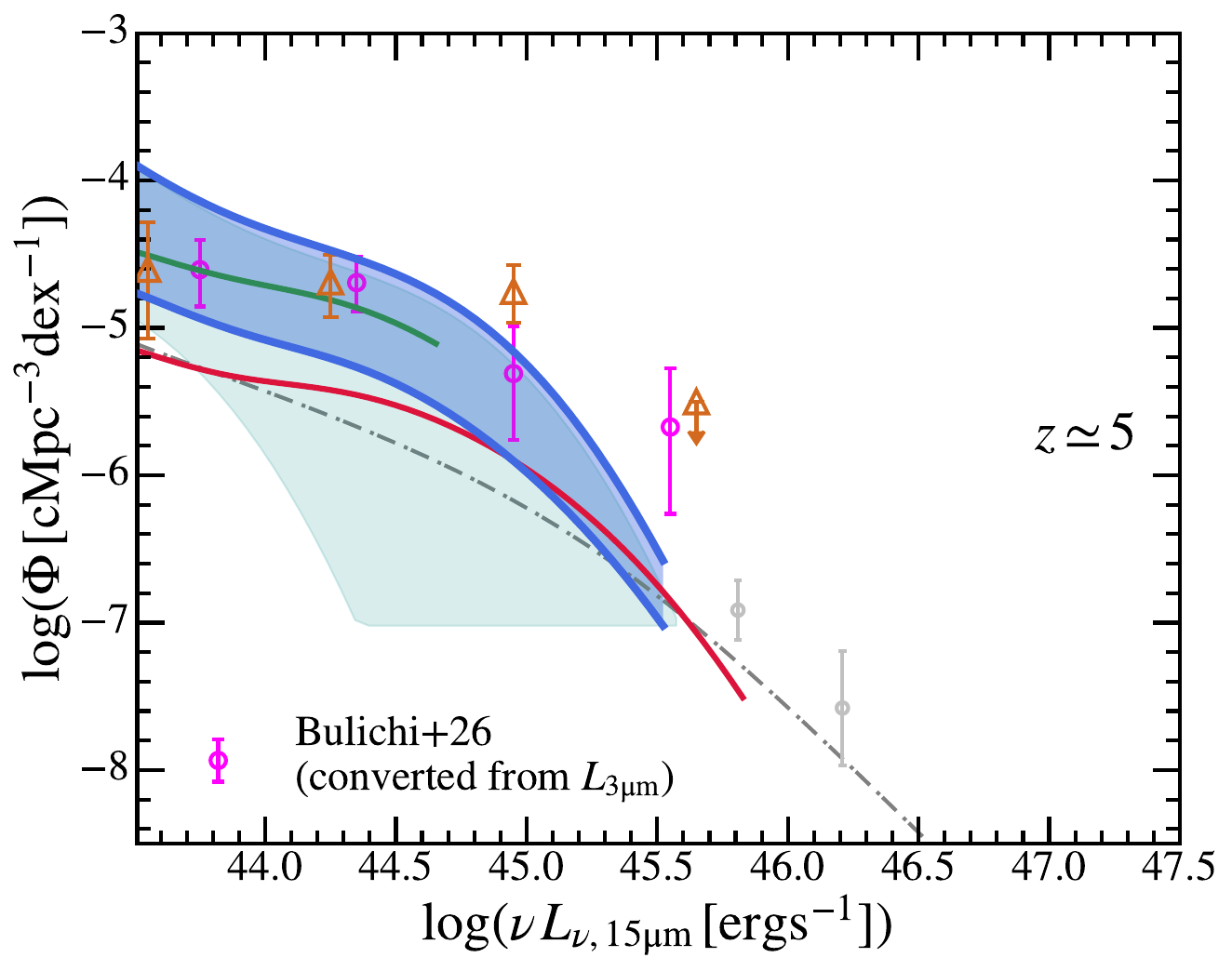}
    \caption{Rest-frame mid-IR band ($\sim 15\micron$) LFs of AGN at $z=3-5$ in \luminasim. Results from \tng and \mtng are shown for reference. For comparison, we include observational constraints compiled in \citet{Shen2020}. We also include \textit{JWST} MIRI constraints from \citet{Bulichi2026}, with data converted from their bolometric LFs or from their $3\micron$ LFs assuming the canonical AGN SED \citep{Krawczyk2013}.}
    \label{fig:irlf}
\end{figure}

\subsection{UV LFs}
\label{subsec:uv}

The rest-frame UV emission in LRDs remains one of their most puzzling features. Several scenarios have been proposed, including leaked or reprocessed AGN light \citep[e.g.,][]{Greene2024,Labbe2024,Kocevski2025,Lin2026-local,Madau2026}, emission from young stars in the host galaxy or compact nuclear starbursts \citep[e.g.,][]{Killi2024,Akins2025a,Asada2026,Inayoshi2026,Baggen2026,Naidu2025,Sun2026}, nebular emission from low-density ionized gas around the central SMBH \citep[e.g.,][]{Chen2025,Chen2025b}, or a mixture of them~\citep{Cloonan2026}. The UV emission is crucial for the photometric identification of LRDs, since the population is often selected through its characteristic ``V-shaped'' SED, with a blue UV continuum and a red optical continuum. It is therefore important to ask whether a coherent picture can simultaneously explain the UV and optical luminosities of LRDs.

In Figure~\ref{fig:uvqlf}, we show the rest-frame UV LFs of AGN at $z\simeq 3-6$ in \luminasim, again comparing our predictions to \tng and \mtng. At the bright end, we include the pre-\textit{JWST} observational constraints compiled by \citet{Shen2020}, together with the measurements from \citet{Niida2020} at $z\simeq 5$ and \citet{Schindler2023} at $z\simeq 6$. The best-fit quasar UV LFs in \citet{Kulkarni2018} and \citet{Shen2020} are also shown for reference. Similar to the optical LFs discussed above, the simulations show reasonable agreement with these pre-\textit{JWST} observations at moderately bright luminosities ($M_{\rm UV} \lesssim -22$).

At the faint end, we compare our predictions to two categories of observational constraints at $z\simeq 5-6$. The first category is based on spectroscopically selected BLAGN, without an explicit colour or morphological pre-selection \citep{Matthee2024,Harikane2023-agn,Maiolino2024,Taylor2025}. The second category consists of LRDs selected with an additional ``V-shape'' SED criterion, applied either explicitly via photometry \citep{Kokorev2024,Kocevski2025} or implicitly through spectroscopic follow-up of photometric LRD samples \citep{Greene2024}. The BLAGN abundance from the first category is generally larger than the LRD abundance from the second category at fixed $M_{\rm UV}$, although the flux-limited blind Grism selection in \citet{Matthee2024} gives results close to the LRD LFs. Recent works \citep[e.g.,][]{Brazzini2026,Madau2026} interpret this difference as evidence that LRDs are a $\sim 10-30\%$ sub-population of the broader low-luminosity BLAGN, with the bluer majority named ``Little Blue Dots'' (LBDs). The potential selection bias and the physical origin of UV emission in LRDs/LBDs are yet to be understood.

Here, we adopt one interpretation recently developed by \citet{Sun2026}, in which the rest-frame UV emission of LRDs is dominated by their host galaxies, with only a $\sim 15\%$ contribution from the AGN at $\lambda \sim 1500$\AA. We therefore multiply the reported total UV luminosities of LRDs by this AGN fraction before comparing them to the AGN-only LF, corresponding to a shift of $-0.8$ dex in luminosity ($+2\mmag$ in $M_{\rm UV}$). This treatment is also consistent with the LRD SED model adopted in Section~\ref{subsec:lrdsed}, which is based on stacked LRD spectra either after subtracting the host-galaxy component \citep{Sun2026} or of sources most strongly dominated by the optical component \citep{PG2026}. After applying this empirical correction, we find that the faint-end predictions from \luminasim agree reasonably well with the inferred AGN contribution to the UV LFs of LRDs. We restrict our quantitative comparison to the LRD observations (second category mentioned above), while showing the BLAGN LFs only for context (applied the same correction factor in $M_{\rm UV}$ just to keep the relative position to LRD constraints). This is because the host-galaxy contribution in BLAGN (that are not classified as LRDs) could be much higher than inferred in \citet{Sun2026}, and it is still possible that the UV emission is produced by leaked or post-processed radiation from the accreting SMBHs~\citep[e.g.,][]{Chen2025,Madau2026}, which our empirical LRD SED model would fail to represent. In both cases, we underestimate the UV luminosities. It could be the reason that the observed BLAGN LFs are systematically above our model predictions as well as the LRD LFs.

Another caveat of our model is that the simulated host galaxies of the selected LRD-like BHs are typically too UV luminous. Their host-galaxy intrinsic UV luminosities are more than an order of magnitude higher than what is required to explain the total observed UV luminosities of LRDs. Dust attenuation is likely not sufficient to reconcile this tension: given the relatively blue UV slopes $\beta_{\rm UV}\sim -2$ of LRDs \citep[e.g.,][]{Greene2024,Kocevski2025}, $A_{\rm UV}\lesssim 0.5\mmag$ if one assumes a typical \citet{Meurer1999} IRX-$\beta$ relation of host galaxies. At the same time, their optical luminosities are also too large for the systems to be identified as LRDs in real observations. For example, the median stellar mass and intrinsic UV magnitude of host galaxies, for LRDs above $L_{\rm bol}>10^{43}\erg\,{\rm s}^{-1}$, are around $10^{9}\msun$ and $-21\mmag$ in \luminasim, compared to $10^{8.3}\msun$ and $-18.5\mmag$ reported in \citet{Sun2026}, $5\times 10^{7}\msun$ reported in \citet{Matthee2025}, and $10^{8.5}\msun$ reported in \citet{Lin2026} using clustering measurements. This issue persists even when we select BHs with the largest optical luminosity ratios relative to their host galaxies. 

We illustrate this tension in Figure~\ref{fig:bh-host}, where we show the distribution of luminosity ratios between BHs and hosts in the UV versus in the B band for the LRDs in \luminasim. We use the LRD bolometric corrections of the ``weak IR'' model here, while the ``strong IR'' model will only make BHs even dimmer in both bands. We roughly divide the space into four regions. If BHs or hosts dominate in both bands, the source should appear as a pure AGN or galaxy. In the picture we described above, only the ones that are dominated by the BH in the optical and by the host galaxy in the UV will appear as LRDs. There are AGN in \luminasim that lie in the LRD regime, but the fraction of them is too small compared to the $f_{\rm duty}=0.3$ implied by our comparisons of LFs. Meanwhile, a substantial population of AGN is moderately dominated by their host galaxies, which in fact could appear as the LBDs. The conclusion here is that host galaxies of LRDs in \luminasim are overmassive and overbright to self-consistently explain the observed LRDs with the ``V-shape'' SED selection criteria. This remains a major unresolved issue in cosmological simulations such as \luminasim. As already highlighted in e.g. \citet{Matthee2025,Habouzit2025}, the host galaxy stellar masses of LRDs are significantly lower than predictions in a range of modern cosmological simulations and the values inferred from local scaling relations between SMBH and host masses. The tensions we find here are independent of the BH mass measurements, yet consistent with these findings.

\begin{figure*}
    \centering
    \includegraphics[width=0.49\linewidth]{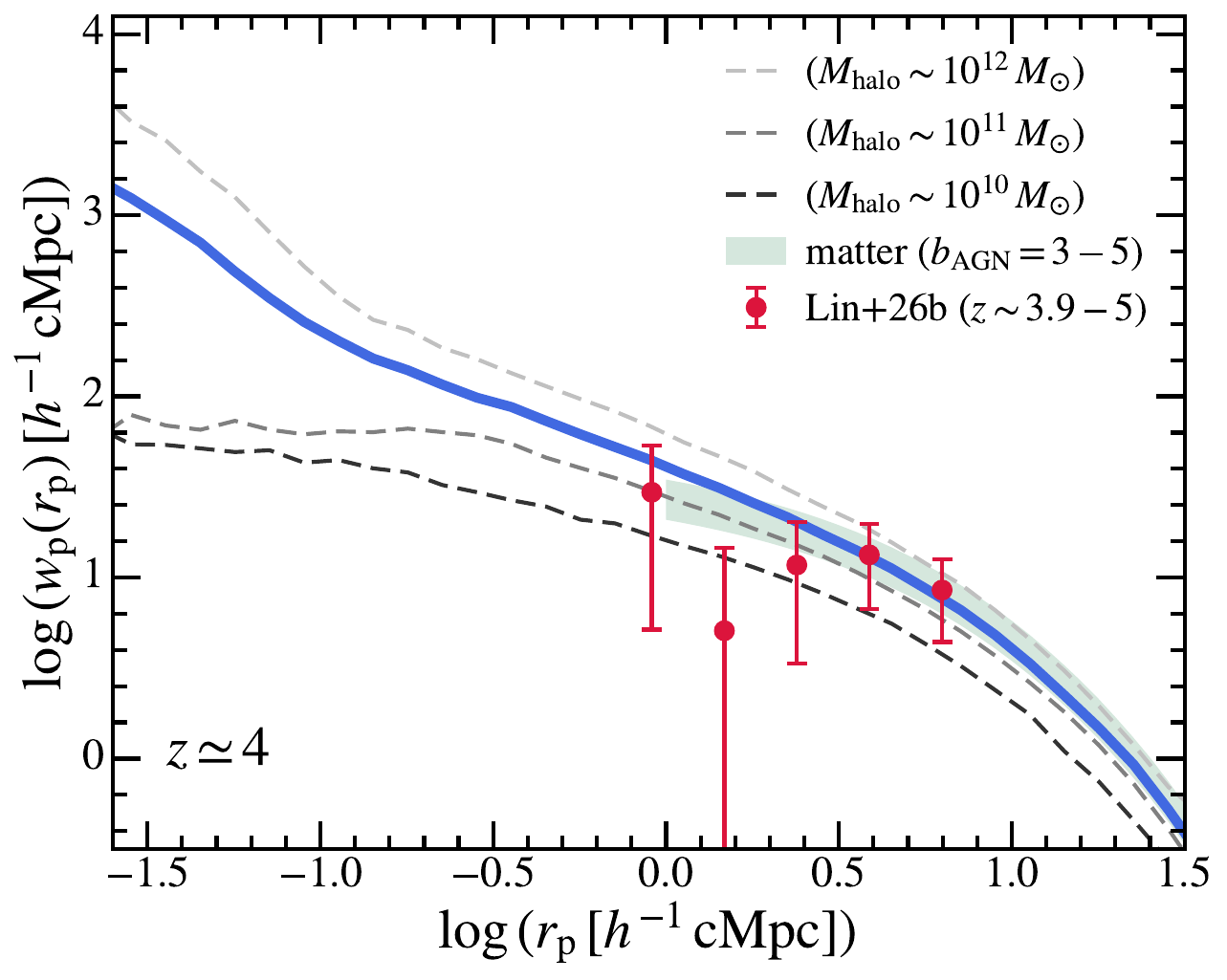}
    \includegraphics[width=0.49\linewidth]{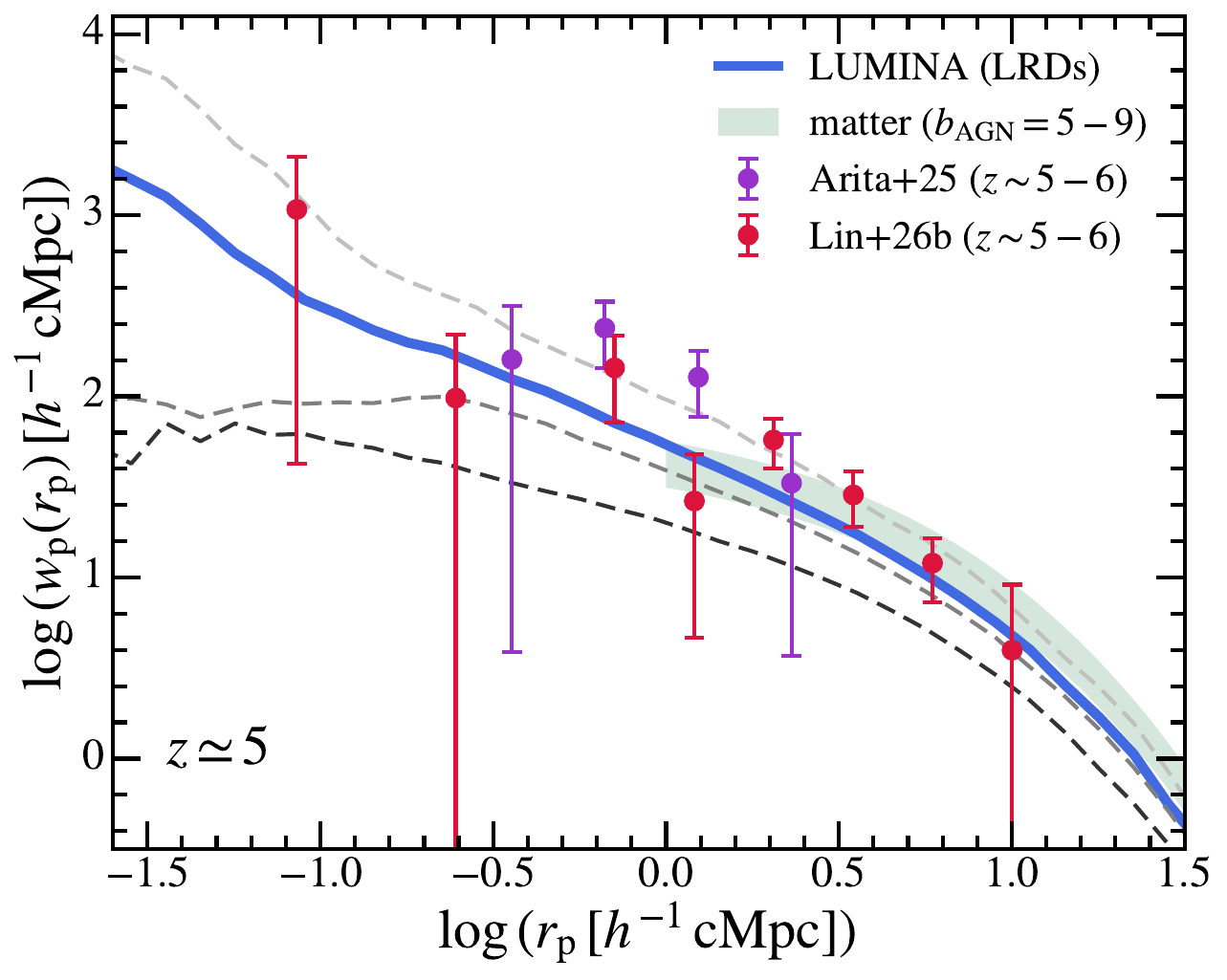}
    \caption{Projected cross-correlation function ($w_{\rm p}$) of LRDs with galaxies at $z\simeq 4$ (left panel) and $z\simeq 5$ (right panel) in \luminasim. As discussed in the main text, we select galaxies with SFR$>1.7\msun\,{\rm yr}^{-1}$ to mimic the tracer galaxy sample used in observations. We further measure the bias factor of the selected galaxy sample and apply a $b^{\rm sim}_{\rm gal}/b^{\rm obs}_{\rm gal}$ correction to the observed $w_{\rm p}$ in \citet{Lin2026-env} and \citet{Arita2025}. For reference, we show the cross-correlation function of different mass haloes with the same galaxy sample in \luminasim. We also show the biased matter auto-correlation function assuming different bias factors of LRD. At $z\simeq 4$, our LRD clustering signal agrees well with \citet{Lin2026-env}, suggesting host halo mass mildly above $10^{11}\msun$ and a bias factor $\sim 4$. At $z\simeq 5$, our LRD sample has a bias factor of $\sim 5$ and still lies in haloes slightly above $10^{11}\msun$, while the observed LRDs \citep{Arita2025,Lin2026-env} have larger bias factors and correlation functions close to our $10^{12}\msun$ haloes. Notably, the data point from \citet{Lin2026-env} at the smallest $r_{\rm p}$ may suggest the one-halo term from clustering of satellite star-forming galaxies within massive haloes that only shows up when halo mass gets above $10^{11}\msun$ and is consistent with our results.}
    \label{fig:lrd_xcorr}
\end{figure*}

\subsection{X-ray LFs}
\label{subsec:xray}

Another critical validation of our model comes from the X-ray LFs of AGN. Hard X-ray emission in canonical AGN is expected to trace the compact corona associated with accretion onto BHs and is relatively less affected by contamination from host-galaxy starlight than the UV or optical bands \citep[e.g.,][]{Osterbrock2006,Brandt2015,Hickox2018}. At the same time, the X-ray band provides an important consistency check on our LRD interpretation. Most LRDs and \textit{JWST}-selected faint AGN appear to be X-ray weak or undetected relative to expectations from standard AGN scaling relations \citep[e.g.][see $\alpha_{\rm ox}$ discussed in Section~\ref{subsec:sed}]{Yue2024,Ananna2024,Maiolino2025}, although the recently identified ``X-ray dot'' \citep{Hviding2026} may represent an important exception in a transitional phase. Nevertheless, we assume that all our selected LRD-like objects are X-ray dark. The key test of our model is whether the remaining canonical AGN population in the simulation can reproduce the observed X-ray LFs.

In Figure~\ref{fig:hxlf}, we show the hard X-ray ($2-10\kev$) LFs of AGN at $z\simeq 3-6$ in \luminasim, \tng, and \mtng. For comparison, we show the observational constraints compiled in \citet{Shen2020} based on pre-\textit{JWST} X-ray surveys. These constraints predate the improved characterization of heavily obscured and CTK AGN uncovered by \textit{NuSTAR} \citep[e.g.,][]{Marchesi2018,Ananna2019,Tanimoto2022}. We have updated our column density distribution model in Section~\ref{subsec:extinction} accordingly. For the comparisons here, we include new observational constraints from \citet{Ananna2019} and \citet{Pouliasis2024}. \citet{Ananna2019} constructed an AGN population-synthesis model designed to reproduce X-ray AGN number counts, the cosmic X-ray background, and the observed CTK AGN fraction, yielding an X-ray LF in the $(z,L_{\rm HX},N_{\rm H})$ space. \citet{Pouliasis2024} derived the X-ray luminosity and absorption functions at $3\leq z\leq 6$ using a large sample of soft X-ray-selected AGN from the Chandra Deep Fields, the Chandra COSMOS-Legacy survey, and the XMM-XXL northern field. We also include the $z\simeq 6$ constraint from \citet{Wolf2021}, based on an eROSITA-detected quasar, although this point lies close to or beyond the effective dynamical range sampled by our simulations.

As in many X-ray luminosity-function studies, the published $L_{\rm HX}$ values in these works are intrinsic, absorption-corrected luminosities. To compare them to our post-absorption predictions, we first generate the intrinsic model LF of \citet{Shen2020} by setting $N_{\rm H}=0$. We then record the offset in $\log\Phi$ between the observational constraints and this intrinsic reference model, and apply the same offset to the post-absorption LF from \citet{Shen2020}. This procedure preserves the relative information carried by the observational measurements while placing them on the same luminosity definition as our model predictions. 

Encouragingly, the simulation predictions reproduce the observed X-ray LFs across the redshift range considered here. At the bright end, the agreement is similar to what we found in the optical and UV bands. Towards the faint end, the predicted X-ray LF flattens, broadly following the trend inferred from observations. The X-ray comparison supports a picture in which the canonical AGN population in \luminasim is broadly consistent with existing hard X-ray constraints, while the LRD-like population is largely hidden from current X-ray surveys. We overpredict the number density of faint X-ray AGN with $L_{\rm HX}\lesssim 10^{44}\erg\,{\rm s}^{-1}$ at $z\lesssim 4$. A similar excess was already seen in the bolometric LFs discussed in Section~\ref{subsec:bol}. However, this discrepancy should not necessarily be interpreted as a failure of the model. As will become clearer from the IR LFs below, current X-ray surveys can remain incomplete for canonical obscured AGN even before accounting for the LRD population. 

\subsection{Mid-IR LFs}
\label{subsec:midir}

The final band we consider is the mid-IR. For canonical AGN, the near- to mid-IR emission is generally interpreted as accretion-disc radiation absorbed and reprocessed by circumnuclear dust, often associated with the dusty torus or more extended dusty polar structures \citep[e.g.,][]{Antonucci1993,Urry1995,Fritz2006,Netzer2015,RamosAlmeida2017,Hickox2018}, while the far-IR can receive substantial contributions from colder dust heated by star-formation in the host galaxy. As discussed in Section~\ref{subsec:lrdsed}, the mid- to far-IR SEDs of LRDs remain highly uncertain, and even the physical origin of their IR emission is still debated. Some studies find little evidence for hot or cold dust emission in individual or stacked LRD samples, challenging simple dust-reddened AGN or dusty-starburst interpretations \citep[e.g.,][]{Setton2025a,Chen2025,Xiao2025}. However, evidence for warm or hot dust emission in LRDs has been reported in stacking analyses \citep{Delvecchio2025} and in individual sources \citep{Ronayne2025,Barro2026}. Therefore, the IR LF provides a complementary consistency check, similar in spirit to the X-ray LF. If LRDs are relatively faint in the mid-IR compared to canonical obscured AGN, then the IR LF should mainly validate whether the canonical AGN population in the simulation remains consistent with observations.

Observational constraints in this regime have recently become possible with deep \textit{JWST}/MIRI surveys. Early MIRI results showed that mid-IR selection can reveal a population of heavily obscured high-redshift AGN missed by X-ray surveys \citep[e.g.,][]{Yang2023,PG2024,Lyu2024}. More recently, the MEOW survey \citep[Leung in prep.,][]{Bulichi2026} used deep MIRI imaging, together with NIRCam, spectroscopic, and HST data, to identify obscured AGN out to $z\sim 6$ and infer a rapidly increasing obscured fraction towards high redshift. We explore two ways of converting the MEOW constraints to $15\micron$ LF. The first is taking their bolometric LFs and applying a typical mid-IR bolometric correction of $\sim 10$ (see Table~\ref{tab:lrd-corr}). The second is taking their $3\micron$ LFs and converting those to $15\micron$ with luminosity differences inferred from the \citet{Krawczyk2013} template we used for canonical AGN SED. They yield broadly consistent results.

In Figure~\ref{fig:irlf}, we show the mid-IR ($\sim 15\micron$) LFs of AGN at $z\simeq 3-5$ in \luminasim, compared to \tng and \mtng. We compare these predictions to the observational constraints compiled by \citet{Shen2020}, as well as to the recent MIRI-selected constraints from \citet{Bulichi2026}. The choice of $15\micron$ as the primary mid-IR band follows the convention of the pre-JWST compilation \citep{Hopkins2007,Shen2020} that anchors our comparisons here. We find that our faint-end IR LF lies above the extrapolation of the best-fitting model from \citet{Shen2020}, but is more consistent with the higher AGN abundance inferred by \citet{Bulichi2026}. This agreement is particularly encouraging because \citet{Bulichi2026} found a substantially larger obscured AGN fraction than previous UV-, optical-, X-ray-, or pre-\textit{JWST} IR-selected surveys. We emphasize that this faint-end enhancement is not driven solely by the LRD-like population. In our ``weak IR'' model (the lower limit of predicted IR LF), the faint-end mid-IR LF remains substantially contributed, and in some luminosity ranges dominated, by canonical AGN in the simulation. Thus, the mid-IR comparison supports the interpretation that \luminasim contains a significant population of obscured but otherwise canonical AGN, in addition to the empirical LRD component introduced above.

\subsection{The large-scale environment of AGN}
\label{subsec:environment}

So far, we have focused primarily on the LFs, and hence the abundances, of AGN. The large-scale environment of AGN provides an important and independent probe of their host DM haloes \citep[e.g.,][]{Eilers2024,Pizzati2024,Pizzati2025}. This is particularly valuable for LRDs. Unlike luminosity- or SED-based arguments, which depend sensitively on bolometric corrections, dust attenuation, host-galaxy subtraction, and the physical origin of the broad-line emission, clustering directly traces the connection between LRDs and the underlying matter density field. Massive haloes originate from rare peaks in the Gaussian random density field and are expected to be more clustered compared to the matter density field \citep{Kaiser1984,Bardeen1986}. Therefore, the large-scale autocorrelation or cross-correlation amplitude of a tracer population and the inferred bias \citep[e.g.,][]{Mo1996,Sheth2001,Cooray2002,Tinker2010} provides information about the characteristic host halo mass of AGN. This test is especially useful for distinguishing between different physical interpretations of LRDs.

Recent \textit{JWST} observations have begun to make this test feasible. For instance, \citet{Matthee2025} used the deep NIRCam Grism ALT survey to study the environments of faint broad H$\alpha$ emitters at $z\simeq 4-5$. By comparing the number of neighbouring emission-line galaxies around LRDs to that around a reference sample of star-forming galaxies, they inferred host stellar masses of order $M_\ast \sim 5\times10^7\msun$. This is much lower than what would be inferred from galaxy-only SED fits and implies large BH-to-stellar-mass ratios. \citet{Arita2025} measured the cross-correlation functions of $27$ low-luminosity BLAGN found by \textit{JWST} with photometrically-selected galaxies at $z\simeq 5-6$. They found that the typical host halo mass of these AGN is $\log(M_{\rm halo}/\msun)\simeq 11.6-11.7$. In a complementary analysis, \citet{Lin2026-env} measured the large-scale environments of $28$ low-luminosity BLAGN (including $5$ V-shape SED selected LRDs) and their cross-correlation functions with H$\alpha$ emitters at $3.9<z<6$ in the GOODS-N field using the CONGRESS and FRESCO survey data. Their inferred characteristic host halo masses are $\log(M_{\rm halo}/\msun)\simeq 11.0-11.2$, with corresponding host stellar masses of $\log(M_\ast/\msun)\simeq 8.4-8.6$. 

In \luminasim, we compute the cross-correlation functions of LRDs and tracer galaxies as
\begin{equation}
    \xi_{\rm ag}(r)=\frac{D_{\rm a}D_{\rm g}(r)-D_{\rm a}R_{\rm g}(r)-D_{\rm g}R_{\rm a}(r)+R_{\rm a}R_{\rm g}(r)}{R_{\rm a}R_{\rm g}(r)} \, ,
\end{equation}
where ``a'' and ``g'' represent AGN and tracer galaxies, respectively. $D_{\rm a}D_{\rm g}(r)$ is the number of cross pairs between the two types of sources at a distance $r$, while $D_{\rm a}R_{\rm g}(r)$, $D_{\rm g}R_{\rm a}(r)$, and $R_{\rm a}R_{\rm g}(r)$ are the corresponding pair counts involving randomly distributed sources. The auto-correlation function of LRDs or galaxies can be obtained in a similar way using the same source type for all terms. One can further define the projected cross-correlation function as
\begin{equation}
    w_{\rm p}(r_{\rm p}) = 2\int_{0}^{r_{\pi,{\rm max}}} \xi(r_{\rm p},r_\pi)\,{\rm d}\,r_\pi \, ,
\end{equation}
where $r_{\rm p}$ and $r_\pi$ are the transverse and line-of-sight separations, respectively. We adopt the canonical choice in literature, $r_{\pi,{\rm max}}=8\,h^{-1}\,{\rm cMpc}$. The code \textsc{corrfunc} \citep{Sinha2019,Sinha2020} is used to facilitate this calculation.

The selection of the tracer galaxies is critical, as the cross-correlation function of AGN and galaxies in the linear perturbation theory can be decomposed as \citep[see the review][]{Desjacques2018}
\begin{equation}
    \xi_{\rm ag}(r) \sim \xi_{\rm mm}(r)\,b_{\rm AGN}\,b_{\rm gal} \, ,
\end{equation}
where $\xi_{\rm mm}$ is the matter auto-correlation function, $b_{\rm AGN}$ and $b_{\rm gal}$ are the bias factors of AGN and tracer galaxies, respectively. \citet{Lin2026-env} used H$\alpha$ emitters above the luminosity threshold $10^{41.5}\erg\,{\rm s}^{-1}$ as the tracer galaxies. We convert this to a cut in SFR of $1.7\msun\,{\rm yr}^{-1}$, assuming the conversion factors from \citet{Murphy2011,Kennicutt2012,Shen2020b}. We first compute the auto-correlation function of galaxies above this SFR threshold from \luminasim and obtain $b_{\rm gal}$ by comparing it to $\xi_{\rm mm}$ at $r=8\,h^{-1}\,{\rm cMpc}$ at the redshift of interest. The calculation of $\xi_{\rm mm}$ is done using \textsc{colossus} \citep{Diemer2018}, assuming the \citet{Eisenstein1998} transfer function. The $b_{\rm gal}$ of tracer galaxies in \luminasim is $4.8$ at $z\simeq 5$ and $3.7$ at $z\simeq 4$. For comparison, the galaxy sample in \citet{Lin2026-env} has $b_{\rm gal}=5.9$ at $5<z<6$ and $b_{\rm gal}=4.11$ at $3.9<z<5$. The galaxy sample selected in \citet{Arita2025} has $b_{\rm gal}=5.2$ at $5<z<6$. Therefore, to account for the remaining differences in tracers and obtain a clean comparison of the clustering power of AGN, we will apply a correction factor $b^{\rm sim}_{\rm gal}/b^{\rm obs}_{\rm gal}$ to the correlation functions reported in observations. 

In Figure~\ref{fig:lrd_xcorr}, we compare the projected cross-correlation functions between LRDs (with $L_{\rm bol}>10^{43}\erg\,{\rm s}^{-1}$) and selected star-forming galaxies in \luminasim at $z\simeq4$ and $z\simeq5$ with the measurements from \citet{Lin2026-env}\footnote{\citet{Lin2026-env} originally measured the volume-weighted projected cross-correlation function. We find that it can be converted to $w_{\rm p}$ with a simple multiplication factor $16$.} and \citet{Arita2025}\footnote{\citet{Arita2025} used $r_{\pi,{\rm max}} = 50\,h^{-1}\,{\rm cMpc}$ when measuring $w_{\rm p}$. Empirically, based on \luminasim, we find that increasing $r_{\pi,{\rm max}}$ from $8\,h^{-1}\,{\rm cMpc}$ to $50\,h^{-1}\,{\rm cMpc}$ has $<0.1$ dex impact on $w_{\rm p}$ at $r_{\rm p}\lesssim 2 \,h^{-1}\,{\rm cMpc}$ where \citet{Arita2025} data reach.}. The simulation predictions agree well with the observed amplitudes of LRD-galaxy clustering at both redshifts, which provides an important validation of the model beyond LFs alone. To translate the clustering amplitude into an approximate LRD host halo mass scale, we further compare the LRD-galaxy cross-correlation to the cross-correlations between DM haloes of different masses and the same star-forming-galaxy sample. The observed and simulated LRD clustering amplitudes are best matched by haloes with characteristic masses slightly above $10^{11}\msun$. This inferred halo mass is consistent with the estimate from \citet{Lin2026-env} and \citet{Arita2025}, supporting a picture in which LRDs are associated with relatively common, moderate-mass haloes rather than rare massive quasar host haloes with $M_{\rm halo}\sim 10^{12}-10^{13}\msun$ \citep[e.g.,][]{Arita2023,Eilers2024,Pizzati2024}. It is also worth noting that the data point from \citet{Lin2026-env} at the smallest $r_{\rm p}$ may reflect the one-halo term arising from the clustering of satellite star-forming galaxies within massive haloes, a feature that only emerges once the halo mass exceeds $10^{11}\msun$, and which is consistent with our results. We note that our LRD-galaxy cross-correlation is almost identical to the galaxy auto-correlation function in our sample, so we show only the former in the figure. This is consistent with what has been found observationally \citep{Arita2025,Lin2026-env}, although it is not too surprising in our case, since the BH seeding and the LRD model depend only on halo/BH mass with no additional requirement on the halo environment. For \luminasim, this characteristic host halo mass of LRDs is also mainly shaped by this mass-only selection. Since BHs are seeded when haloes first pass $5\times 10^{10} {h}^{-1} \msun$ (Section~\ref{subsec:smbh}) and a scaling relation between BH mass and halo mass emerges as BHs grow in the \illustrisTNG model \citep{Weinberger2018,Pillepich2018Model}, the typical LRD mass we choose naturally corresponds to the host halo mass $\sim 10^{11}\msun$. The clustering measurements here serve as an important consistency check of the assumptions we made in seeding and LRD selection.

In observational studies, halo mass is usually inferred from clustering by comparing it to $\xi_{\rm mm}$, measuring the bias, and applying an analytical bias-mass relation \citep[e.g.,][]{Tinker2010}. Motivated by this, we show the LRD-galaxy cross-correlation function by multiplying $\xi_{\rm mm}$ with the $b_{\rm gal}$ measured in our galaxy sample and different values of $b_{\rm AGN}$. The LRDs selected in \luminasim have bias factor $\simeq 4$ at $z\simeq 4$ and $\simeq 5$ at $z\simeq 5$. For comparison, \citet{Lin2026-env} found $b_{\rm AGN}\simeq 3$ at $3.9<z<5$ and $\simeq 8$ at $5<z<6$. \citet{Arita2025} found $b_{\rm AGN}\simeq 6.6$ at $5<z<6$. In general, these values are mildly higher than our predictions. It is interesting that the host halo masses they infer are consistent with ours, suggesting that the discrepancy may arise from inaccuracies in the tracer galaxy bias estimates or in the bias-halo mass relation they adopt.

At the same time, clustering should not be interpreted purely as a one-to-one measure of halo mass. The clustering of galaxies and haloes is not determined by halo mass alone: halo assembly bias \citep{Gao2005,Gao2007} can modify the clustering amplitude at fixed mass if LRD formation preferentially selects haloes with particular secondary properties \citep{Wang2026}, such as early formation time, high concentration, or low spin. This issue is especially relevant because several proposed LRD formation channels are tied to unusual halo environments or assembly histories, such as the metal-free haloes exposed to strong Lyman-Werner radiation from nearby companions in direct-collapse BH scenarios \citep[e.g.,][]{Bromm2003,Begelman2006}, and concentrated early-forming haloes in DM-seeding scenarios \citep[e.g.,][]{Xiao2021,ShenT2025,Jiang2026,Roberts2026}. Therefore, the agreement in clustering should be interpreted with some caution. We cannot rule out the possibility that LRDs reside in somewhat less massive DM haloes with secondary properties that enhance their large-scale bias. In that case, the duty cycle required to match the luminosity-function constraints would likely be smaller than the value inferred under a mass-only halo-bias interpretation. 

\subsection{AGN in the cosmological context}
\label{subsec:cosmo}

The comparisons above focus on LFs in individual bands and on the large-scale environments of AGN. We now place the model in a broader cosmological context by examining the redshift evolution of several integrated properties of SMBHs. 

\begin{figure}
    \centering
    \includegraphics[width=\linewidth]{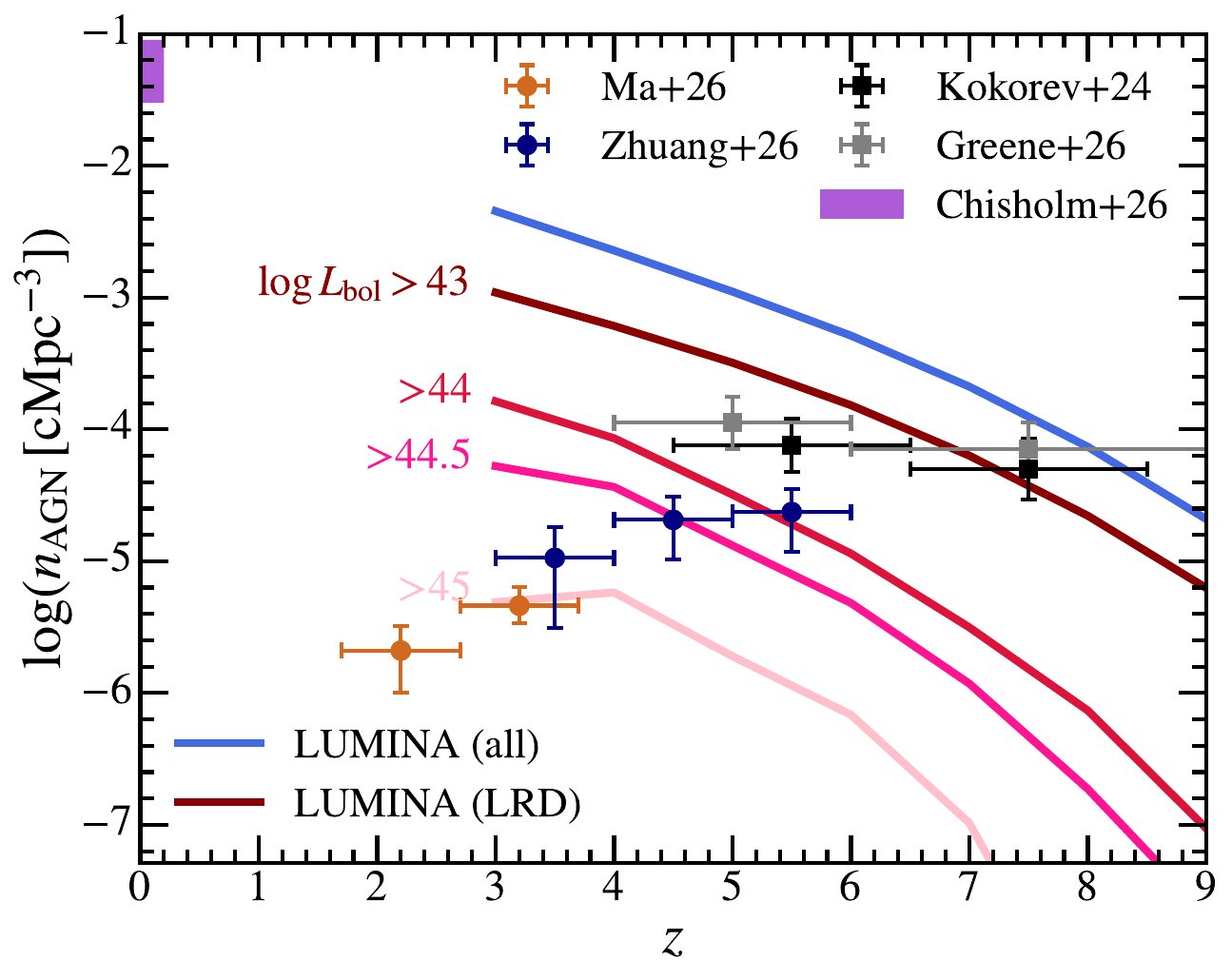}
    \caption{Number density of AGN versus redshift. We show the number density of all AGN with $L_{\rm bol}\geq 10^{43}\erg\,{\rm s}^{-1}$, and LRDs with different thresholds of $L_{\rm bol}$ as labelled. For comparison, we show observational constraints for LRDs from \citet{Kokorev2024,Greene2026} at $z\gtrsim 5$ and \citet{Ma2026,Zhuang2026} at lower redshifts. The constraints from \citet{Kokorev2024,Greene2026} reach down to $L_{\rm bol}\sim 10^{43}-10^{43.5}\erg\,{\rm s}^{-1}$ with updated bolometric corrections, while \citet{Ma2026,Zhuang2026} reach $L_{\rm bol}\sim 10^{44}-10^{44.5}\erg\,{\rm s}^{-1}$. At $z=0$, we show the LRD remnant estimation from \citet{Chisholm2026} (we apply the same $f_{\rm duty}$ factor just for comparison with our high-redshift LRD predictions), which considers LRDs as globular clusters in formation.}
    \label{fig:n_vs_z}
\end{figure}

\subsubsection{Redshift evolution of AGN number density}

In Figure~\ref{fig:n_vs_z}, we show the comoving number density of all AGN with $L_{\rm bol}\geq 10^{43}\erg\,{\rm s}^{-1}$ as a function of redshift. We show the LRD population selected by our empirical model with different $L_{\rm bol}$ thresholds as labelled. On the observational side, \citet{Ma2026,Zhuang2026} reported low-redshift searches of LRDs using ground-based and NIRCam/WFSS Grism surveys to an observational limit of $M_{\rm 5500}\sim -20.5$, which corresponds to $L_{\rm bol}\simeq 10^{44.5}-10^{45}\erg\,{\rm s}^{-1}$ assuming our LRD bolometric corrections. In addition, for LRD searches at $z\gtrsim 5$, we take the sum of the binned estimations in \citet{Kokorev2024,Greene2026} to obtain LRD number densities. These high-redshift observations typically have lower detection limits. In \citet{Greene2026}, the lowest $L_{\rm bol}$ bin extends to $\sim 10^{43.5}\erg\,{\rm s}^{-1}$. In \citet{Kokorev2024}, the lowest $L_{\rm bol}$ bin centers on $10^{44}\erg\,{\rm s}^{-1}$ and would reduce by a factor of $\sim 10$ to $10^{43}\erg\,{\rm s}^{-1}$ if one adopts updated bolometric corrections \citep[see the comparison in][]{Greene2026}.

In \luminasim, the total AGN number density increases monotonically towards lower redshift. The LRD population follows a similar evolution pattern, while the brightest LRD abundance declines faster. At $z\simeq 3-5$, the predicted number density of observable LRDs is consistent with the measurements from \citet{Zhuang2026} and higher than the ones from \citet{Ma2026} assuming a consistent $L_{\rm bol}$ limit. At $z\gtrsim 5$, the LRD abundance reported in other \textit{JWST} surveys is more consistent with the LRDs in \luminasim, assuming a consistent $L_{\rm bol}$ limit of $10^{43}-10^{44}\erg\,{\rm s}^{-1}$ \citep{Kokorev2024,Greene2026}. We do not find the rise-and-decline trend of LRD number density versus redshift, as found in observational studies and predicted in other theoretical works \citep[e.g.,][]{Pacucci2025-spin,Inayoshi2025b}. From our analysis, it is possible that observational bias (deeper detection limits of \textit{JWST} compared to ground-based surveys targeting LRDs at lower redshift) contributes to the decline of LRD abundance at $z\lesssim 4$, and the observed samples may represent only the bright tail of a larger underlying population. However, the full story is likely more complicated since LRD analogues are rare in the local Universe \citep[e.g.,][]{LinR2025,Lin2026-local,Ji2026b}, while the number density of SMBHs below $M^{\rm crit}_{\rm BH}$ should keep increasing dramatically towards lower redshifts. 

This apparent tension indicates several key missing ingredients in our BH seeding prescription. In \luminasim, BHs continue to be seeded in low-mass haloes once they cross the seeding threshold, even at relatively low redshifts. This may be somewhat unphysical if BH seeding occurs only under special physical conditions at high redshifts, such as the extremely metal-poor environment that is necessary for the formation of direct-collapse BHs \citep[e.g.,][]{Rees1984,Bromm2003,Begelman2006,Lodato2006} and/or supermassive stars \citep[e.g.,][]{Baumgarte1999,Volonteri2010,Hosokawa2013,Woods2017}. Meanwhile, the LRD phenomenon itself, i.e., the activation of the LRD phase in low-mass SMBHs, may be directly tied to the early-Universe environment. Failing to capture these conditions in the model could lead to an overprediction of low-mass, faint AGN at later times. A recent work \citep{Chisholm2026} links LRDs to the formation of globular clusters and estimates a local LRD remnant number density of $\sim 0.1-0.3\,{\rm cMpc}^{-3}$, which we include in the figure for comparison, applying the same $f_{\rm duty}=0.3$ as for our high-redshift LRDs. Their inferred number density lies above even the extrapolation of our results. The key difference here is that they associate LRDs with the early-formation phase of globular clusters, which implies that $f_{\rm duty}$ should evolve to a substantially lower value than we adopt.

\begin{figure}
    \centering
    \includegraphics[width=\linewidth]{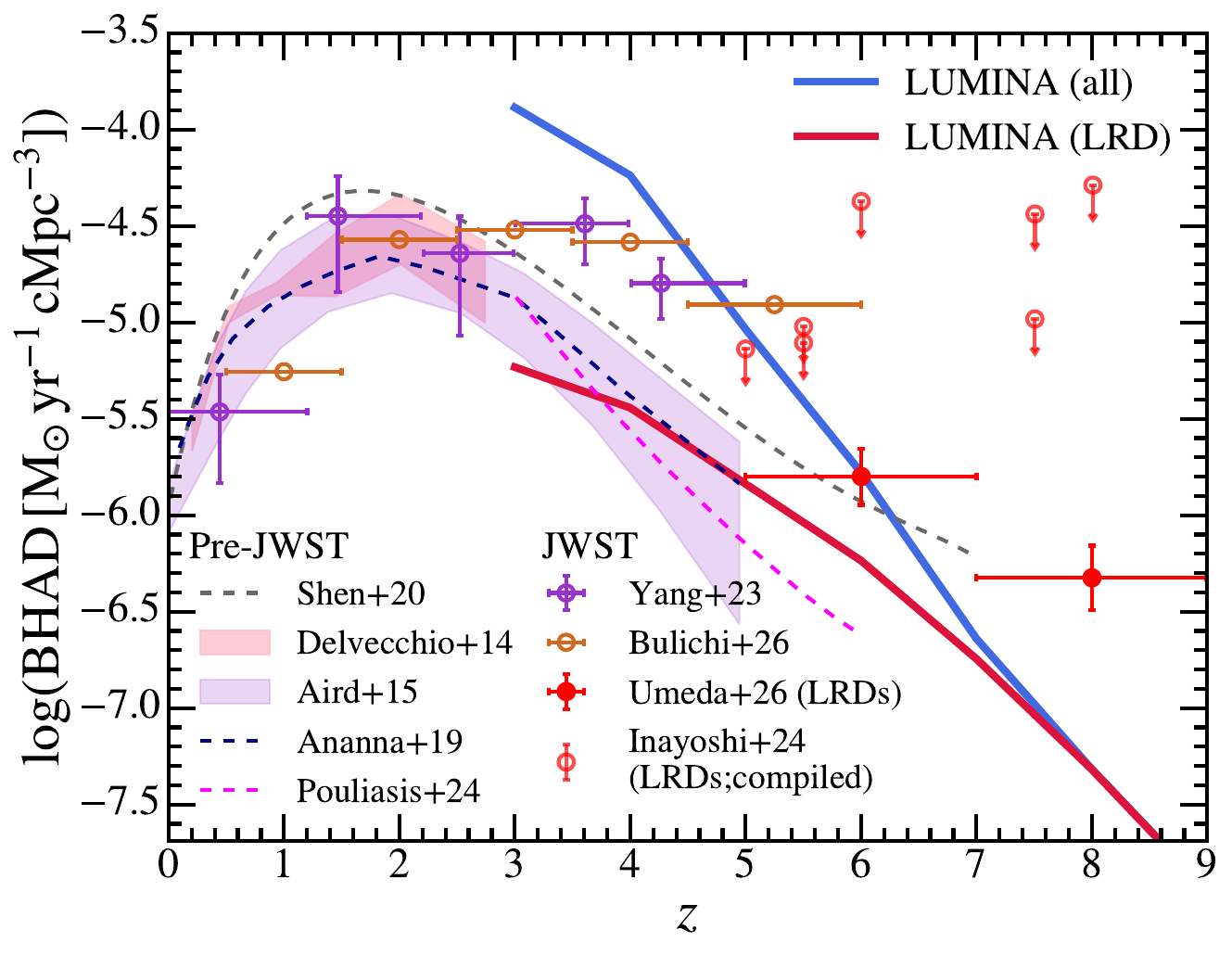}
    \caption{SMBH accretion rate density (BHAD) versus redshift, obtained through a summation of accretion rates of BHs in \luminasim in the bolometric luminosity range $L_{\rm bol}=10^{43}-10^{48}\erg\,{\rm s}^{-1}$. We show the BHAD contributed by all AGN and by LRDs in \luminasim and compare them to LRD constraints compiled in \citet{Inayoshi2024} (originally taken from \citealt{Matthee2024,Greene2024,Kokorev2024,Akins2025}, treated as upper limits because canonical quasar bolometric corrections have been applied) and  \citet{Umeda2026}. At lower redshifts, X-ray constraints from \citet{Aird2015a, Ananna2019,Pouliasis2024} and mid-IR constraints from \citet{Delvecchio2014,Yang2023,Bulichi2026} are shown for reference. Many of these works assume $\epsilon_{\rm rad}=0.1$, and we have corrected them to be consistent with $\epsilon_{\rm rad}=0.2$ assumed in \luminasim.}
    \label{fig:rhomdot}
\end{figure}

\begin{figure}
    \centering
    \includegraphics[width=\linewidth]{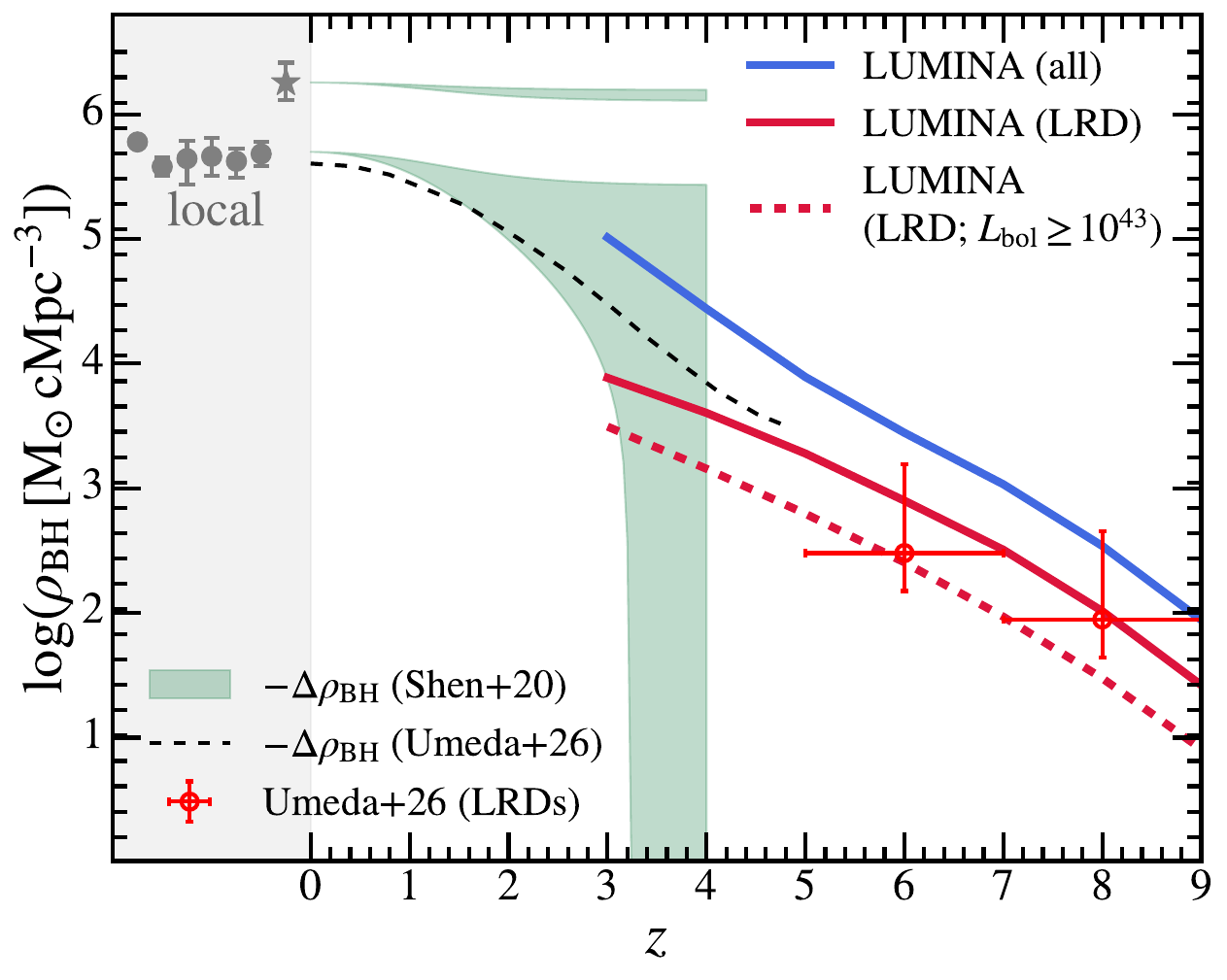}
    \caption{SMBH mass density versus redshift. We show the mass density of all SMBHs, LRDs, and LRDs with $L_{\rm bol}\geq10^{43}\erg\,{\rm s}^{-1}$. The Local Universe constraints are taken from \citet{Shankar2004,Shankar2009,Marconi2004,Graham2007,Hopkins2007,Yu2008} (grey circles) and \citet{Liepold2024} from the MASSIVE survey (grey star), which suggests a substantially higher local BH mass density. The green band shows a back propagation of the BH mass density by subtracting the contribution by quasars in \citet{Shen2020} through redshifts from the local value, assuming $\epsilon_{\rm rad}=0.1-0.2$. We show both the results assuming the canonical local $\rho_{\rm BH}\simeq 5 \times 10^5\msun\,{\rm cMpc}^{-3}$ and assuming the updated $\rho_{\rm BH}\simeq 1.8 \times 10^6\msun\,{\rm cMpc}^{-3}$ in \citet{Liepold2024}. We show the results derived in a similar way from \citet{Umeda2026} using X-ray LFs for reference. For LRDs, we show the observational estimates from \citet{Umeda2026}.}
    \label{fig:rhombh}
\end{figure}

\subsubsection{The Soltan argument}

The Soltan argument attributes the SMBH mass density in the local Universe to the integrated radiatively efficient accretion (manifest as AGN) over cosmic time \citep[e.g.,][]{Soltan1982,Yu2002,Marconi2004,Hopkins2007,Shankar2009,Shen2020}. It is a critical integral constraint for SMBH models. In Figure~\ref{fig:rhomdot}, we show the SMBH accretion rate density (BHAD) as a function of redshift. Commonly, in LF studies, this quantity can be computed by integrating the bolometric LF and converting luminosity to BH mass growth as
\begin{equation}
    \dot{\rho}_{\rm BH} =
    \frac{1-\epsilon_{\rm rad}}{\epsilon_{\rm rad}\,c^2}
    \int^{L_{\rm max}}_{L_{\rm min}} L_{\rm bol}\,\Phi(L_{\rm bol})\,{\rm d}\log L_{\rm bol} \, ,
\end{equation}
where $L_{\rm min}$ and $L_{\rm max}$ are usually taken to be $10^{43}$ and $10^{48}\erg\,{\rm s}^{-1}$. From simulations, however, we can directly compute BHAD as the summation of the accretion rates of BHs within the bolometric luminosity range. A finite-volume correction would be required at the bright end, where the \luminasim box cannot directly sample the rarest quasars up to $L_{\rm bol}=10^{48}\erg\,{\rm s}^{-1}$. We estimate this missing contribution by integrating the bolometric quasar LF of \citet{Shen2020} from the maximum BH luminosity in \luminasim to the integration limit. We find that it accounts for less than $10\%$ of the total BHAD even at $z\simeq 7$ and is smaller at lower redshifts. Neglecting it, therefore, does not impact our conclusions.

We find that the total BHAD in \luminasim rises rapidly from $z\gtrsim 8$ to $z\simeq 3$. Over the redshift range where comparisons are available, the predicted total BHAD is higher than constraints derived from X-ray \citep{Aird2015a,Ananna2019,Pouliasis2024} and pre-\textit{JWST} mid-IR surveys \citep{Delvecchio2014}. Our predictions are more consistent with more recent \textit{JWST}/MIRI survey results \citep{Yang2023,Bulichi2026}. Many of these works assume $\epsilon_{\rm rad}=0.1$. Since the inferred $\dot{\rho}_{\rm BH}\propto (1-\epsilon_{\rm rad})/\epsilon_{\rm rad}$, we have applied a correction factor to these observationally inferred BHAD to be consistent with the $\epsilon_{\rm rad}=0.2$ assumed in \luminasim. The LRD contribution is lower than the total AGN BHAD but remains non-negligible at $z\gtrsim 4$, indicating that the LRD phase can contribute appreciably to early SMBH growth without dominating the entire accretion budget. We also compare to observational estimates of the LRD accretion rate density obtained in \citet{Inayoshi2024}, which are based on measurements from \citet{Matthee2024,Greene2024,Kokorev2024,Akins2025}. These estimates typically adopt canonical quasar bolometric corrections and should be regarded as an approximate upper limit. \citet{Umeda2026} provides an updated estimate using a BH-envelope model, with a similar resulting LRD SED to us. The LRD BHAD in \luminasim lies below the upper-limit compilation and closer to the lower values inferred by \citet{Umeda2026}. This suggests that LRDs can be numerous and optically prominent without requiring an unrealistically large global SMBH growth rate or creating a severe tension with the BH mass density budget \citep{Inayoshi2024}.

In Figure~\ref{fig:rhombh}, we show the corresponding SMBH mass density as a function of redshift. We compute the total SMBH mass density in \luminasim, the mass density of BHs in LRD phase, and the mass density in actively visible LRDs with $L_{\rm bol}\geq 10^{43}\erg\,{\rm s}^{-1}$. We examine the consistency of the model with the Soltan argument. We start from the local SMBH mass-density estimates from \citet{Shankar2004,Marconi2004,Graham2007,Hopkins2007,Yu2008,Shankar2009}, although the recent MASSIVE survey suggests a substantially higher local SMBH mass density than past measurements \citep{Liepold2024}, which is independently supported by the large amplitude of the nanohertz gravitational-wave background reported by the pulsar timing arrays \citep{Agazie2023,SatoPolito2025}. We first take a canonical estimate of $\rho_{\rm BH}\sim 5 \times 10^{5}\msun\,{\rm cMpc}^{-3}$, and back-propagate this local SMBH mass density by subtracting the integrated quasar accretion history from \citet{Shen2020}, assuming $\epsilon_{\rm rad}=0.1$ and $0.2$ to obtain lower and upper limits of $\rho_{\rm BH}$ at high redshifts. A similar exercise was carried out by \citet{Umeda2026} using X-ray LFs, and they obtained similar results to us. The total SMBH mass density in \luminasim remains below the local value and evolves in a manner broadly compatible with the back-propagated constraints. The LRD population contributes a substantial fraction of the high-redshift SMBH mass density, but itself lies comfortably below any SMBH mass budget constraints. The predicted LRD mass density is consistent with the observational estimates in \citet{Umeda2026}.

However, if we take the updated local SMBH mass density $\rho_{\rm BH}\sim 1.8 \times 10^{6}\msun\,{\rm cMpc}^{-3}$ from \citet{Liepold2024}, the back-propagated mass density at high redshifts is substantially higher than the value from \luminasim, and is in fact too large to be reconciled with existing AGN LF constraints. Closing this gap may require a substantial fraction of SMBH mass to be assembled through obscured or radiatively inefficient accretion that is missed by the optical/UV/X-ray AGN census at low redshift. This is qualitatively consistent with our findings in Section~\ref{subsec:midir} that both the observed mid-IR LF and the predictions from \luminasim are higher than the pre-\textit{JWST} census, but our SMBH density comparison here shows that it is still not sufficient to explain the deficit. One cannot rule out the possibility that the \citet{Liepold2024} result itself is biased by selection effects in the dynamically-measured SMBH samples that anchor the massive end of the local scaling relations. 

\subsubsection{The role of AGN in \HeII reionization}

\begin{figure}
    \centering
    \includegraphics[width=\linewidth]{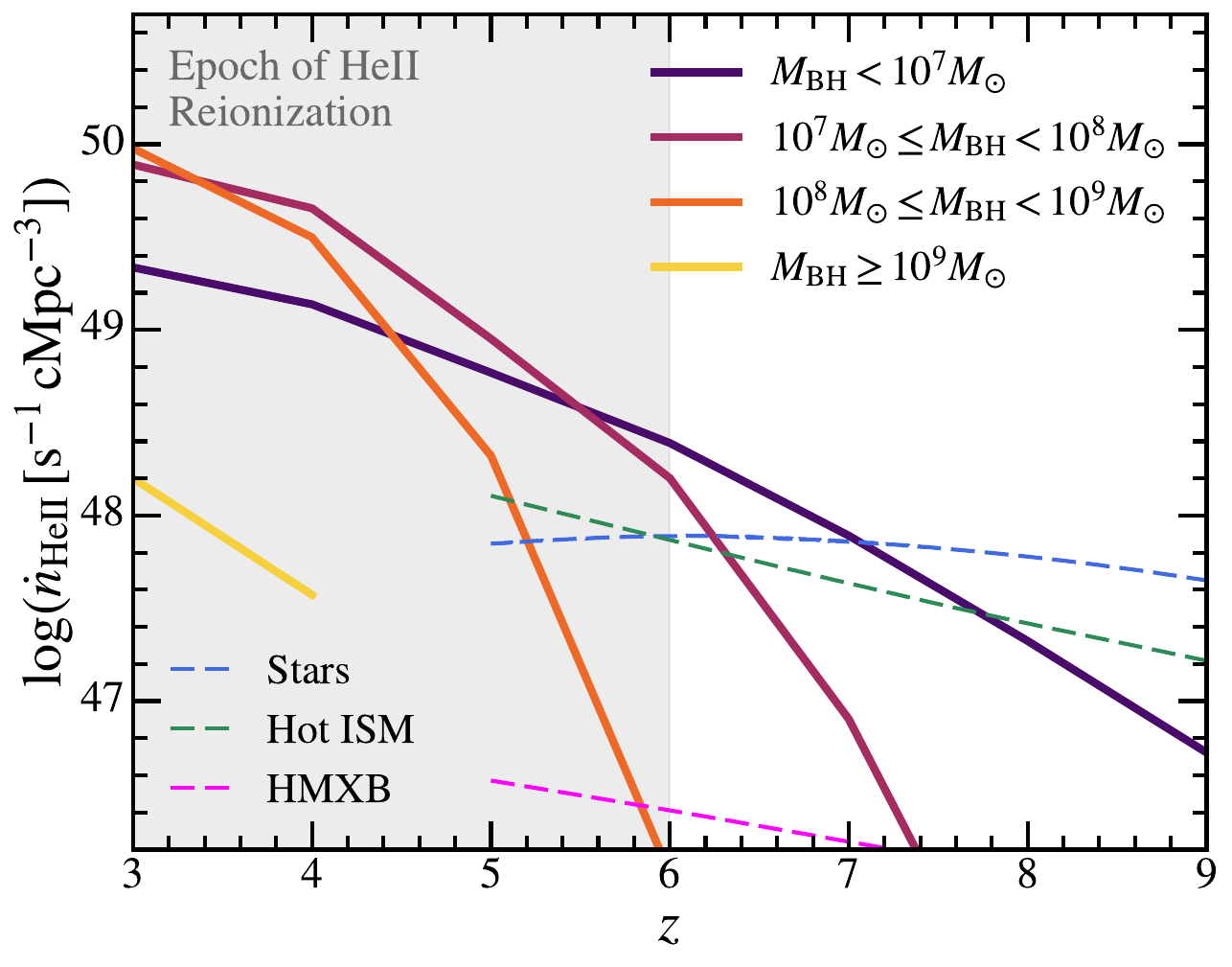}
    \caption{Comoving production rate density of \HeII-ionizing photons at $54.42-100\,{\rm eV}$, $\dot{n}_{\rm ion}^{\rm HeII}$, as a function of redshift in \luminasim. We show decomposed emissivity from SMBHs in four mass bins. The vertical shaded band marks the \HeII-reionization epoch in \luminasim, $z\simeq 3-6$. We also show the contribution from other sources to $z\simeq 5$, which are subdominant for \HeII reionization. They are therefore not tracked at $z<4.75$ (see Section~\ref{subsec:rt}). The main drivers for \HeII reionization in \luminasim are SMBHs in the mass range $10^{7}-10^{9}\msun$.}
    \label{fig:nphoton}
\end{figure}

While stars in low-mass galaxies are widely believed to dominate the reionization of hydrogen \citep[e.g.,][]{FG2008Letter,Kuhlen2012,Robertson2015,Onoue2017,Parsa2018}, AGN have long been considered the principal drivers of
the reionization of singly ionized helium, owing to their hard EUV spectra extending well beyond $4\,{\rm Ryd}\simeq 54.42\,{\rm eV}$ \citep[e.g.,][]{Miralda1998,Madau1999,FG2008,
McQuinn2009,Haardt2012} and is supported by \HeII Lyman$\alpha$ forest observations \citep[e.g.,][]{Worseck2014,Worseck2016}. The large abundance of LRDs discovered by \textit{JWST} at $z\gtrsim 4$ has revived the long-standing question of whether AGN could also contribute appreciably to hydrogen reionization \citep[e.g.,][]{MadauHaardt2015,Haardt2015,Madau2024,Asthana2025}, although models in which AGN drive both phase transitions remain in tension with the observed Ly$\alpha$ forest and IGM thermal history \citep{Asthana2025}. 

In \luminasim, the joint cosmic radiation field of stars and AGN is followed self-consistently with on-the-fly radiative transfer (Section~\ref{subsec:rt}). As described in \citet{Zier2026}, \luminasim predicts a \HeII reionization history that begins around $z\simeq 6$ and completes near $z\simeq 3$, broadly consistent with existing constraints from the \HeII Ly$\alpha$ forest \citep[e.g.,][]{Worseck2014,Worseck2016,Worseck2019} and the IGM thermal history \citep{Lidz2010,Gaikwad2021}. Here, we leverage this self-consistent picture to dissect the AGN contribution to \HeII reionization in greater detail, asking how SMBHs of different masses and evolutionary phases share the photon budget.

We compute the comoving production rate density of \HeII-ionizing photons by summing the bolometric luminosities of all accreting SMBHs in the simulation volume,
\begin{equation}
    \dot{n}_{\rm ion}^{\rm HeII} \simeq \dfrac{\mathcal{F}_{\rm HeII}}{\langle \epsilon_{\rm HeII} \rangle\,V_{\rm box}}\,\sum_{i} f_{\rm esc}(L^{i}_{\rm bol})\, L^{i}_{\rm bol} \, ,
    \label{eq:nion-HeII}
\end{equation}
where $\mathcal{F}_{\rm HeII}=0.0455$ is the fraction of the bolometric luminosity emerging at the \HeII-ionizing band ($54.42-100\,{\rm eV}$) given the EUV spectral shape adopted in our canonical AGN SED model (Section~\ref{subsec:sed}; \citealt{Lusso2015,Shen2020}), $\langle \epsilon_{\rm HeII} \rangle \simeq 74.93\,{\rm eV}$ is the mean energy in the band, $V_{\rm box}$ is the comoving volume of the simulation, and the sum runs over active SMBH particles. $f_{\rm esc}(L_{\rm bol})$ is the luminosity-dependent escape fraction introduced in Section~\ref{subsec:rt}. 

In Figure~\ref{fig:nphoton}, we show the resulting \HeII-ionizing photon emissivity in \luminasim, decomposed into four BH mass bins. We also show the contribution from stars, hot ISM, and HMXBs at $z\gtrsim 5$ for comparison. It is clear that AGN dominate the production of \HeII-ionizing photon since $z\gtrsim 7$ and justifies our choice to deactivate other sources at $z<4.75$ (see a more thorough discussion in \citealt{Zier2026}). During the bulk of \HeII reionization at $z\lesssim 6$, the photon budget is dominated by SMBHs in the range $10^{7}-10^{9}\msun$, with only a modest contribution from BHs above $10^{9}\msun$ that reflects their rarity at these epochs. At the beginning of \HeII reionization, emissivity is dominated by BHs with mass $10^{7}-10^{8}\msun$ while more massive BHs at $10^{8}-10^{9}\msun$ become equally important closer to the end of reionization. Only at $z\gtrsim 6$, when the most massive SMBHs have not yet assembled, do BHs below $10^{7}\msun$ contribute an appreciable share of the ionizing emissivity. The characteristic bolometric luminosity of $\sim 10^{8}\msun$ BHs is $L_{\rm bol} \sim L_{\rm Edd}\sim 10^{46}\,\erg\,{\rm s}^{-1} \sim 10^{12.5}\Lsun$, which is aligned with the conventional wisdom that BHs at the knee of the quasar LF \citep{Shen2020} drive the \HeII reionization \citep[e.g.,][]{Sokasian2002,FurlanettoOh2008,McQuinn2009,Yung2021}.

This mass dichotomy has a useful consequence for our empirical AGN model. The mass scale below which BHs do not contribute substantially to the ionizing emissivity coincides with our adopted LRD threshold, $M^{\rm crit}_{\rm BH}\sim 10^{7}\msun$ (Section~\ref{subsec:split}). The EUV spectra of LRDs are observationally poorly constrained: most LRDs are remarkably X-ray faint or undetected, and the dense gas inferred from their optical spectra suggests that their intrinsic EUV continuum may differ substantially from the canonical quasar template adopted in Section~\ref{subsec:rt}. Reassuringly, Figure~\ref{fig:nphoton} demonstrates that LRD-mass BHs contribute negligibly to the \HeII-ionizing photon budget over the redshift range where \HeII reionization actually proceeds in \luminasim. Our conclusion that AGN drive \HeII reionization is therefore essentially independent of how the EUV emission of LRDs is modelled and rests on the well-characterized population of $\gtrsim 10^{7}\msun$ canonical AGN. 

\section{Discussion}
\label{sec:discussion}

\subsection{AGN variability is essential for early SMBH growth}
\label{subsec:variability}

As discussed in Section~\ref{subsec:lbol-scatter} above, one fundamental limitation of any cosmological simulation is that many relevant physical processes related to gas inflow and BH accretion operate many orders of magnitude below the resolution limit. The Bondi--Hoyle prescription effectively delivers a ``smooth'' accretion rate set by the kpc-scale gas properties, while real AGN are observed to vary over timescales of hours to decades or longer. Quantitatively, in large optical quasar samples, the asymptotic structure-function amplitudes (which quantify the total light-curve variability) are typically $\sim 0.1-0.4$ mag on rest-frame timescales of $\sim 10^{2}-10^{3}$ days \citep[e.g.,][]{VandenBerk2004,Macleod2010,Caplar2017,Burke2021}. In hard X-rays, where variability traces the corona on much shorter timescales, fractional rms amplitudes of a few tens of percent are commonly observed on timescales of hours to days \citep[e.g.,][]{Vaughan2003,GonzalezMartin2012,Ponti2012}, corresponding to $\sim 0.1$ dex scatter in luminosity. On the much longer timescales relevant to the lifetime of an individual quasar ($\sim 0.1-100$ Myr), the observed ERDF can span several orders of magnitude in $\lambda_{\rm Edd}$ \citep[e.g.,][]{Hopkins2009,Hickox2014,Aird2018,Sartori2018}, with effective log-normal dispersions of $\sim 0.3-0.4$ dex \citep{Kollmeier2006,Kauffmann2009,Kelly2013,Lusso2012}. Our adopted $\sigma_{\rm bol}=0.3$ dex therefore sits naturally between the short-timescale optical/X-ray variability amplitude and the long-timescale ERDF dispersion.

The $\sigma_{\rm bol}=0.3$ dex log-normal scatter applied in our empirical model (Section~\ref{subsec:lbol-scatter}) is an essential and physically motivated ingredient. It captures unresolved accretion variability and translates the simulated mean accretion rates into a realistic distribution of observed AGN luminosities. However, we caution that this treatment remains simplified in several aspects: (\textit{i}) The scatter is applied to the bolometric luminosity, so it does not just reflect intrinsic accretion-rate variability but may also absorb fluctuations in the obscuration properties. (\textit{ii}) The assumed log-normal kernel is itself a simplification, whereas the observed ERDFs of AGN are often better described by a Schechter form, with a power-law tail towards low Eddington ratios and an exponential cutoff near the Eddington limit \citep[e.g.,][]{Hopkins2009,Schulze2010,Aird2018}. (\textit{iii}) We adopt a constant scatter across the full luminosity range, whereas a luminosity- or mass-dependent kernel may provide a better description (see Figure~\ref{fig:alt1} for a hint).

Variability is also likely vital for early SMBH growth and quenching galaxies. Recent zoom-in simulations have shown that allowing bursty super-Eddington episodes, physically motivated by photon-trapping in slim discs, can grow $\sim 10^{4} \msun$ seeds to $\sim 10^{8-9} \msun$ within several hundred Myr, in regimes where Eddington-limited Bondi--Hoyle accretion would fail by orders of magnitude \citep[e.g.,][]{Lupi2024,PacucciNarayan2024,Husko2025}. The duty cycle of these super-Eddington bursts is short (often $\lesssim 10$ Myr), but the integrated mass growth is dominated by them. It offers a more efficient pathway to assembling the most massive quasars known at $z\gtrsim 6-7$ with $M_{\rm BH}\gtrsim 10^{9}\msun$ \citep[e.g.,][]{Mortlock2011,Banados2018,Wang2018b,Wu2022}, whose existence has long challenged growth from $\lesssim 10^{4-5}\msun$ seeds under standard Eddington-limited accretion \citep[e.g.,][]{Volonteri2010,Smith2019,Inayoshi2020}. The recent \textit{JWST} detections of BLAGN out to $z>10$ \citep{Bogdan2024,Maiolino2025} only sharpen this issue. In addition, in many modern AGN feedback prescriptions (including the \illustrisTNG model on which \luminasim is based), AGN feedback transitions from a radiatively-coupled thermal mode to a substantially more efficient kinetic mode once the BH mass exceeds a characteristic threshold of a few times $10^{8}\msun$ \citep{Sijacki2007,Weinberger2017,Shen2025}, which quenches massive galaxies in these models \citep[e.g.,][]{Terrazas2020,Piotrowska2022}. A stochastic early growth could let BHs enter the low-accretion rate kinetic mode earlier and more frequently (even below the mass threshold), and accelerate the quenching of massive host galaxies and potentially help to explain the population of quiescent massive galaxies now identified by \textit{JWST} at $z\gtrsim 4$ \citep[e.g.,][]{Carnall2023,Looser2024,Glazebrook2024,DeGraaff2025a}. Exploring these possible consequences of a bursty early growth phase of SMBHs will require more self-consistent adjustments to the BH accretion and feedback recipes in simulations rather than the simple empirical scatter of luminosity, and is deferred for future work. 

The amplification of bright-quasar abundances by intrinsic variability is closely analogous to the amplification of the high-redshift galaxy UV LF by bursty star-formation. \textit{JWST} surveys at $z\gtrsim 10$ revealed an excess of UV-bright galaxies relative to pre-\textit{JWST} extrapolations \citep[e.g.,][]{Finkelstein2023,Harikane2023,Finkelstein2024,Naidu2026}, and a leading explanation is that the rest-UV luminosities of low-mass high-redshift galaxies vary stochastically by $\sigma_{\rm UV}\simeq 1-2$ mag driven by the short feedback-modulated SFR cycles seen in zoom-in simulations \citep[e.g.,][]{Mirocha2023,Shen2023,Shen2024,Shen2026,Mason2023,Sun2023,McClymont2025}. Despite potentially very different relevant astrophysical timescales, the conceptual lesson is the same: in regimes where the underlying source population is steeply declining at the bright end and where short-timescale fluctuations are large, time variability is a central component of the observable LF.

\subsection{Summary of explicit \& implicit assumptions about LRDs}
\label{subsec:model-assumptions}

We have described above how \luminasim, paired with an empirical model connecting SMBHs to observable AGN, can reproduce several key properties of high-redshift quasars and LRDs. Here, we provide a more direct summary of the explicit and implicit assumptions entering this framework. The most important explicit assumptions are:
\begin{enumerate}
    \item In \luminasim, we place $8\times 10^5\,h^{-1}\msun$ seed BH particles in every halo that exceeds $5\times 10^{10}\,h^{-1}\msun$ and does not already contain a BH. It provides the endpoint of the unresolved BH growth below the seed mass.
    
    \item For BH particles below $M^{\rm crit}_{\rm BH}=\kappa\,M_{\rm seed}=1.18\times 10^7\msun$ ($\kappa = 10$), a fraction $f_{\rm duty}=0.3$ is assumed to appear observationally as LRDs. This intuitively associates LRDs with the early growth phase of SMBHs.
    
    \item We assume Bondi--Hoyle accretion onto BH particles, without an additional boost factor, limited by the Eddington rate, and adopt a constant radiative efficiency $\epsilon_{\rm rad}=0.2$ for all BHs.
    
    \item We assume an intrinsic unresolved scatter in instantaneous BH luminosities of $\sigma_{\rm bol}=0.3$ dex, which phenomenologically accounts for accretion variability below the resolution scale of the simulation.
\end{enumerate}
The remaining ingredients are mainly observationally motivated, including the SEDs and bolometric corrections of canonical AGN and LRDs, and the column-density distribution of canonical AGN, which determines absorption and dust extinction corrections. In this sense, our model separates the problem into two parts: the simulation provides the abundance, growth, and environments of BH particles, while the empirical model specifies how these BH particles appear in different wavelength bands.

The physical interpretation of $f_{\rm duty}$ is broader than its implementation suggests. For example, because \luminasim seeds BHs in every halo above the adopted threshold, $f_{\rm duty}$ effectively absorbs the true occupation fraction of BH seeds in early haloes. Alternatively, a timescale-based interpretation of $f_{\rm duty}$ emerges if we assume that essentially every newly seeded BH passes through a single LRD-like episode and is observable while in that phase. In that limit, $f_{\rm duty}\sim t_{\rm LRD}/t_{\rm low}$, where $t_{\rm LRD}$ is the duration of the LRD-emitting phase and $t_{\rm low}$ is the total time a BH spends below $M^{\rm crit}_{\rm BH}$. Both time scales can be set from physical models. For example, in the late-stage quasi-star scenario \citep{Begelman2008,Begelman2026}, the LRD phase coincides when a central BH swallowing a convectively powered envelope and lasts of order the Salpeter time $t_{\rm Sal} \equiv \epsilon_{\rm rad}\,M_{\rm BH} c^{2}/[(1-\epsilon_{\rm rad}) L_{\rm Edd}] \simeq 112.5 \Myr$ (assuming $\epsilon_{\rm rad}=0.2$), modulo a factor that depends on the envelope-to-BH mass ratio at envelope dispersal \citep{Begelman2010,Volonteri2010,Coughlin2024}. The lifetime $t_{\rm low}$ is set by the time required for the BH to grow from $M_{\rm seed}$ to $\kappa M_{\rm seed}$ at its average accretion rate. For fast but Eddington-limited growth with $\langle\lambda_{\rm Edd}\rangle\sim 0.5-1$ and our fiducial $\kappa=10$, we have $t_{\rm low}\sim t_{\rm Sal} \ln\kappa /\langle\lambda_{\rm Edd}\rangle\simeq 260\Myr$ in \luminasim. Therefore, one predicts $f_{\rm duty}\sim 0.2 - 0.4$, comfortably bracketing our fiducial choice. Alternative scenarios, in which LRDs correspond to recurrent super-Eddington accretion bursts each lasting only a few Myr \citep{Inayoshi2025b}, are also consistent with $f_{\rm duty}=0.3$ provided that the bursts repeat a few times during the LRD-eligibility window, as those models in fact predict.

The BH particle in \luminasim should be viewed as a sub-grid tracer of an unresolved proto-nuclear ecosystem that may include the seed BH itself, a surrounding gas reservoir, and possibly a compact nuclear star cluster. Several physical models have been proposed for what such systems look like, including but not limited to the quasi-star picture mentioned above, and an intermediate-mass BH growing rapidly inside a dense nuclear star cluster (or a cluster of clusters) fed by runaway stellar collisions and mergers \citep[e.g.,][]{Devecchi2009,Davies2011,Katz2015,Shi2024,Dekel2025}. \luminasim resolves the large-scale environment, halo assembly, and global gas supply common to all of these scenarios, but does not resolve the internal radiative and dynamical processes that distinguish them. Our empirical model for AGN is deliberately agnostic to which picture applies. The broad consistency with the quasi-star timescale estimates discussed above is encouraging in this context, but is not by itself sufficient to discriminate between the underlying scenarios. Note that this interpretation also implies the BH mass density presented in Figure~\ref{fig:rhombh} is an upper limit before the gas reservoir (and maybe also the nuclear star cluster) is consumed by the central BH. This will not break the Soltan argument given the large uncertainties in the Local BH mass density and the BH radiative efficiency at low redshifts.

Several implicit assumptions also enter the empirical model. The transition between the LRD-like and canonical AGN phases is treated as a sharp mass cut, although in reality it likely depends on additional properties such as gas density, metallicity, and halo assembly history. The discrepancy with the observed decline of the LRD number density at lower redshifts (Section~\ref{subsec:cosmo}) may be one consequence. We further assume that the LRD SED is approximately universal, independent of redshift, BH mass, luminosity, and host-galaxy properties, and that the same effective parameters apply across $z\simeq 3-6$. These choices are deliberately simple as the main goal of this work is to test whether a plausible LRD-like phase can be embedded in a large-volume cosmological simulation while remaining consistent with a wide range of AGN observables and constraints. The agreement found above demonstrates that such a region of parameter space exists, although it may not be unique.

\subsection{The impact of model parameters and their degeneracies}
\label{subsec:parameters}

The empirical model presented in Section~\ref{sec:smbh-agn} introduces a small set of free parameters: the unresolved bolometric luminosity scatter $\sigma_{\rm bol}$, the LRD mass threshold $\kappa = M^{\rm crit}_{\rm BH}/M_{\rm seed}$, and the LRD duty cycle $f_{\rm duty}$. In this subsection, we examine how each of these parameters individually shapes the predicted multi-band AGN LFs, and discuss potential degeneracies between them and with other implicit assumptions of the simulation. For brevity, we focus on the bolometric LF at $z\simeq 3$, where \luminasim predictions overlap most strongly with the bright-end observational constraints, and on the optical and hard-X-ray LFs at $z\simeq 5$, where most LRD observational constraints are. 

\begin{figure}
    \centering
    \includegraphics[width=\linewidth]{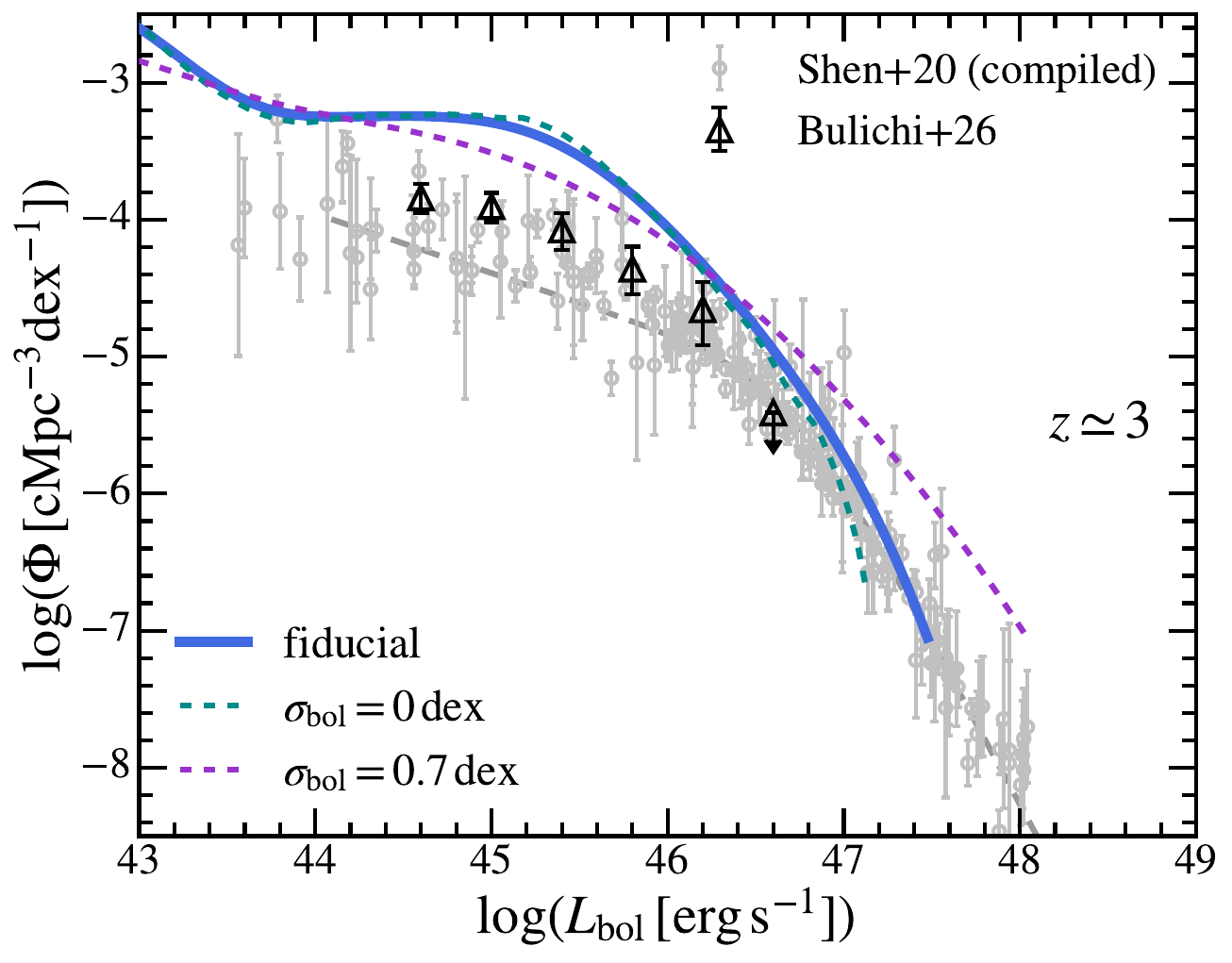}
    \caption{Bolometric LFs of AGN at $z\simeq 3$ in \luminasim, assuming $\sigma_{\rm bol}=0$ and $0.7$ dex, compared to the fiducial model. We choose $z\simeq 3$ for this comparison because it is the redshift where \luminasim predictions have most overlap with the bright-end observational constraints. A lower (higher) $\sigma_{\rm bol}$ results in under- (over-) prediction of the bright-end LF, while the faint end is less affected. A luminosity- or mass-dependent $\sigma_{\rm bol}$ may result in a better agreement at both the bright end and $L_{\rm bol}\sim 10^{45}\erg\,{\rm s}^{-1}$, where \luminasim overpredicts the AGN number density.}
    \label{fig:alt1}
\end{figure}

In Figure~\ref{fig:alt1}, we show the bolometric LFs at $z\simeq 3$ for three values of $\sigma_{\rm bol}$ bracketing our fiducial value of $0.3$ dex. The bright end of the LF responds strongly to $\sigma_{\rm bol}$. Removing the scatter ($\sigma_{\rm bol}=0$) underpredicts the abundance of luminous AGN at $L_{\rm bol}\sim 10^{47}\,{\rm erg\,s^{-1}}$. This is a minor effect here for \luminasim, but becomes critical for larger-volume simulations such as \mtng, which we included in the comparisons above. On the other hand, adopting a higher $\sigma_{\rm bol}=0.7$ dex severely overpredicts the bright-end LF. The faint end, however, is much less sensitive to $\sigma_{\rm bol}$. This asymmetry is anticipated as a log-normal scatter applied to a steeply declining bright-end LF preferentially scatters faint, abundant sources up to high luminosities, while a shallow faint-end LF is essentially invariant under the same convolution. 

\begin{figure}
    \centering
    \includegraphics[width=\linewidth]{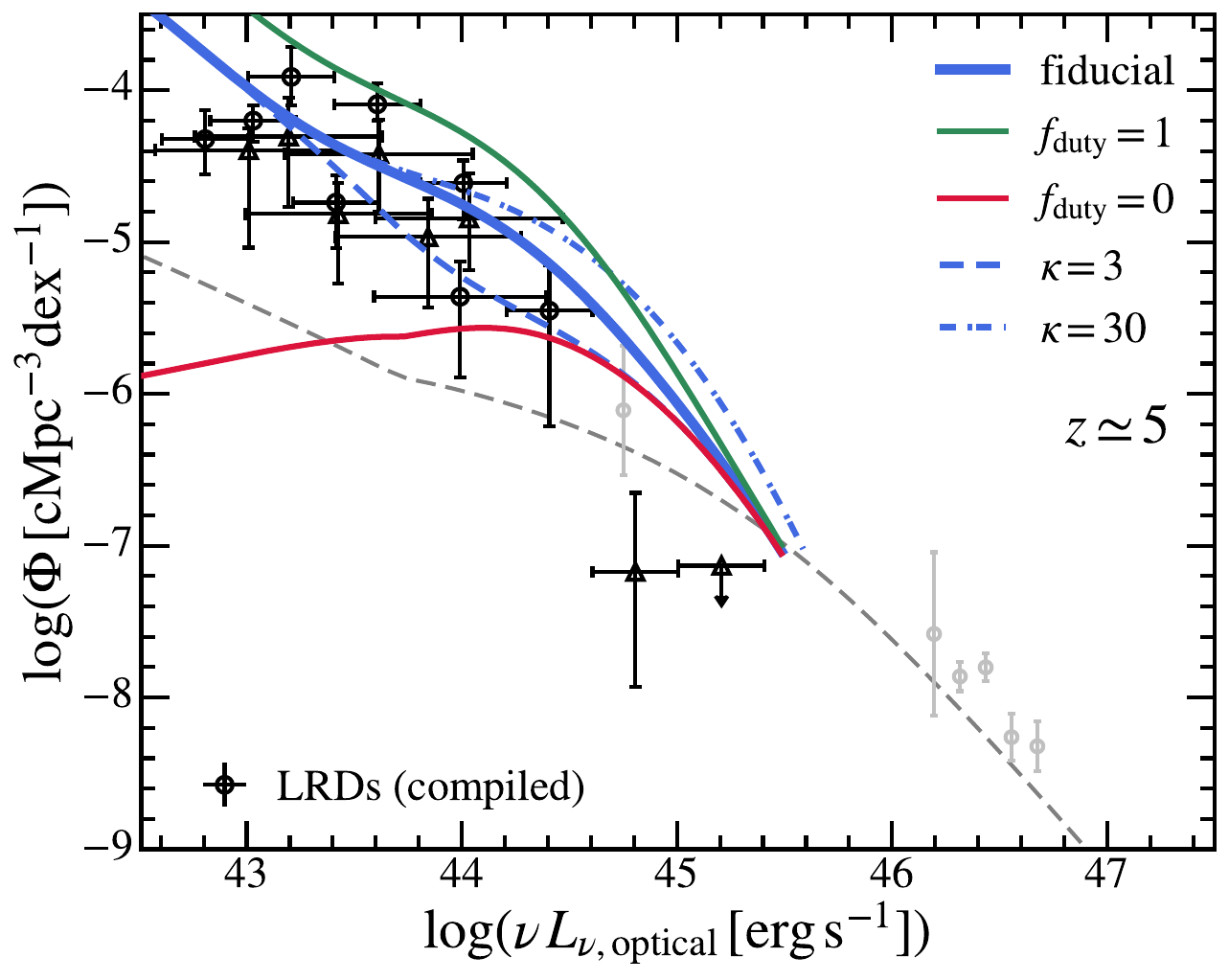}
    \includegraphics[width=\linewidth]{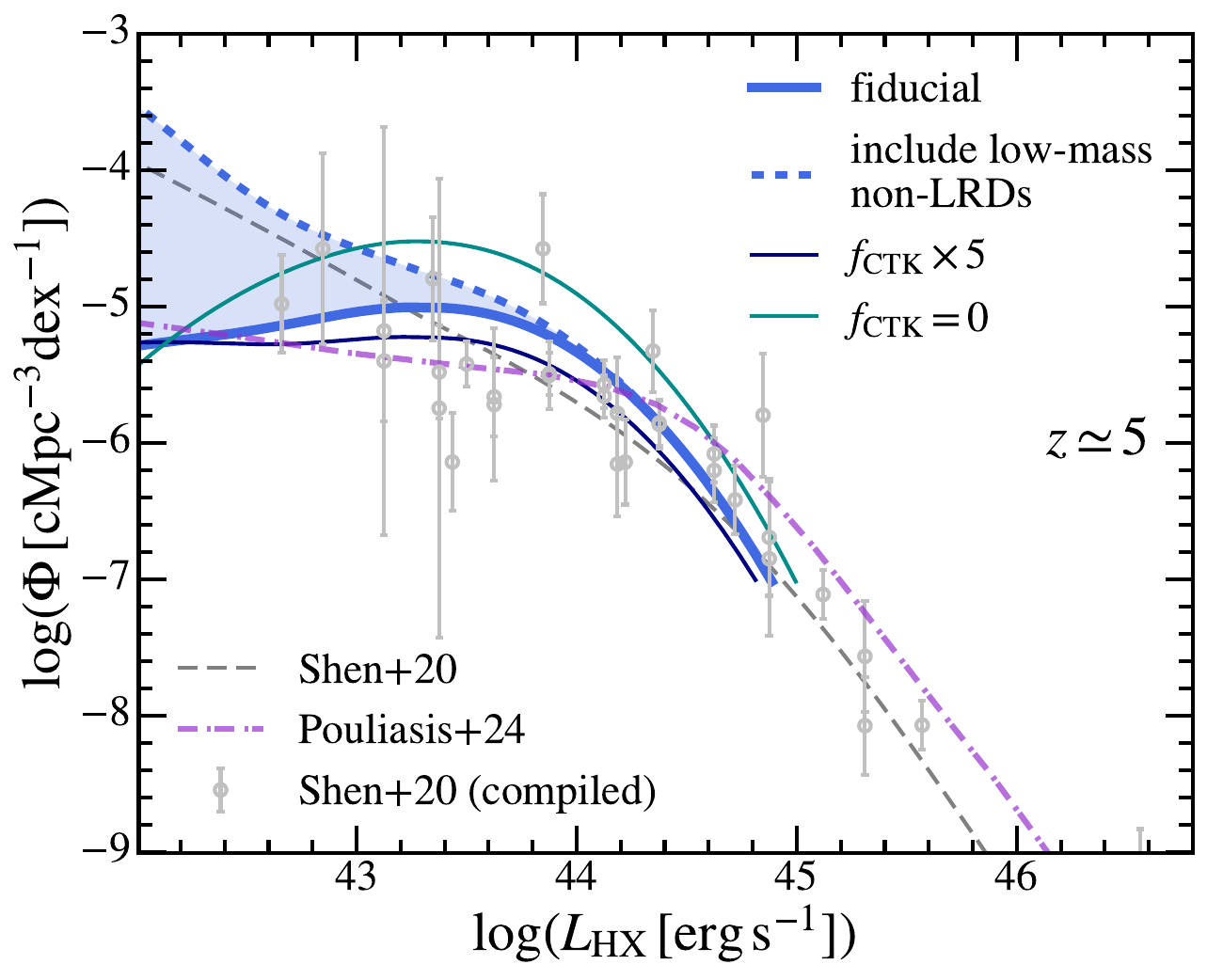}
    \caption{{\it Top:} AGN B-band optical LF at $z\simeq 5$ in \luminasim, for several combinations of $f_{\rm duty}$ and $\kappa$. For clarity, we show only the predictions using the ``weak IR'' bolometric correction of LRDs. The observational constraints of LRDs compiled in Section~\ref{subsec:opt} are shown here for comparison. $f_{\rm duty}$ controls the normalization of the faint-end LF that is dominated by LRDs, while $\kappa$ mainly influences the transitional luminosity between LRD and canonical AGN LFs. {\it Bottom:} AGN hard X-ray LF at $z\simeq 5$ in \luminasim. In contrast to the optical, $f_{\rm duty}$ has no impact on the X-ray LF, because LRDs are intrinsically X-ray weak in our model. The dominant uncertainty here is whether the $(1-f_{\rm duty})$ population of BHs below $M^{\rm crit}_{\rm BH}$ is included as canonical AGN or treated as electromagnetically silent: this choice substantially affects the predicted X-ray LF at $L_{\rm HX}\lesssim 10^{43}\erg\,{\rm s}^{-1}$. We also show the predictions assuming different values of $f_{\rm CTK}$, which mildly affects the normalization of the X-ray LF.}
    \label{fig:alt2}
\end{figure}

In the top panel of Figure~\ref{fig:alt2}, we show the rest-frame B-band LF at $z\simeq 5$ for several variations of $\kappa$ and $f_{\rm duty}$ around our fiducial values. For clarity, we show only the predictions using the ``weak IR'' model. The two parameters are physically distinct and shape the predicted LF in complementary ways. The threshold $\kappa$ controls which simulated BHs are eligible for staying in the LRD phase. Through the LRD bolometric correction at roughly fixed Eddington ratio, it sets the maximum LRD luminosity in the optical, as well as the transition luminosity of the LRD- and canonical-AGN-dominated branches of the LF. Our fiducial value $\kappa=10$ ($M^{\rm crit}_{\rm BH}\sim 10^{7}\msun$) is chosen to coincide with the sharp cut-off of LRD number densities at brighter luminosities \citep[e.g.,][]{Ma2025}. Raising $\kappa$ to $30-100$ extends the LRD branch into the bright-AGN regime and produces a high-luminosity bump, while lowering $\kappa$ truncates the LRD branch at lower luminosities, both of which are less consistent with the data. On the other hand, the duty cycle $f_{\rm duty}$ controls the fraction of eligible BHs that are in the LRD phase at any given epoch and therefore modulates the normalization of the LRD-dominated faint end without affecting the maximum LRD luminosity. Variations in either parameter leave the bright end of the LF, dominated by canonical AGN, essentially unchanged.

In the bottom panel of Figure~\ref{fig:alt2}, we show the hard X-ray LF at $z\simeq 5$. Because LRDs in our model are assumed X-ray dark (Section~\ref{subsec:lrdsed}), $f_{\rm duty}$ itself has no impact on the X-ray LF. The major uncertainty in X-rays is whether the $(1-f_{\rm duty})$ population of BHs below $M^{\rm crit}_{\rm BH}$ that are not in the LRD phase at a given epoch is assigned the canonical AGN SED or is instead treated as electromagnetically silent. This treatment of the off-LRD duty cycle low-mass BHs can substantially alter the predicted X-ray LFs at $L_{\rm HX}\lesssim 10^{43}\erg\,{\rm s}^{-1}$. Assigning them the canonical AGN SED produces a significantly larger faint-end abundance. Nevertheless, the current X-ray observational constraints cannot distinguish the two scenarios. We note that this population has almost no influence on the optical/UV LFs since LRDs strongly dominate over dust-obscured canonical AGN at the faint end. The physical nature of these off-LRD duty cycle BHs in our model is highly uncertain. Future deep, high-angular-resolution X-ray surveys, in particular with the AXIS \citep{Reynolds2023AXIS} and NewAthena \citep{Cruise2025} mission concepts, will provide the most direct route to constraining the transition between the LRD phase and the broader low-mass AGN population at $z\gtrsim 4$. In this figure, we also show the X-ray LFs assuming different CTK AGN fraction ($f_{\rm CTK}$) in our $N_{\rm H}$ distribution model. Removing the CTK population ($f_{\rm CTK}=0$) or increasing the $f_{\rm CTK}$ boost to $5$ has a moderate impact on the X-ray LF at $\sim 0.2-0.5$ dex level.

Beyond the parameters of the empirical model, the predicted AGN LFs also depend on the BH accretion physics built into \luminasim itself. In particular, the Bondi--Hoyle accretion rate adopted in the simulation does not include the multiplicative ``boost factor'' $\alpha$ that is commonly applied in lower-resolution cosmological simulations, e.g., $\alpha\sim 100$ in \illustris \citep{Springel2005,Sijacki2015} and ASTRID \citep{Bird2022,Ni2022}. An ad-hoc accretion boost above unity (along with the removal or relaxation of the Eddington cap) would shift the predicted BH mass and bolometric LFs to the more massive and more luminous end \citep[e.g.,][]{Booth2009,Negri2014,Bennett2024}. To preserve a match to the observed LRD LFs, both $M^{\rm crit}_{\rm BH}$ and $f_{\rm duty}$ would then need to be reduced, and more luminous BHs would correspondingly be placed in lower-mass DM haloes and host galaxies. As noted in Section~\ref{subsec:uv}, the simulated host galaxies of LRD-mass BHs in \luminasim are often too optically luminous to satisfy the observational LRD selection criteria, and also too UV luminous, overshooting the UV LF. Therefore, this re-calibration may help relieve the AGN-host relation tension. It would also reduce the clustering bias of LRDs, but only modestly, and is unlikely to spoil the clustering comparison in Section~\ref{subsec:environment} given the large statistical uncertainties in current observational measurements \citep[e.g.,][]{Matthee2024,Arita2025,Lin2026-env}. 

Another relevant numerical assumption is the treatment of BH dynamics. As described in Section~\ref{subsec:smbh}, \luminasim inherits the \illustrisTNG repositioning scheme, which repeatedly moves BHs towards the local gravitational-potential minimum. This effectively assumes efficient BH sinking, keeping BHs close to dense nuclear gas with small relative velocities and facilitating prompt BH mergers. Without repositioning, BH growth can be strongly suppressed in this class of models \citep{Bahe2022}. This assumption may be particularly important at high redshifts, where clumpy and turbulent proto-galaxies can substantially delay the orbital decay of $\sim10^{6}-10^{8}\msun$ BHs \citep{Pfister2019,Ma2021}. Less efficient sinking would therefore tend to suppress both gas accretion and merger-driven growth, lowering the predicted faint high-redshift AGN abundance. Meanwhile, BHs moving through such clumpy environments may undergo intermittent episodes of rapid, potentially super-Eddington accretion. Our phenomenological luminosity scatter is an attempt to capture such unresolved variability but only qualitatively. A more complete future treatment of high-redshift AGN demographics will likely require a physically motivated model for BH seeding, dynamics, and accretion combined with a self-consistent treatment of super-Eddington phases (as we discussed in Section~\ref{subsec:variability}).

\subsection{Comparison to other cosmological simulations}
\label{subsec:model-comparisons}

\begin{figure}
    \centering
    \includegraphics[width=\linewidth]{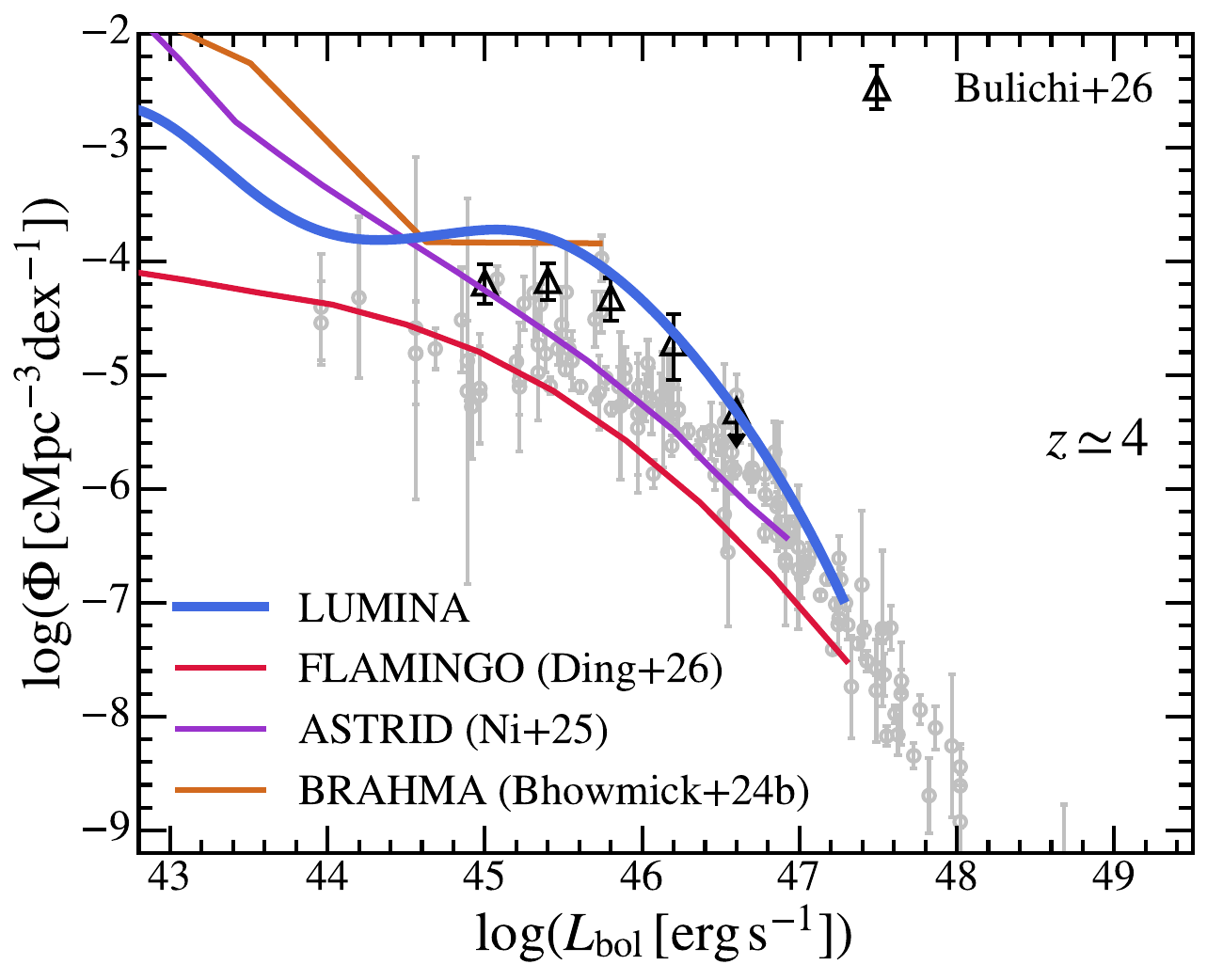}
    \includegraphics[width=\linewidth]{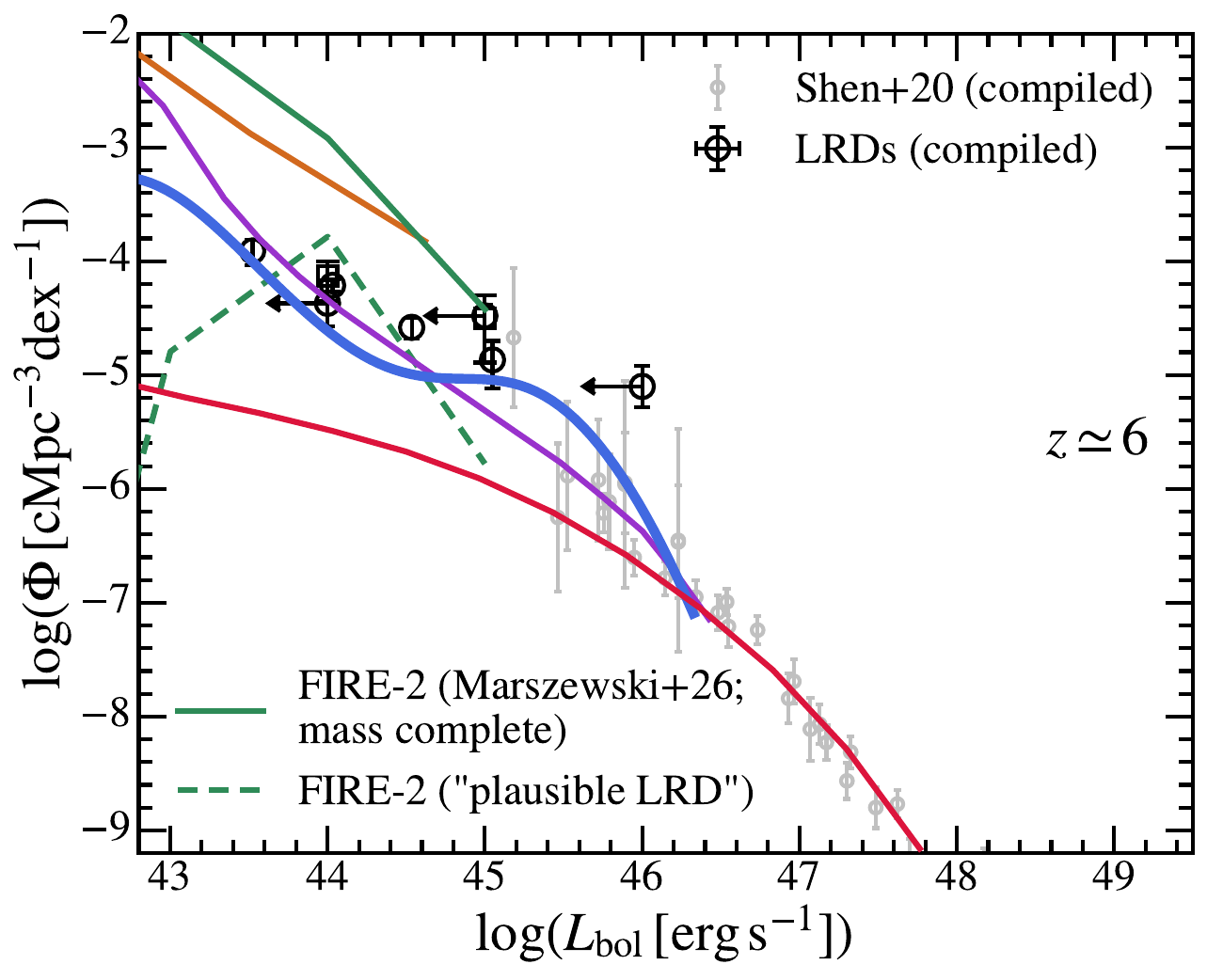}
    \caption{Comparison of AGN bolometric LFs at $z\simeq 4$ and $6$ between \luminasim and ASTRID \citep{Ni2022,Ni2025}, FLAMINGO \citep{Schaye2023flamingo,Ding2026}, BRAHMA \citep{Bhowmick2024-overmassive}, and FIRE-2 (a ``passive'' BH model; \citealt{Marszewski2026}). We include the observational data compiled in Section~\ref{subsec:bol} for comparison.}
    \label{fig:sim-comparison}
\end{figure}

Cosmological simulations adopt substantially different sub-grid prescriptions for SMBH seeding, accretion, and feedback, yielding distinct predictions for properties of SMBHs and their host galaxies both in the local Universe and at higher redshifts \citep[see e.g. the comparisons in][]{Habouzit2021,Habouzit2022}. In this section, we compare the bolometric AGN LF predicted by \luminasim with results from a representative set of recent cosmological simulations covering the high-redshift AGN population: ASTRID \citep{Bird2022,Ni2022,Ni2025}, FLAMINGO \citep{Schaye2023flamingo,Ding2026}, BRAHMA \citep{Bhowmick2024,Bhowmick2024-overmassive}, and the post-processing FIRE-2 analysis of \citet{Marszewski2026}. These simulations differ substantially in volume, mass resolution, BH seeding, accretion, and feedback prescriptions. We briefly summarize each model below before turning to the comparison.

ASTRID \citep{Bird2022,Ni2022} is a $(250\,h^{-1}\,{\rm cMpc})^3$ smoothed-particle hydrodynamics simulation with $2\times 5500^3$ resolution elements run with MP-Gadget \citep{Feng2018}. BHs are seeded in haloes with $M_{\rm halo}\gtrsim 5\times 10^9\,h^{-1}\msun$ and $M_\star\gtrsim 2\times 10^6\,h^{-1}\msun$, and the BH seed masses are drawn probabilistically from a power-law distribution between $3\times 10^4\,h^{-1}\,\msun$ and $3\times 10^5\,h^{-1}\msun$. BHs grow via Bondi--Hoyle accretion with an $\alpha=100$ boost factor to account for the underestimation of the accretion rate due to the unresolved cold and hot phase of the sub-grid ISM nearby. Mildly super-Eddington accretion is allowed, but the accretion rate is still capped at two times the Eddington rate. AGN feedback is modelled as an isotropic thermal feedback until $z\sim 2.3$ when the kinetic mode feedback is turned on. The LFs shown in Figure~\ref{fig:sim-comparison} are taken from the most recent ASTRID release \citep{Ni2025}, from which we take the results assuming $\epsilon_{\rm rad}=0.2$. 

FLAMINGO \citep{Schaye2023flamingo} is a suite of large-volume SPH simulations run with \textsc{swift} \citep{swift}. The flagship run covers a $(2.8\,{\rm Gpc})^3$ volume with $5040^3$ baryonic particles and a baryon mass resolution of $1.07\times 10^9\msun$, with smaller boxes at higher resolution. The sub-grid prescriptions for stellar and AGN feedback are calibrated using machine learning to reproduce the low-redshift galaxy stellar mass function and cluster gas fractions. BHs are seeded with $M_{\rm seed}=10^5\,\msun$ in FoF haloes with $M_{\rm halo}\gtrsim 2.757\times 10^{11} \msun$. BHs accrete at a Bondi--Hoyle rate with a density-dependent boost factor~\citep{Booth2009} and are capped at the Eddington rate. In their fiducial model, AGN feedback is modelled thermally. They assume $\epsilon_{\rm rad}=0.1$. \citet{Ding2026} apply an additional log-normal scatter of $0.75$ dex in post-processing to the bolometric luminosities before constructing the AGN LFs shown here, motivated by the comparison to the \citet{Shen2020} compilation at $z\lesssim 3$.

BRAHMA \citep{Bhowmick2024} is a suite of small-volume cosmological simulations run with \arepo using a sub-grid galaxy formation model close to that of \illustrisTNG. The recent addition~\citep{Bhowmick2024-overmassive} is designed specifically to systematically explore heavy-seed (direct-collapse-like) channels for the formation of the first massive BHs. Heavy seeds with $M_{\rm seed}\simeq 10^5\,h^{-1}\msun$ are placed in haloes that satisfy combinations of gas-density, metallicity, spin, Lyman-Werner-flux, and environmental criteria, in $(18\,{\rm cMpc})^3$ boxes. To be specific, we take their ``SM5\_LW10'' model for comparison. The small simulation volume restricts the BRAHMA LFs to $L_{\rm bol}\lesssim 10^{45}\,{\rm erg\,s^{-1}}$ in Figure~\ref{fig:sim-comparison}.

The FIRE-2 high-redshift suite~\citep{Ma2018,Ma2019} consists of zoom-in simulations of individual haloes at $z\gtrsim 5$ using the FIRE-2 model~\citep{Hopkins2018}. Active BHs are not followed self-consistently. Instead, in \citet{Marszewski2026}, hypothetical BHs are placed at the centres of resolved galaxies in post-processing (``passive'' BH model), and their accretion rates are computed from the central gas distribution using either a gravitational-torque-driven accretion prescription \citep{Hopkins2011,AA2017a,AA2017b} or a simple free-fall accretion model. Bolometric luminosities are derived assuming $\epsilon_{\rm rad}=0.1$. We take two illustrative LFs from their gravitational-torque-driven accretion model: a ``mass-complete'' LF that places a BH at the centre of every resolved host galaxy above the simulation completeness threshold, and a ``plausible LRD'' scenario that further restricts to BHs that are super-Eddington-accreting, Eddington-luminosity-limited in host galaxies with $M_{\ast} > 2\times 10^7\msun$.

This paper based on \luminasim differs from these efforts in the following aspects. First, \luminasim strikes a favourable balance between simulation volume and numerical resolution: it resolves low-mass galaxies and faint AGN more reliably than the FLAMINGO flagship run, while sampling rare luminous quasars more completely than BRAHMA, FIRE-2 zoom-in volumes, and even ASTRID. Second, the on-the-fly multi-frequency radiative transfer through \HI and \HeII reionization couples the AGN ionizing emissivity to the IGM thermal and ionization state self-consistently. The \HeII reionization is driven by the same AGN population studied here. Third, rather than restricting the comparison to BH accretion rates and bolometric LFs, we develop an empirical model that maps simulated SMBHs onto multi-band AGN observables, and demonstrate that a coherent picture of the high-redshift AGN, including LRDs, can be established within a single framework.

Returning to the bolometric LFs in Figure~\ref{fig:sim-comparison}, as discussed in earlier sections, \luminasim agrees well with the \citet{Shen2020} compilation at the bright end and with the \textit{JWST} constraints at the faint end. ASTRID also reproduces the \citet{Shen2020} compilation at $z\simeq 4$ and matches \luminasim closely at $z\simeq 6$, but underpredicts compared to the \textit{JWST}/MIRI constraints at $z\simeq 4$. We note that the ASTRID results plotted here adopt a post-processing radiative efficiency $\epsilon_{\rm rad}=0.2$ rather than the $\epsilon_{\rm rad}=0.1$ used in the original simulation, which improves the bright-end agreement at the expense of internal consistency. FLAMINGO underpredicts the bolometric LF across the full luminosity range at $z\simeq 4$, plausibly reflecting the resolution limitations highlighted in our earlier \luminasim--\mtng comparison. At $z\simeq 6$, FLAMINGO recovers the bright end but still underpredicts at the faint-end. BRAHMA is restricted by its small simulation volume and only gives predictions at the faint end. A large abundance of low-luminosity AGN at both redshifts is found in this simulation, which sits above the observational constraints of LRDs at $L_{\rm bol}\sim 10^{44}\erg\,{\rm s}^{-1}$. Finally, the mass-complete FIRE-2 LF overpredicts the faint-AGN abundance by roughly a decade, similar to BRAHMA, which indicates a rapid fuelling of gas to central BHs even in bursty and clumpy star-forming galaxies in FIRE-2. Their subsample restricted to super-Eddington-accreting, Eddington-luminosity-capped BHs in relatively massive hosts (the ``plausible LRD'' sample) reproduces the \textit{JWST} LRD LF at around $L_{\rm bol}\sim 10^{44}-10^{45} \erg\,{\rm s}^{-1}$. 

\section{Conclusions}
\label{sec:conclusions}

In this work, we use the large-volume cosmological radiation-hydrodynamic simulation \luminasim to study the high-redshift AGN population. The balance of large volume and high resolution offered by \luminasim allows us to simultaneously model some rare luminous quasars, the more abundant moderate-luminosity AGN, and the low-mass SMBH population that may be connected to the recently discovered LRDs. We construct an empirical mapping between the simulated SMBH population and observed AGN across multiple wavelength bands, and test whether the resulting AGN population is consistent with the available observational constraints from LFs, large-scale environments, and integrated properties of AGN. As described in Section~\ref{sec:smbh-agn}, this empirical model starts from the SMBH population in \luminasim and applies a modest unresolved log-normal scatter to the instantaneous bolometric luminosity, with fiducial value $\sigma_{\rm bol}=0.3$ dex (Section~\ref{subsec:lbol-scatter}). The simulated BHs are then split into two classes by mass: those above a critical mass threshold $M^{\rm crit}_{\rm BH}=\kappa M_{\rm seed}$ are assigned a canonical quasar SED with luminosity-dependent absorption/extinction corrections (Sections~\ref{subsec:sed} and~\ref{subsec:extinction}), while those below the threshold may instead enter an LRD-like early-growth phase characterized by an empirical universal LRD SED (Section~\ref{subsec:lrdsed}) and a duty-cycle parameter $f_{\rm duty}$ (Section~\ref{subsec:split}). Our fiducial model adopts $\kappa=10$ and $f_{\rm duty}=0.3$. Our main results are as follows.

\begin{enumerate}
    \item \textbf{Bolometric LFs} (Section~\ref{subsec:bol}). The bolometric AGN LFs predicted by \luminasim agree well with pre-\textit{JWST} multi-wavelength constraints at $z\simeq 3-6$ for moderately luminous quasars. The bright end is broadly converged between \luminasim and \mtng, while the faint end remains mildly resolution-dependent when compared to \tng. The stochastic luminosity scatter is essential for matching the observed abundance of luminous AGN and can be interpreted as an effective treatment of unresolved accretion variability below the resolution scale of the simulation. The faint-end bolometric LF in \luminasim matches more recent \textit{JWST} constraints that include the contribution from LRDs and MIRI-selected dust-obscured AGN.

    \item \textbf{Optical and UV LFs} (Sections~\ref{subsec:opt} and~\ref{subsec:uv}). \luminasim reproduces the LFs of both the canonical AGN population and the LRDs in the rest-frame optical and UV. In the optical, the model prediction agrees with the large abundance of LRDs at $L_{\rm opt}\sim 10^{43}-10^{44}\,\erg\,{\rm s}^{-1}$ while smoothly connecting to the pre-\textit{JWST} constraints on canonical AGN at the bright end. The comparison of optical LFs motivates our fiducial choice of $\kappa$ and $f_{\rm duty}$. In the UV, the simulation prediction again agrees with pre-\textit{JWST} constraints at the bright end, while the comparison to LRDs depends sensitively on the interpretation of the physical origin of the UV emission. We explore one scenario in which the LRD UV emission is significantly contributed by the host galaxies. An empirical decomposition of AGN and host-galaxy light in UV inferred from recent observations results in good agreement between \luminasim predictions and observed LRD UV LFs. However, the simulated hosts of LRD-mass BHs are often too UV- and optically luminous to satisfy observational LRD selection criteria, highlighting an important limitation of theoretical models in reproducing the AGN-host galaxy scaling relations at high redshifts.

    \item \textbf{X-ray and mid-IR LFs} (Sections~\ref{subsec:xray} and~\ref{subsec:midir}). The hard X-ray LFs of canonical AGN (assuming no LRD contribution) in \luminasim broadly reproduce the observed LFs at $z\simeq 3-6$, including the flattening towards the faint end. The model slightly overpredicts faint X-ray AGN at $z\lesssim 4$, although this should be interpreted with caution given that current X-ray surveys may still miss a larger population of heavily obscured or intrinsically X-ray weak AGN than previously accounted for. In the mid-IR, \luminasim appears to contain a more complete sample of AGN compared to extrapolations of pre-\textit{JWST} LF constraints, in agreement with recent \textit{JWST}/MIRI constraints that capture even the X-ray weak and previously missed sources.

    \item \textbf{Large-scale environments of LRDs} (Section~\ref{subsec:environment}). The projected cross-correlation between LRDs and star-forming galaxies in \luminasim at $z\simeq 4$ and $z\simeq 5$ agrees well with recent \textit{JWST} survey results. The clustering amplitudes are consistent with characteristic LRD host halo masses slightly above $10^{11} \msun$, similar to those inferred observationally. Therefore, the LRD-like population appears to be associated with relatively common, moderate-mass haloes rather than the rare massive haloes hosting luminous quasars. This serves as an important cross-check of our empirical LRD model, which is largely calibrated based on optical LRD LFs. However, halo assembly bias may complicate a direct mapping between clustering amplitude and halo mass if LRD formation preferentially selects haloes with unusual secondary properties.

    \item \textbf{AGN in the cosmological context} (Section~\ref{subsec:cosmo}). The predicted number density of observable LRDs is broadly consistent with current constraints at $z\simeq 4-6$. The predicted LRD abundance continues to rise towards lower redshifts while the observed number densities show an apparent decline at $z\lesssim 4$. Although we show that observational bias and selection effects can play a role, this may indicate limitations of our BH seeding or LRD model, which could miss additional criteria uniquely associated with the high-redshift environment. The total BHAD in \luminasim is broadly consistent with recent \textit{JWST}/MIRI constraints at $z\lesssim 5$ and is higher than the value inferred from previous X-ray and mid-IR surveys. The LRD contribution is non-negligible at $z\simeq 4-6$ but remains below the total AGN contribution and is closer to estimates based on updated LRD bolometric corrections than to upper limits derived from canonical quasar corrections. The accumulated SMBH mass density at high redshifts remains compatible with canonical local SMBH mass density constraints when we subtract the integrated mass gain from low-redshift quasars (based on the Soltan argument). The full AGN population in \luminasim, including both canonical AGN and the LRD phase, is therefore consistent with the global SMBH growth budget. 

    \item \textbf{AGN-driven \HeII reionization} (Section~\ref{subsec:cosmo}). With multi-frequency radiative transfer followed on the fly, \luminasim predicts a \HeII reionization history that begins around $z\simeq 6$ and completes by $z\simeq 3$, driven essentially by AGN and broadly consistent with observational constraints. The \HeII-ionizing photon budget during the bulk of this epoch is dominated by SMBHs with $M_{\rm BH}\simeq 10^{7}-10^{9}\msun$. Crucially, BHs below our adopted LRD mass threshold contribute negligibly to the \HeII-ionizing emissivity over this redshift range. 
\end{enumerate}

Taken together, these results show that \luminasim provides a coherent cosmological framework for interpreting the high-redshift AGN population. The simulation can simultaneously reproduce the abundance of classical quasars alongside the emerging population of faint AGN revealed by \textit{JWST}, the large-scale environments of AGN, the integrated SMBH growth rates, and the reionization of \HeII in the IGM driven by the same SMBH population, all within a single self-consistent framework. In this picture, LRDs are considered as a possible early-growth phase of SMBHs. The observational evidence collected for LRDs through LFs at multiple wavelengths and large-scale clustering is mutually consistent, with minimal assumptions motivated empirically from observations. Meanwhile, the remaining canonical AGN component stays essential for matching the X-ray, mid-IR, and bright-end AGN optical/UV LF constraints, as well as supplying the bulk of the \HeII-ionizing photon budget. Several uncertainties remain, as discussed in detail in Section~\ref{subsec:model-assumptions} and \ref{subsec:parameters}. The few free parameters in the AGN empirical model and the BH seeding and accretion prescriptions are simplified treatments of unresolved physics with some level of degeneracies. The role of \luminasim in this work is to provide a cosmological backbone within which a physically plausible, effective LRD phase can be tested against the abundance, environments, and other observables of SMBHs simultaneously. Future progress will require a larger and more complete observational sample and more predictive theoretical models.


\section*{Acknowledgements}
We thank Pablo G. P\'{e}rez-Gonz\'{a}lez, Xiaojing Lin, Jenny Greene, David Setton, and Wendy Sun for sharing their stacked SED data with us and for insightful discussions. We thank Claude-Andr\'{e} Faucher-Gigu\`{e}re, Philip Hopkins, and Gordon Richards for their valuable feedback in discussions. We also thank Zihao Wang for sharing the scripts of clustering measurements.
An award of computer time was provided by the INCITE program. This research used resources of the Oak Ridge Leadership Computing Facility at the Oak Ridge National Laboratory, which is supported by the Advanced Scientific Computing Research programs in the Office of Science of the U.S. Department of Energy under Contract No.\ DE-AC05-00OR22725.
Support for programs JWST-AR-04814 (XS, MV) and JWST-AR-08709 (AS) was provided by NASA through a grant from the Space Telescope Science Institute, which is operated by the Association of Universities for Research in Astronomy, Inc., under NASA contract NAS 5-03127.
Support for OZ was provided by Harvard University through the Institute for Theory and Computation Fellowship.
RK acknowledges support from the Natural Sciences and Engineering Research Council of Canada (NSERC) through a Discovery Grant and a Discovery Launch Supplement, funding reference numbers RGPIN-2024-06222 and DGECR-2024-00144, and from York University's Global Research Excellence Initiative.
MV acknowledges support through NASA ATP Grant 23-ATP23-149 and NSF AAG Grant AST-2307699.
This work was supported by the LOEWE programme of the State of Hesse (LOEWE-Spitzen-Professur).
VS and LH acknowledge support from the Simons Foundation through the ``Learning the Universe" initiative.
LCH was supported by the National Science Foundation of China (12233001), the National Key R\&D Program of China (2022YFF0503401), and the China Manned Space Program (CMS-CSST-2025-A09).

\section*{Data Availability}
All \luminasim data will be made publicly available in 2027.
Some of the data presented in the figures of this article are already available at \url{https://www.lumina-simulation.com}.
Additional data are available from the corresponding author upon reasonable request.



\bibliographystyle{mnras}
\bibliography{ref} 




\appendix

\section{Optical LF at 5100\,\AA}

\begin{figure}
    \centering
    \includegraphics[width=\linewidth]{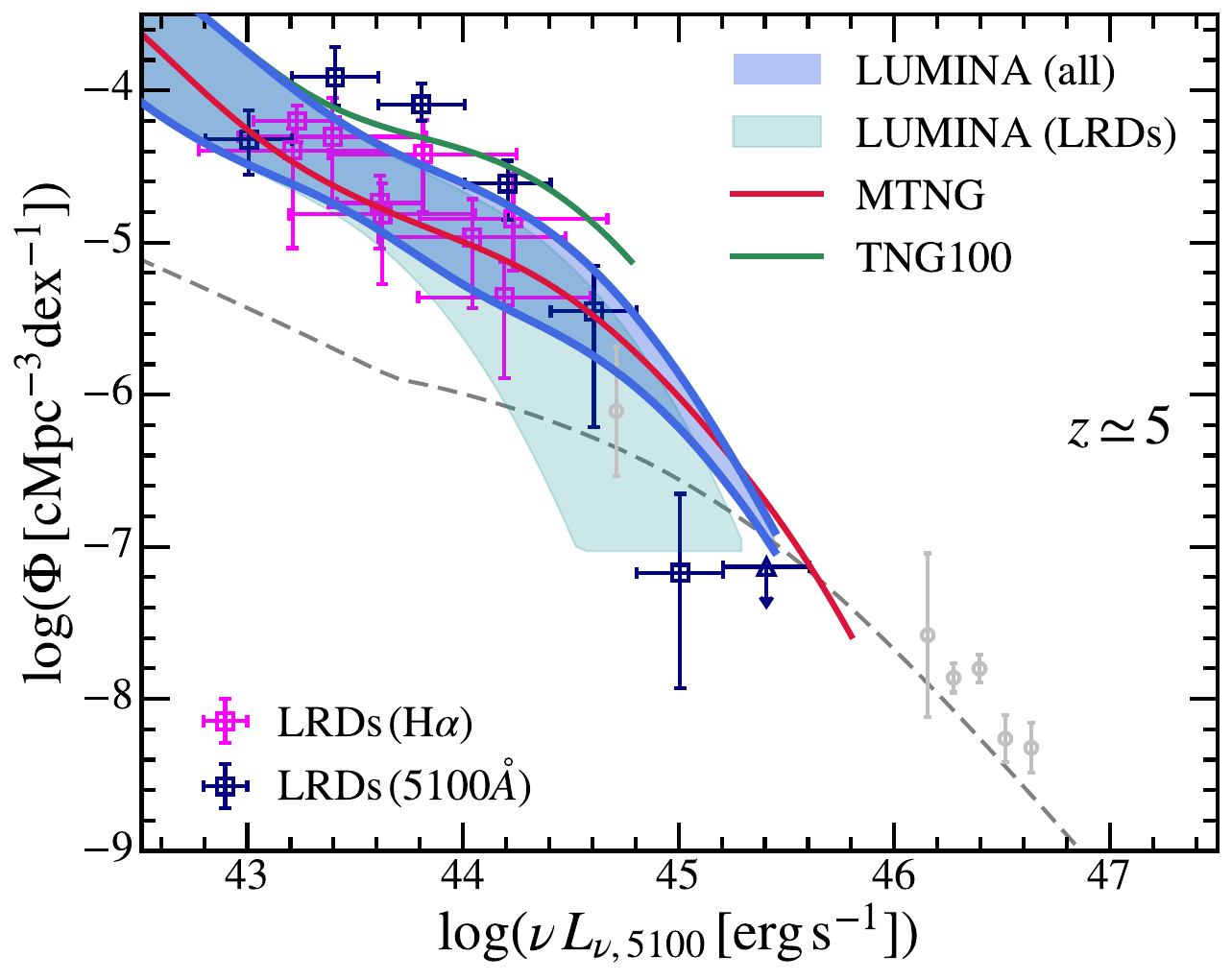}
    \caption{Optical ($\sim 5100$\AA) AGN LFs in \luminasim, \tng, and \mtng. We compare the simulation predictions with the same set of observations shown in Figure~\ref{fig:optlf} in the main text. Our conclusions about optical LF are essentially unchanged.}
    \label{appfig:5100lf}
\end{figure}

In Section~\ref{subsec:opt}, we present the AGN LFs in the rest-frame optical B band. Here we show the same LFs instead at $\lambda\sim 5100$\,\AA, the wavelength at which many observational studies of LRDs report their LFs \citep{Kokorev2024,Ma2025}. Since both our model predictions and the compiled quasar LF data \citep{Shen2020} are defined in the B band, we apply a small constant offset of $-0.04$ dex to convert them to $5100$\,\AA, derived from our quasar SED template. Figure~\ref{appfig:5100lf} shows the resulting $5100$\,\AA\ AGN LF in \luminasim, compared to \tng and \mtng. We include the original measurements from \citet{Kokorev2024,Ma2025} together with the H$\alpha$-based constraints of \citet{Matthee2024,Lin2024,Lin2026}, the latter converted to $L_{5100}$ using the same scaling relation adopted in the main text. The conclusions reached in Section~\ref{subsec:opt} remain unchanged.

\bsp	
\label{lastpage}
\end{document}